\documentclass[aps, pra, twocolumn, nofootinbib, superscriptaddress]{revtex4-2}

\usepackage{comment}
\usepackage{color}
\usepackage{cancel}
\usepackage{ulem}

\usepackage{graphicx}

\usepackage{amsfonts}
\usepackage{amsmath}

\usepackage{physics}
\usepackage{mathrsfs}  

\begin{document}

	
	\title{Topological study of a Bogoliubov-de Gennes system of pseudo-spin-$1/2$ bosons with conserved magnetization in a honeycomb lattice } 
	
	\author{Hong Y. Ling}
	\affiliation{Department of Physics and Astronomy, Rowan University, Glassboro, New Jersey 08028, USA}
	\author{Ben Kain}
	\affiliation{Department of Physics, College of the Holy Cross, Worcester, Massachussets 01610, USA}
	\date{\today}
	
	\begin{abstract}
		
		We consider a Bogoliubov-de Gennes (BdG) Hamiltonian, which is a non-Hermitian Hamiltonian with pseudo-Hermiticity, for a system of (pseudo) spin-$1/2$ bosons in a honeycomb lattice under the condition that the population difference between the two spin components, i.e., magnetization, is a constant.  Such a system is capable of acting as a topological amplifier, under time-reversal symmetry, with stable bulk bands but unstable edge modes which can be populated at an exponentially fast rate.  We quantitatively study the topological properties of this model within the framework of the 38-fold way for non-Hermitian systems. We find, through the symmetry analysis of the Bloch Hamiltonian, that this model is classified either as two copies of symmetry class $\mbox{AIII}+\eta_-$ or two copies of symmetry class $\mbox{A}+\eta$ depending on whether the (total) system is time-reversal symmetric, where $\eta$ is the matrix representing pseudo-Hermiticity and $\eta_-$ indicates that pseudo-Hermiticity and chiral symmetry operators anticommute. We prove, within the context of non-Hermitian physics where eigenstates obey the bi-orthonormality relation, that a stable bulk is characterized by a single topological invariant, the Chern number for the Haldane model, independent of pairing interactions.  We construct a convenient analytical description for the edge modes of the Haldane model in semi-infinite planes, which is expected to be useful for models built upon copies of the Haldane model across a broad array of disciplines.  We adapt the theorem in our recent work [Ling and Kain, Phys. Rev. A {\bf 104}, 013305 (2021)] to pseudo-Hermitian Hamiltonians that are less restrictive than BdG Hamiltonians and apply it to highlight that the vanishing of an unconventional commutator between number-conserving and number-nonconserving parts of the Hamiltonian indicates whether a system can be made to act as a topological amplifier.
	\end{abstract}
	
	\pacs{Valid PACS appear here}
	\maketitle
	
	
	\section{introduction}

	Symmetry principles play a pivotal role in our understanding of topological matter \cite{hasan10RevModPhys.82.3045,qi2011RevModPhys.83.1057}, i.e., systems whose edge modes are guaranteed to be robust against disorder by bulk topological invariants.
	Haldane in 1988 \cite{haldane88PhysRevLett.61.2015} first recognized that it is not so much a net magnetic field but rather the breaking of time-reversal symmetry that is responsible for the integer quantum Hall (IQH) effect \cite{klitzing1980PhysRevLett.45.494,thouless1982PhysRevLett.49.405}. 
	Kane and Melee in 2005 \cite{kane2005PhysRevLett.95.146802,kane2005PhysRevLett.95.226801} showed that the anomalous quantum spin Hall (QSH) effect \cite{murakami04PhysRevLett.93.156804} can arise from a time-reversal invariant system of spin-$1/2$ fermions.  The formulation of the QSH effect in terms of topology \cite{kane2005PhysRevLett.95.146802} along with the experimental realization of the ideas behind the Kane-Melee model \cite{kane2005PhysRevLett.95.146802} in HgTe quantum wells \cite{konig2007Science.318.766} following the theoretical proposal by Bernevig et al. \cite{bernevig2006Science.314.1757} has motivated the creation of a periodic table, called the 10-fold way \cite{schnyder2008PhysRevB.78.195125,kitaev2009AIP.1134.22,Ryu2010NewJournalOfPhysics.12.065010}, of topological insulators (and superconductors) for closed systems of fermions described by quadratic Hermitian Hamiltonians \cite{shen2012,bernevig2013}.
	
	A  topological insulator has a bulk characterized by a topologically non-trivial invariant, which is determined globally by the bulk Bloch band structure.  Since Bloch bands are universal in periodic systems, topological phenomena are ubiquitous. This realization has generated an enormous interest in topological phases of matter across a wide variety of systems, including photons \cite{haldane08PhysRevLett.100.013904,wang2008PhysRevLett.100.013905,wang2009Nature.461.772,hafezi2011NaurePhysics.7.907,kejie2012NatPhoton2012},
	phonons (and mechanical metamaterials) \cite{kane2013NaturePhysics.10.39,prodan2009PhysRevLett.103.248101,fleury2014Science.343.516,yang2015PhysRevLett.114.114301,peano2015PhysRevX.5.031011,he2016NatPhys.12.1124}, magnons \cite{shindou2013PhysRevB.87.174402,shindou2013PhysRevB.87.174427}, and ultracold atoms \cite{aidelsburger2013NatPhys.9.795,aidelsburger2013PhysRevLett.111.185301,hirokazu2013PhysRevLett.111.185302,jotzu2014Nature.515.237,Stuhl2015Science.349.1514}
	(see recent review articles \cite{cooper2019RevModPhys.91.015005,ozawa2019RevModPhys.91.015006,zhang2018CommunPhys.1.97,kondo2020} for additional references).  In fact, a large body of studies in recent years has been devoted to topological phenomena in non-Hermitian Hamiltonians for systems that may not be closed, fermionic, or even in thermal equilibrium \cite{rudner2009PhysRevLett.102.065703,esaki2011PhysRevB.84.205128,liang2013PhysRevA.87.012118,lee2016PhysRevLett.116.133903,leykam2016PhysRevLett.117.143901,xiong2018JPhysCommun.2.035043,shen2018PhysRevLett.120.146402,yao2018PhysRevLett.121.086803} (see Refs. \cite{gong2018PhysRevX.8.031079,kawabata2019PhysRevX.9.041015,zhou2019PhysRevB.99.235112,ghatak2019JOfPhysCondMatt.31.263001,ashida2020} and references therein). 
	
	The prospect of topological classes unique to non-Hermitian Hamiltonians \cite{esaki2011PhysRevB.84.205128,kawabata2019NatureCommunications.10.297,gong2018PhysRevX.8.031079} calls for systematic classification of topological symmetry classes beyond the 10-fold way paradigm. 
	The 10-fold way is founded on the celebrated Altland-Zirnbauer (AZ) theory \cite{altland1997PhysRevB.55.1142}  which groups quadratic Hermitian Hamiltonians according to three fundamental symmetries, i.e., time-reversal, particle-hole, and chiral symmetry, into 10 classes \cite{schnyder2008PhysRevB.78.195125,kitaev2009AIP.1134.22,Ryu2010NewJournalOfPhysics.12.065010}, a direct generalization of the Wigner-Dyson symmetry classes \cite{dyson1962JMathPhys.3.1199} in random matrix theory  \cite{mehta1991}.  There exists, also in random matrix theory, a non-Hermitian analog of the AZ classification, due to Bernard and LeClair (BL) \cite{bernardAmdLeClair2002.207}, based on four fundamental symmetries \cite{sato2012ProgressOfTheoreticalPhysics.127.937,lieu2018PhysRevB.98.115135} --- three AZ symmetries plus a so-called Q symmetry which distinguishes between Hermitian and non-Hermitian matrices. As such, the BL classification contains two analogs of time-reversal and particle-hole symmetries (C and K symmetries) which are identical for Hermitian matrices but are distinct for non-Hermitian matrices, leading, in large part, to the proliferation of the symmetry classes in non-Hermitian systems.  In fact, following recent studies \cite{esaki2011PhysRevB.84.205128,lieu2018PhysRevB.98.115135,gong2018PhysRevX.8.031079} which traced some symmetry classes of non-Hermitian Hamiltonians to the BL scheme, Kawabara et al. \cite{kawabata2019PhysRevX.9.041015} and Zhou and Lee \cite{zhou2019PhysRevB.99.235112} categorized non-Hermitian Hamiltonians into a total of 38 classes, thereby giving rise to the so-called (BL-based) 38-fold way, a non-Hermitian generalization of the (AZ-based) 10-fold way \cite{schnyder2008PhysRevB.78.195125,kitaev2009AIP.1134.22,Ryu2010NewJournalOfPhysics.12.065010}.
	
	In this paper, we focus on a quadratic bosonic Hamiltonian, which, as in other quadratic bosonic models  \cite{barnett13PhysRevA.88.063631,galilo15PhysRevLett.115.245302,engelhardt2016PhysRevLett.117.045302,peano2016PhysRevX.6.041026}, contains pairing terms (or interactions) that break particle number conservation, making it a bosonic analog of the Bogoliubov-de Gennes (BdG) Hamiltonian for fermions. However, in contrast to the fermionic BdG model, where the matrix to be diagonalized is Hermitian, the matrix to be diagonalized in the bosonic BdG model is non-Hermitian.   As we shall see in later sections,  our bosonic BdG Hamiltonian and the sub-Hamiltonians that are descended from it belong to a special class of non-Hermitian Hamiltonians which obey  pseudo-Hermiticity, whereby a Hamiltonian differs from its Hermitian conjugate by a similarity transformation characterized by a unitary and Hermitian matrix $\eta$.  This is a special version of the Q symmetry in the BL classification and thus our Hamiltonian and its associated sub-Hamiltonians are grouped according to the 38-fold way classification of Hamiltonians with pseudo-Hermiticity.

	The purpose of the present work is to conduct, within the framework of the 38-fold way, a systematic study of the topological properties of such a quadratic bosonic Hamiltonian.  Our paper is organized as follows. In Sec. \ref{sec:Model}, we present our model Hamiltonian --- a generic BdG Hamiltonian for spin-$1/2$ bosons in a honeycomb lattice with both periodic boundaries (Sec. \ref{sec:The Model with Periodic Boundaries}) and open boundaries (Sec. \ref{sec:The Model with Open Boundaries}).  We give a brief description of a cold atom realization in an optical lattice with a spin-1 condensate, while relegating the details to Appendix \ref{sec: detailed description of our model}.
	
	In Sec. \ref{sec:Haldane Model}, we visit the Haldane model to which our system traces its topological origin.  We review the Chern number in Sec. \ref{sec:Bulk Topology: Chern Number}. We derive in \ref{eq:Edge Modes Haldane Model} a set of (relatively) simple analytical results describing edge modes in the Haldane model in semi-infinite planes, which are expected to be useful for understanding edge modes not only in our model but also in many other models that are built on copies of the Haldane model. 
	
	In Sec. \ref{sec:our model without time reversal symmetry}, we generalize the study in Sec. \ref{sec:Haldane Model} to our model in the absence of time-reversal symmetry. A unitary symmetry due to magnetization conservation in our model partitions the Hilbert space into two independent sectors so that the total Hamiltonian inherits the topologies of  the sub-Hamiltonians in the smaller sectors.  In Sec. \ref{sec:symmetries and topological classification}, we show that these sub-Hamiltonians are class A$+\eta$ systems with pseudo-Hermiticity. We also show, with the help of the bi-orthonormality relation involving left- and right-eigenstates, that the bulk topology is indeed characterized by a single bulk invariant, the Chern ($\mathbb{Z}$) number, as predicted by the 38-fold way.    Furthermore, we prove that for a stable bulk, this single Chern number is independent of pairing interactions and corresponds to the Chern number for the Haldane model.
	In Sec. \ref{sec:edge modes for our model}, we investigate the energy spectrum of a sub-Hamiltonian for systems with open boundaries and apply the simple results in Sec. \ref{eq:Edge Modes Haldane Model} to gain analytical insights into the edge modes that lie inside the gaps of particle and hole bulk bands.
	
	In Sec. \ref{sec:our model with time reversal symmetry}, we include time-reversal symmetry in our system, which has been investigated previously in connection with exponential amplification of the edge modes with \cite{galilo15PhysRevLett.115.245302} and without \cite{ling2021PhysRevA.104.013305} inversion symmetry. In Ref. \cite{ling2021PhysRevA.104.013305}, we used this model to illustrate a theorem for creating topological amplifiers with stable bulks in BdG Hamiltonians. In Appendix \ref{sec:a generalized theorem}, we generalize this theorem to  sub-Hamiltonians which are not BdG Hamiltonians, and apply it in Sec. \ref{sec:our model with time reversal symmetry} to prove that, as long as time-reversal symmetry is preserved, our system can act as a topological amplifier with its bulk spectrum free of instabilities.  Furthermore, we offer a detailed explanation of why the total system should now be classified as two copies of class AIII$+\eta_-$ systems, where $\eta_-$ indicates that the matrix representing pseudo-Hermiticity anticommutes with the matrix representing chiral symmetry.  Here, the bulk topology is described again by the Chern ($\mathbb{Z}$) number for the Haldane model, independent of pairing interactions.
	
	In Sec. \ref{sec:conclusion}, we conclude our work.
	
	\section{Our Model: Tight binding Hamiltonians}
	\label{sec:Model}
	\begin{figure}[ht]
		\centering
	\includegraphics[width=0.42\textwidth]{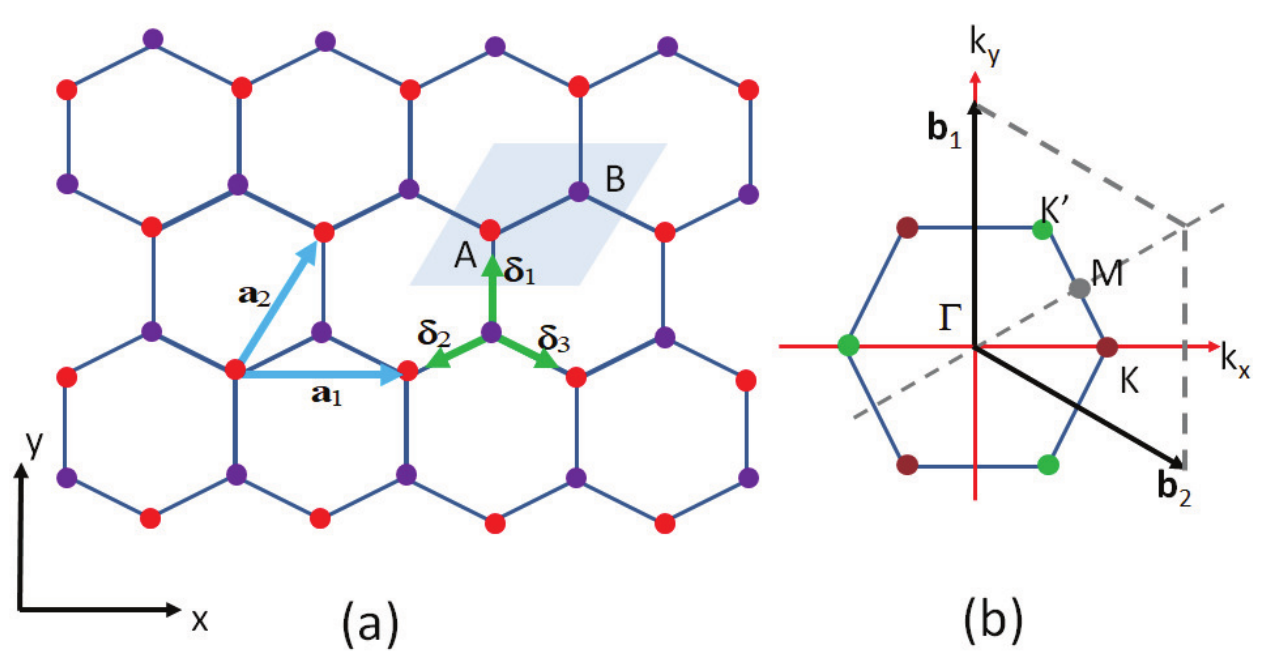}
		\caption{(a) A honeycomb lattice model consisting of an A sublattice and a B sublattice where $\vb{a}_1 $ and $\vb{a}_2$ are the base vectors for a unit cell, $\delta_1, \delta_2 $, $\delta_3$ are the vectors between a site and its three nearest neighbors. The separation between two adjacent sites on the same sublattice is chosen as the distance unit. (b) The Brillouin zone of a honeycomb lattice where $\vb{b}_1$ and $\vb{b}_2$ are the base vectors of a unit cell in the reciprocal (momentum) space. Indicated are high symmetric points in the Brillouin zone: $\Gamma$, $M$, and two inequivalent Dirac points $K$ and $K'$.
		}
		\label{fig:fig1}
	\end{figure}

	We consider a system of pseudo-spin-$1/2$ bosons in a honeycomb lattice described, in position space, by the tight-binding Hamiltonian,
	\begin{equation}
		\label{eq:H_half}
		\begin{split}
			\hat{H}_{1/2} = & -t_1 \sum_{\left< \vb{ij} \right>} \hat{b}_{\vb{i}}^\dag s_0 \hat{b}_{\vb{j}} + t_2\sum_{\left<\left< \vb{ij} \right> \right>} e^{-i\nu_{\vb{ij}} \phi} \hat{b}_{\vb{i}}^\dag s_z \hat{b}_{\vb{j}} \\
			&+ q \sum_{\vb{i}} \hat{b}_{\vb{i}}^\dag s_0 \hat{b}_{\vb{i}} 
			+ m \sum_{\vb{i}} \xi_{\vb{i}}  \hat{b}_{\vb{i}}^\dag s_0 \hat{b}_{\vb{i}}\\
			&+\frac{c}{2} \sum_{\vb{i}}(\hat{b}_{\vb{i}} s_x \hat{b}_{\vb{i}} +\hat{b}^\dag_{\vb{i}} s_x \hat{b}^\dag_{\vb{i}}),
		\end{split}
	\end{equation}
	where $\hat{b}_{\vb{j}} = (\hat{b}_{\vb{j},\uparrow},\hat{b}_{\vb{j},\downarrow})$ is the two-component field operator at site $\vb{j}$ and $s_{x,y,z}$ and $s_0$ are, respectively, the Pauli matrices and the identity matrix in (pseudo) spin space,
	\begin{equation}
		\label{eq:spin space}
		\mathscr{H}_s = (\ket{\uparrow},\ket{\downarrow}).
	\end{equation}
	In Eq. (\ref{eq:H_half}),
	$t_1$ is the nearest-neighbor hopping amplitude, $t_2e^{-i\nu_{\vb{ij}}\phi}$ is the next-nearest-neighbor hopping amplitude introduced by Haldane \cite{haldane88PhysRevLett.61.2015}, where $\nu_{\vb{ij}}=\pm 1$ alternates periodically in the manner of Kane and Mele \cite{kane2005PhysRevLett.95.146802}, $c$ denotes the strength of pairing interactions, and $q$ and $m$ measure, respectively, the nonstaggered and staggered onsite potential with $\xi_{\vb{i}}=+1$ for sites on sublattice A and $-1$ for sites on sublattice B.
	
	Equation (\ref{eq:H_half}) is a generic Hamiltonian and may be engineered in different physical systems.  Galilo et al. \cite{galilo15PhysRevLett.115.245302}  proposed in 2015 a realization in an ultracold atomic gas of interacting spin-1 bosons in a honeycomb lattice, which was inspired, in large part, by the cold atom implementation of the Haldane model by Esslinger group \cite{jotzu2014Nature.515.237}. The present study is based on a generalization of their model described in Appendix \ref{sec: detailed description of our model}.
	Here, the spin-$1/2$ system is a subsystem consisting of spin-$\pm 1$ excitations which are generated by quenching the onsite potential in such a manner that it does not affect the spin-$0$ component, which is stable and decoupled from the spin-$1/2$ system.  In this context, Eq. (\ref{eq:H_half}) is the postquench Hamiltonian for the spin-$1/2$ system, which we recently employed to test a criterion for creating a topological amplifier with a bulk free of dynamical instability \cite{ling2021PhysRevA.104.013305}.

	\subsection{The model with periodic boundaries}
	\label{sec:The Model with Periodic Boundaries}
	To explore the bulk band topology, we consider a honeycomb lattice with periodic boundaries and move from position space in Fig. \ref{fig:fig1}(a) to momentum $\vb{k}$ space in Fig. \ref{fig:fig1}(b) where the Hamiltonian is diagonal.
	
	In the absence of pairing interactions, the (Bloch) Hamiltonian in momentum space simply consists of two independent copies of the Haldane model, with
	\begin{equation}
		\label{eq:hadane model for spin up} 
		h_\uparrow(\vb{k}) =  h_0(\vb{k})\sigma_0 + \vb{h}(\vb{k})\cdot \vb{\sigma}
	\end{equation}
	for the spin-$\uparrow$ copy and 
	\begin{equation}
		\label{eq:hadane model for spin down} 
		h_\downarrow(\vb{k}) = h'_0(\vb{k})\sigma_0 + \vb{h'}(\vb{k})\cdot \vb{\sigma}
	\end{equation}
	for the spin-$\downarrow$ copy, where $\sigma_{x},\sigma_{y},\sigma_{z}$, and $\sigma_0$ are the Pauli and identity matrices, respectively, in sublattice space,
	\begin{equation}
		\label{eq:sublattice space}
		\mathscr{H}_\sigma = (\ket{A}, \ket{B}).
	\end{equation}
	In Eqs. (\ref{eq:hadane model for spin up}) and (\ref{eq:hadane model for spin down}),  
	\begin{equation}
		\label{eq:h_0(k)}
		h_0(\vb{k}) =  q + \epsilon(\vb{k}), \quad h'_0(\vb{k}) =  q - \epsilon(\vb{k}),
	\end{equation}
	are purely real, where
	\begin{equation}
		\epsilon(\vb{k})=2t_2\cos\phi \Re\Lambda_{\vb{k}},
	\end{equation}
	and 
	\begin{equation}
		\label{eq:three h components}
		\begin{split}
			h_x(\vb{k}) = &h'_x(\vb{k})= -t_1 \Re \Gamma_{\vb{k}},\\
			h_y(\vb{k}) = &h'_y(\vb{k})= +t_1 \Im \Gamma_{\vb{k}},\\
			h_z(\vb{k}) = & h'_z(-\vb{k}) = m -2t_2\sin\phi \Im \Lambda_{\vb{k}},
		\end{split}
	\end{equation}
	are the components of $\vb{h}(\vb{k})$ and $\vb{h}'(\vb{k})$, which are also purely real,
	where
	\begin{align}
		\Gamma_{\vb{k}} &=e^{i\frac{k_y}{\sqrt{3}}} + 2\cos \frac{k_x}{2}e^{-i\frac{k_y}{2\sqrt{3}}},\label{eq:Gamma(k)}\\
		\Lambda_{\vb{k}} &=  e^{ik_x}+2e^{-i\frac{k_x}{2}}\cos\frac{\sqrt{3}k_y}{2}.
	\end{align}
	
	In the presence of pairing interactions, the Hamiltonian can no longer be separated into two independent spin copies.  Let $\hat{\psi}_{\vb{k}} = (\hat{b}_{\vb{k}},\hat{b}^\dag_{-\vb{k}})$ be the Nambu spinor, where $\hat{b}_{\vb{k}}$  and its Hermitian conjugate are the vector fields in the tensor product of spin and sublattice spaces: $\mathscr{H}_s \otimes \mathscr{H}_\sigma$.  The total Hamiltonian up to a constant reads
	\begin{equation}
		\label{eq:quadratic H_half for bosons} 
		\hat{H}_{1/2}= \frac{1}{2}\sum_{\vb{k}} \hat{\psi}^\dag_{\vb{k}} \Sigma_z H_{BdG}(\vb{k}) \hat{\psi}_{\vb{k}},
	\end{equation}
	where $\Sigma_z$ is a matrix defined directly below Eq. (\ref{eq:paraunitary transformation}) and $H_{BdG}(\vb{k})$ is given explicitly in Eq. (\ref{eq:H_BdG first}). In this work, we call matrix $H_{BdG}(\vb{k})$ the bosonic BdG Hamiltonian, which is typically non-Hermitian, and matrix
	$\Sigma_z H_{BdG}(\vb{k})$ the ``first quantized" Hamiltonian, which is guaranteed to be Hermitian by Bose statistics.
	
	$\hat{H}_{1/2}$ in Eq. (\ref{eq:quadratic H_half for bosons}), which is a bosonic analog of the fermionic BdG Hamiltonian that lies at the heart of the Bardeen-Cooper-Schrieffer description of superconductivity \cite{schrieffer64Book}, can be diagonalized in terms of the quasiparticle field operator $\hat{\phi}_{\vb{k}}$ defined as
	\begin{equation}
		\hat{\phi}_{\vb{k}} = T \hat{\psi}_{\vb{k}}   
	\end{equation}
	with the help of the generalized Bogoliubov transformation $T$.  The requirement that $T$ be canonical, i.e., $T$ leaves the Bose commutator invariant, dictates that $T$ be paraunitary, i.e.,
	\begin{equation}
		\label{eq:paraunitary transformation}
		T \Sigma_z T^\dag = \Sigma_z,
	\end{equation}
	where $\Sigma_z = \tau_z \otimes s_0\otimes \sigma_0$.  Here, we have defined $\tau_{x}, \tau_{y}, \tau_{z}$, and $\tau_0$ as the Pauli and identity matrices, respectively, in Nambu space.
	
	The quadratic bosonic Hamiltonian (\ref{eq:quadratic H_half for bosons}) is thus diagonalized by a paraunitary transformation, in sharp contrast with its fermionic counterpart which is diagonalized by a unitary transformation.  As a consequence, it is the matrix
	\begin{equation}
		\label{eq:H_BdG first}
		{H}_{BdG}(\vb{k}) =
		\begin{pmatrix}
			A(\vb{k}) & B(\vb{k}) \\
			-B^*(-\vb{k}) & -A^*(-\vb{k})
		\end{pmatrix},
	\end{equation}
	that is to be diagonalized, where
	\begin{equation}
		A(\vb{k}) = \mathbb{P}_\uparrow \otimes h_\uparrow(\vb{k}) +\mathbb{P}_\downarrow \otimes  h_\downarrow(\vb{k}), \quad B(\vb{k}) = cs_x \otimes \sigma_0,
	\end{equation}
	and
	\begin{equation}
		\label{eq:projection operators}
		\mathbb{P}_{\uparrow/\downarrow} = \frac{1\pm s_z}{2}, \quad \mathbb{P}_{+/-} = \frac{1\pm\tau_z}{2},
	\end{equation}
	are, respectively, the projection operators in spin and Nambu space (we shall use $\mathbb{P}_{+/-}$ in later sections). One may easily verify that
	\begin{equation}
		A(\vb{k}) = A^\dag(\vb{k}), \quad B(\vb{k}) = B^T(-\vb{k}),
	\end{equation} which are guaranteed by the Hermiticity of $\Sigma_z H_{BdG}(\vb{k})$.

	\subsection{The model with open boundaries}
	\label{sec:The Model with Open Boundaries}

	Topologically protected edge modes for a system with open boundaries follow from the existence of a topologically nontrivial invariant, which itself follows from the bulk band structure.  To examine edge properties, we consider the same honeycomb lattice but in a stripe geometry with open boundaries along $y$ and periodic (zigzag) boundaries along $x$ so that the $x$-momentum $k_x$ remains a good quantum number.  Let $N_x$ ($N_y$) be the number of unit cells along $x$ ($y$) and $X_{\vb{j}}$ be the x-component of the position vector at site $\vb{j}$. Applying a partial Fourier transformation along $x$,
	\begin{equation}
		\hat{b}_{\vb{j}\equiv(j_x,j_y)} =\sum_{k_x} \hat{b}_{k_x,j_y} \frac{e^{ik_x X_{\vb{j}}}}{\sqrt{N_x}}, \quad j_y = 1, 2, \cdots, N_y,
	\end{equation}
	we change Eq. (\ref{eq:H_half}) to the corresponding Hamiltonian in the stripe geometry,
	\begin{equation}
		\label{eq:h half kx}
		\hat{H}_{1/2} =\frac{1}{2}\sum_{k_x}\hat{\psi}^\dag_{k_x} \Sigma_z {H}_{BdG}(k_x) \hat{\psi}_{k_x},
	\end{equation}
	where $\hat{\psi}_{k_x} = (\hat{b}_{k_x},\hat{b}^\dag_{-k_x})$ is the Nambu spinor with $\hat{b}_{k_x}$ and its Hermitian conjugate the vector fields in $ \mathscr{H}_I\otimes \mathscr{H}_s \otimes \mathscr{H}_\sigma$, where 
	\begin{equation}
		\label{eq:unit cell space}
		\mathscr{H}_I= (\ket{1},\ket{2},\cdots, \ket{N_y})
	\end{equation}
	is the ($y$) unit-cell space and $\mathscr{H}_s$ and $\mathscr{H}_\sigma$ are  spin and sublattice spaces previously defined in Eq. (\ref{eq:spin space}) and Eq. (\ref{eq:sublattice space}), respectively.  In Nambu space, $H_{BdG}(k_x) $ in  Eq. (\ref{eq:h half kx}) takes the form, 
	\begin{equation}
		\label{eq:HBdG(k_x)}
		{H}_{BdG}(k_x) = \begin{pmatrix}
			{A}(k_x) & {B}(k_x)\\
			-{B}(-k_x) & -{A}(-k_x)
		\end{pmatrix},
	\end{equation}
	where ${B}(k_x)$ and ${A}(k_x)$ are two real symmetric $4N_y \times 4N_y$  matrices in $ \mathscr{H}_I\otimes \mathscr{H}_s \otimes \mathscr{H}_\sigma$ defined as,
	\begin{equation}
		\label{eq:mathcal B}
		{B}(k_x) = c I_0 \otimes \sigma_0 \otimes s_x,
	\end{equation}
	and
	\begin{equation}
		{A}(k_x) = \mathbb{P}_\uparrow \otimes {h}_\uparrow(k_x)  + \mathbb{P}_\downarrow \otimes {h}_\downarrow(k_x), 
	\end{equation}
	where ${h}_{s=\uparrow,\downarrow}(k_x)$ are matrices in $\mathscr{H}_I\otimes \mathscr{H}_\sigma$ given by 
	\begin{equation}
		\label{eq:A_spin_up(k_x)}
		{h}_{s=\uparrow,\downarrow}(k_x) = I_- \otimes \alpha_s(k_x) + I_0\otimes \beta_s(k_x) + I_+ \otimes \gamma_s(k_x).
	\end{equation}
	In Eq. \ref{eq:A_spin_up(k_x)} , $I_0$, $I_-$ and $I_+$ are the main-, sub- and super-diagonal identities in  $\mathscr{H}_I$, e.g.,
	\begin{equation}
		I_0=\begin{pmatrix}
			1 & 0 &0\\
			0 & 1 & 0\\
			0 & 0 & 1
		\end{pmatrix}, I_- = \begin{pmatrix}
			0 & 0 &0\\
			1 & 0 & 0\\
			0 & 1 & 0
		\end{pmatrix},I_+=\begin{pmatrix}
			0 & 1 &0\\
			0 & 0 & 1\\
			0 & 0 & 0
		\end{pmatrix},
	\end{equation}
	for the case where $N_y=3$, and $\alpha_s,\beta_s$, and $\gamma_s$ are $2\times 2$ (real) matrices in $ \mathscr{H}_\sigma$ defined as
	\begin{equation}
		\label{eq: alpha beta gamma stripe up}
		\begin{split}
			\alpha_{\uparrow}(k_x) = & 2t_2\cos\phi_+(k_x) \sigma_z - t_1\sigma_+, \\
			\beta_{\uparrow}(k_x) = & [2t_2\cos\phi_-(k_x) +m] \sigma_z\\ 
			&+ q\sigma_0 - 2t_1\cos\frac{k_x}{2} \sigma_x, \\
			\gamma_{\uparrow}(k_x) =& 2t_2\cos \phi_+(k_x)  \sigma_z - t_1\sigma_-,
		\end{split}
	\end{equation}
	and 
	\begin{equation}
		\label{eq: alpha beta gamma stripe down}
		\begin{split}
			\alpha_{\downarrow}(k_x) = & -2t_2\cos\phi_+(k_x) \sigma_z - t_1\sigma_+, \\
			\beta_{\downarrow}(k_x) = & [-2t_2\cos\phi_-(k_x) +m] \sigma_z \\
			&+ q\sigma_0 - 2t_1\cos\frac{k_x}{2} \sigma_x, \\
			\gamma_{\downarrow}(k_x) =& -2t_2\cos \phi_+(k_x)  \sigma_z - t_1\sigma_-,
		\end{split}
	\end{equation}
	where $\sigma_{\pm} = (\sigma_x \pm i \sigma_y)/2 $ and
	\begin{equation}
		\phi_+(k_x) = \phi + \frac{k_x}{2}, \quad \phi_-(k_x) = \phi - k_x.
	\end{equation}
	
	\section{Haldane model}
	\label{sec:Haldane Model}
	As explained in  Sec. \ref{sec:Model}, our system described in the absence of pairing interactions is reduced to two independent copies of the Haldane model, a spin-$\uparrow$ copy and a spin-$\downarrow$ copy. As a preparation for the topological study of our system which includes pairing interactions, we review the Chern number in Sec. \ref{sec:Bulk Topology: Chern Number}, and conduct an analytical study of the edge modes in  semi-infinite plane geometries in Sec. \ref{eq:Edge Modes Haldane Model}, using,  without loss of generality, the spin-$\uparrow$ copy of the Haldane model.

	\subsection{Bulk Topology: Chern Number}
	\label{sec:Bulk Topology: Chern Number}
	In 1988, Haldane  \cite{haldane88PhysRevLett.61.2015} demonstrated, by subjecting a 2D graphene model to a periodic local magnetic field with no net magnetic flux, that the IQH effect can appear as a result of time reversal symmetry breaking, rather than from an external magnet field (for that matter, Landau levels). The Haldane model according to the 10-fold way is a class A insulator whose bulk is characterized by the Chern number.  The Chern number is an intrinsic property of the bulk band structure, which is determined from the eigenvalue equation of $h_\uparrow(\vb{k})$,
	\begin{equation}
		h_\uparrow(\vb{k}) \ket{n} = \left[\omega_n(\vb{k}) + \epsilon(\vb{k})\right]\ket{n},
	\end{equation}
	where 
	\begin{equation}
		\label{eq:omega_1(k) and omega_2(k)}
		\omega_1(\vb{k}) = q - h(\vb{k}), \quad \omega_2(\vb{k}) = q + h(\vb{k}),
	\end{equation}
	are the bulk band energies [relative to $\epsilon(\vb{k})]$ with 
	\begin{equation}
		\label{eq:h(k)}
		h(\vb{k}) = \sqrt{h^2_x(\vb{k}) +h^2_y(\vb{k}) + h^2_z(\vb{k})},
	\end{equation}
	and 
	\begin{equation}
		\begin{split}
			\label{eq:two states}
			\ket{1}= & \cos\theta_{\vb{k}} \ket{A} - \sin\theta_{\vb{k}} e^{i\phi_{\vb{k}}} \ket{B},\\
			\ket{2}= & \sin\theta_{\vb{k}} e^{-i\phi_{\vb{k}}} \ket{A} + \cos\theta_{\vb{k}} \ket{B},
		\end{split}
	\end{equation}
	are the corresponding eigenvectors in sublattice space with
	\begin{equation}
		\begin{split}
			\theta_{\vb{k}}= &\tan^{-1}\sqrt{\frac{h(\vb{k}) + h_z(\vb{k})}{h(\vb{k}) - h_z(\vb{k})}},\\
			e^{i\phi_{\vb{k}}} = &\frac{h_x(\vb{k}) + i h_y(\vb{k}) }{\sqrt{h^2_x(\vb{k}) +  h^2_y(\vb{k})}}.
		\end{split}
	\end{equation}

	Phase transitions cannot take place unless the upper band, $\omega_2(\vb{k})$, and the lower band, $\omega_1(\vb{k})$,  cross, which may only occur at Dirac points $K$ and $K'$ defined in the first Brillouin zone as shown in Fig. \ref{fig:fig1}(b).  The low-energy physics near  $K$ and $K'$ are modeled by the Dirac Hamiltonians, which are defined, respectively, as 
	\begin{align}
		h_{\uparrow}(\vb{K} + \delta \vb{k}) &= \tilde{m}_{+}\sigma_z +  \frac{\sqrt{3}t_1}{2}(\delta k_x \sigma_x +\delta k_y \sigma_y),  \\
		h_{\uparrow}(\vb{K}' + \delta \vb{k}) &=\tilde{m}_{-}\sigma_z-\frac{\sqrt{3}t_1}{2}(\delta k_x \sigma_x -\delta k_y \sigma_y) \label{eq:K'},
	\end{align}
	(apart from an overall constant) where
	\begin{equation}
		\tilde{m}_{\pm} = m \pm 3\sqrt{3}t_2 \sin \phi.
	\end{equation} Note that, in arriving at Eq. (\ref{eq:K'}), we used, instead of $K'$ marked in Fig. \ref{fig:fig1} (b), a Dirac point equivalent to $K'$ but that lies opposite to $K$.
	
\begin{figure}[ht]
		\centering
	\includegraphics[width=0.48\textwidth]{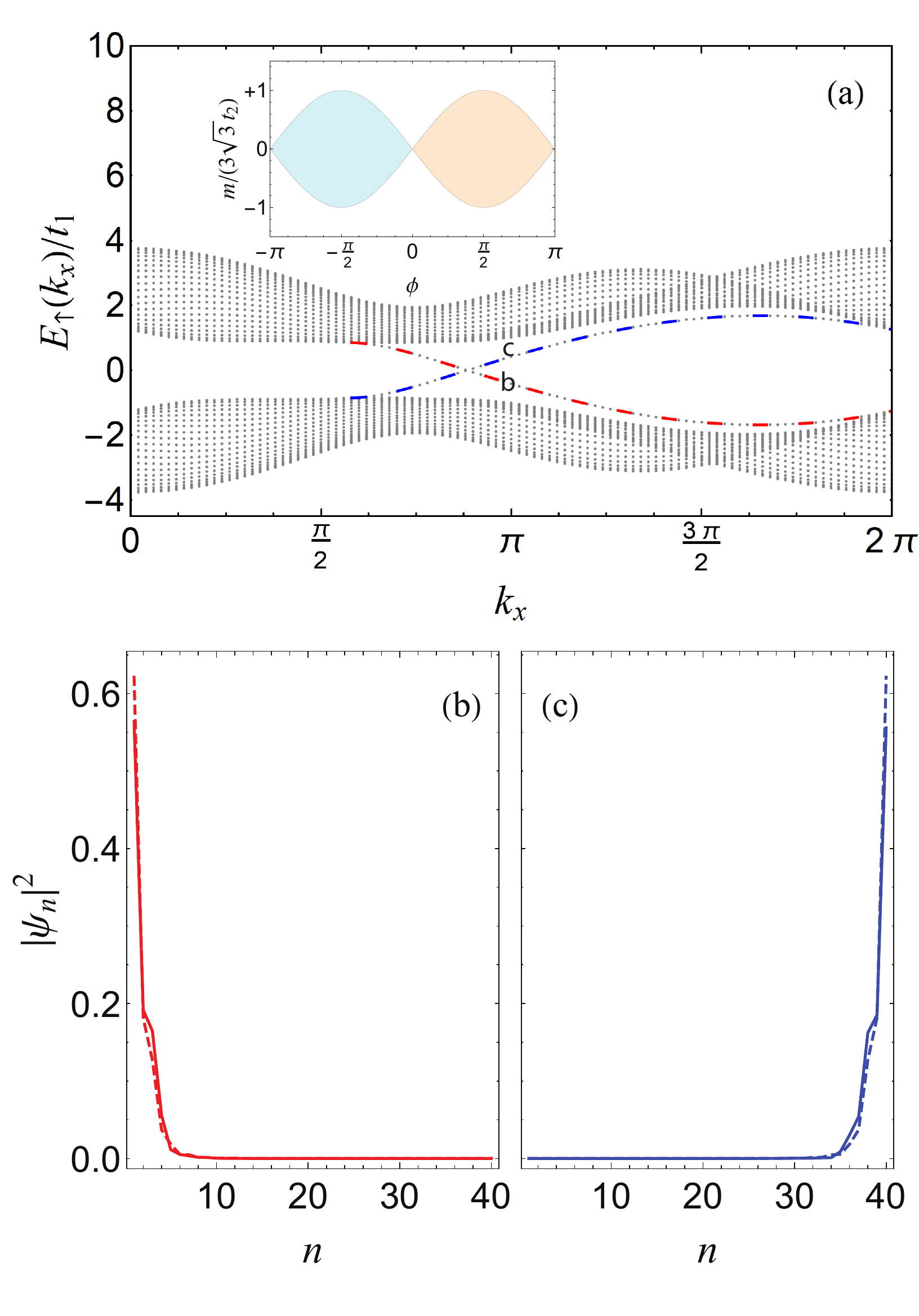}
		\caption{
			(a) The energy spectrum,  $E_\uparrow(k_x)$, of the Hamiltonian, $h_\uparrow(k_x)$ [Eq. (\ref{eq:A_spin_up(k_x)})], for the Haldane model in the stripe geometry for a state located at  $m= 3^{-1}\times 3\sqrt{3}t_2$ and $\phi=\pi/4$ in the $\phi-m$ phase diagram (inset), where regions with the Chern number  $-1$, $+1$, and $0$ are indicated with blue, orange, and white colors, respectively.   
			Panel (a) compares numerical edge mode dispersions (dotted black lines) with their analytical counterparts in semi-infinite geometries from Eqs. (\ref{eq:edge mode dispersions for A up}) and  (\ref{eq:condition p b}) with $i=b$ [which are reformulations of Eqs. (\ref{eq:bottom edge mode dispersion}) and (\ref{eq:bottom edge mode restrictions})] for the bottom edge mode (red dashed line) and from Eqs. (\ref{eq:edge mode dispersions for A up}) and Eq. (\ref{eq:condition p t}) with $i=t$ for the top edge mode (blue dashed line).
			Panels (b) and (c) compare numerically normalized population distributions, $\abs{\psi_n}^2$ (solid lines), and their analytical counterparts (dashed lines) for the edge modes at point $b$ and point $c$ with $k_x=\pi$ marked in panel (a).  Other parameters are $t_1=t_2=1$, $q=0$, and $N_y = 20$.
		}
		\label{fig:fig2}
	\end{figure}

	Using the Streda formula \cite{streda1982JPhysC.15.L717}, Haldane found, from the Hall conductance of this effective model, the Chern number,
	\begin{equation}
		\label{eq:haldane's Chern number}
		C = \frac{1}{2}\left[ \mathrm{sgn} (\tilde{m}_+) - \mathrm{sgn}(\tilde{m}_-) \right],
	\end{equation}
	which gives rise to the iconic phase diagram with two loops in the $\phi-m$ space as shown in the inset of Fig. \ref{fig:fig2}(a). 
	The same result can also be obtained from Eq. (\ref{eq:chern number integral in the k space}) in Sec. \ref{subsec:Bulk Topology}, where we describe a generic approach for determining the Chern number from a surface integral of Berry curvatures, which, for the Haldane model, are determined by the eigenstates of  $h_\uparrow(\vb{k})$ summarized in Eq. (\ref{eq:two states}).

	\subsection{Edge modes}
	\label{eq:Edge Modes Haldane Model}
	
	For an open system operating such that its bulk is characterized with a nontrivial Chern number, the bulk-boundary correspondence \cite{hatsugai1993PhysRevLett.71.3697,asboth2016} dictates that topologically protected modes, robust to defects and disorder, exist along the system boundaries. Good insight into edge modes can be gained by analyzing the Haldane model in semi-infinite planes with zigzag boundaries. For illustrative purposes, we consider the edge mode at the bottom of an upper-half plane that is localized at $n=1$ where $n$ is the index to the $y$ unit cell.  This edge mode, within the context of the present model, is an eigenstate of $h_\uparrow(k_x)$, a tridiagonal system described by Eq. (\ref{eq:A_spin_up(k_x)}). Let $\psi_n$ describe the amplitude of this state at the $n$th unit cell, where $\psi_n$ is itself a two-component vector in sublattice space. This state with eigenvalue $\epsilon(k_x)$ is governed in the bulk where $n>1$ by 
	\begin{equation}
		\label{eq:bulk schrodinger equation}
		\alpha_\uparrow \psi_{n-1} + \beta_\uparrow \psi_n +  \gamma_\uparrow \psi_{n+1} = \epsilon \psi_n,
	\end{equation}
	and in the boundary where $n=1$ by
	\begin{equation}
		\label{eq:boundary equation}
		\beta_\uparrow \psi_1 + \gamma_\uparrow \psi_{2} = \epsilon \psi_1,
	\end{equation}
	where we made use of the open boundary condition $\psi_0=0$. Note that when no confusion is likely to arise, we do not indicate dependence on $k_x$. We choose
	\begin{equation}
		\psi_n \propto \lambda^{n}\psi_1
	\end{equation}
	as the ansatz for the bulk \cite{creutz1994PhysRevD.50.2297,konig2008JPhysSocJapan.77.031007},
	where $\abs{\lambda}<1$, as required by the boundary condition $\psi_n \rightarrow 0$ in the semi-infinite limit $n\rightarrow \infty$. This ansatz changes Eq. (\ref{eq:bulk schrodinger equation}) into a set of homogeneous equations for $\psi_1$,
	\begin{equation}
		\label{eq:homogeneous equation for psi_1}
		\left[  \lambda^{-1}\alpha_\uparrow + \beta_\uparrow + \lambda \gamma_\uparrow \right]\psi_1 = \epsilon \psi_1.
	\end{equation}
	Just as in the case when $m=0$ \cite{wang2009PhysRevB.80.115420}, the condition for non-trivial solutions to Eq. (\ref{eq:homogeneous equation for psi_1}) gives rise immediately to a quadratic function of $\lambda + \lambda^{-1}$ (not shown) which guarantees the existence of two roots, which we call $\lambda_\pm$, smaller in magnitude than 1.  This implies that a general edge mode is a linear superposition of the solution with $\lambda_+$ and $\lambda_-$.  Together with the boundary condition at $n=0$, a general edge mode solution takes the form,
	\begin{equation}
		\label{eq:the bottom edge mode of an upper semi-infinite plane}
		\psi_n \propto (\lambda_+^n - \lambda_-^n)f,
	\end{equation}
	where $f$ is the solution to Eq. (\ref{eq:homogeneous equation for psi_1}).
	In terms of $f$, the equation at the boundary $n=1$,  Eq. (\ref{eq:boundary equation}), becomes
	\begin{equation}
		\label{eq:equations at the bundary n=1}
		\left[\beta_\uparrow + \gamma_\uparrow(\lambda_+ + \lambda_-) \right] f = \epsilon f,
	\end{equation}
	which is a matrix equation in $\mathscr{H}_\sigma$,
	\begin{widetext}
		\begin{equation}
			\label{eq:equations at the bundary n=1 matrix}
			\begin{pmatrix}
				2t_2\cos\phi_- + m + 2t_2(\lambda_+ + \lambda_-)\cos\phi_+  & -2t_1\cos\frac{k_x}{2}\\
				-2t_1\cos\frac{k_x}{2} - t_1(\lambda_+ + \lambda_-) & 
				-2t_2\cos\phi_- - m  -2t_2(\lambda_+ + \lambda_-)\cos\phi_+ 
			\end{pmatrix}
			\begin{pmatrix}
				f_1 \\
				f_2
			\end{pmatrix} = (\epsilon-q)\begin{pmatrix}
				f_1 \\
				f_2
			\end{pmatrix}.
		\end{equation}
	\end{widetext}
	Combining Eq. (\ref{eq:equations at the bundary n=1}) with Eq. (\ref{eq:homogeneous equation for psi_1}) where $\psi_1$ is replaced with $f$ gives rise to
	\begin{equation}
		\label{eq:simpler one}
		(\alpha_\uparrow - \lambda_+\lambda_-\gamma_\uparrow) f = 0,
	\end{equation}
	which is another matrix equation in $\mathscr{H}_\sigma$,
	\begin{equation}
		\label{eq:lambda1 lambda2 f}
		\begin{pmatrix}
			1- \lambda_+\lambda_- & -\frac{t_1}{2t_2\cos\phi_+}\\
			+\frac{t_1\lambda_+\lambda_-}{2t_2\cos\phi_+} & \lambda_+\lambda_--1
		\end{pmatrix} 
		\begin{pmatrix}
			f_1 \\
			f_2
		\end{pmatrix}
		=0.
	\end{equation}
	The edge state and its energy are obtained by solving Eqs. (\ref{eq:equations at the bundary n=1 matrix}) and (\ref{eq:lambda1 lambda2 f}) simultaneously. 
	We first consider Eq. (\ref{eq:lambda1 lambda2 f}) which is independent of $\epsilon(k_x)$.  We find that under the condition, 
	\begin{equation}
		\label{eq:quadratic equation}
		(2t_2\cos\phi_+)^2(\lambda_+\lambda_--1)^2 - t_1^2(\lambda_+\lambda_- - 1) - t_1^2=0,
	\end{equation}
	Eq. (\ref{eq:lambda1 lambda2 f}) supports a nontrivial solution,
	\begin{equation}
		\label{eq:f f_1 f_2}
		f \propto \begin{pmatrix}
			-t_1\\
			(\lambda_+\lambda_- -1)2t_2\cos\phi_+
		\end{pmatrix},
	\end{equation}
	where
	\begin{equation}
		\label{eq:lambda_1 times lambda_2}
		\begin{split}
			\lambda_+\lambda_- & =  1 + a_1,
		\end{split}
	\end{equation}
	is the solution of the quadratic equation (\ref{eq:quadratic equation}) with $\abs{\lambda_+\lambda_-}<1$, where 
	\begin{equation}
		a_1(k_x) = \frac{t_1^2}{8t_2^2\cos^2\phi_+} - \sqrt{\left(1 + \frac{t_1^2}{8t_2^2\cos^2\phi_+}\right)^2 -1}.
	\end{equation}
	
	We next consider Eq. (\ref{eq:equations at the bundary n=1 matrix}), which, because $f_1$ and  $f_2$ are already known through Eq. (\ref{eq:f f_1 f_2}), is actually an inhomogeneous equation of $\epsilon-q$ and $\lambda_+ + \lambda_-$,
	\begin{widetext}
		\begin{equation}
			\label{eq:inhomogeneous equation}
			\begin{pmatrix}
				f_1 &, &-2t_2\cos \phi_+ f_1\\
				f_2 &,& 2t_2\cos \phi_+ f_2 - t_1 f_1
			\end{pmatrix}
			\begin{pmatrix}
				\epsilon -q \\
				\lambda_+ + \lambda_-
			\end{pmatrix}
			= \begin{pmatrix}
				-2t_1\cos\frac{k_x}{2}f_2 + (2t_2\cos\phi_- + m)f_1 \\
				- 2t_1\cos\frac{k_x}{2}f_1-(2t_2\cos\phi_- + m)f_2
			\end{pmatrix}.
		\end{equation}
	\end{widetext}
	Solving Eq. (\ref{eq:inhomogeneous equation}), we arrive at the desired edge state dispersion,
	\begin{equation}
		\label{eq:bottom edge mode dispersion}
		\epsilon(k_x) = q+\epsilon'(k_x),
	\end{equation}
	where 
	\begin{equation}
		\label{eq:bottom edge mode dispersion prime}
		\epsilon'(k_x) =\abs{t_1}
		\frac{2t_2\cos\phi_- + m - 8t_2\cos\frac{k_x}{2}\cos\phi_+}{\sqrt{t_1^2 + 16t_2^2\cos^2\phi_+}},
	\end{equation}
	and
	\begin{equation}
		\label{eq:lambda_1 + lambda_2}
		\lambda_+ + \lambda_- = a_1 a_2,
	\end{equation}
	where 
	\begin{equation}
		a_2(k_x)=\frac{2t_1^2 \cos\frac{k_x}{2} +4t_2\cos\phi_+\left( 2t_2\cos\phi_- +m\right)}{\abs{t_1}\sqrt{t_1^2 + 16t_2^2\cos^2\phi_+}}.
	\end{equation}
	Finally, we have by solving simultaneously Eq. (\ref{eq:lambda_1 times lambda_2}) and Eq. (\ref{eq:lambda_1 + lambda_2}),
	\begin{equation}
		\lambda_{\pm}(k_x) =\frac{1}{2}\left(a_1a_2 \pm \sqrt{a_1^2a_2^2 - 4(1+a_1)} \right).
	\end{equation}
	If the edge mode exists, it must be limited to the momentum space in which
	\begin{equation}
		\label{eq:bottom edge mode restrictions}
		\abs{\lambda_+(k_x)} <1 \mbox{ and } \abs{\lambda_-(k_x)}<1
	\end{equation}
	hold true simultaneously.
	
	Equation (\ref{eq:bottom edge mode dispersion prime}), which we have just obtained in the same spirit as the transfer matrix method \cite{hatsugai1993PhysRevLett.71.3697}, reduces to the form by Galilo et al. \cite{galilo15PhysRevLett.115.245302} when $\phi=\pi/2$ and $m=0$ and the form in our earlier publication   \cite{ling2021PhysRevA.104.013305} when $\phi=\pi/2$ but $m$ remains arbitrary. There also exists in the literature other analytical studies under limited conditions \cite{huang2012arXiv:1205.6266,doh2013PhysRevB.88.245115,pantaleon2017JPhysCondensMatter.29.295701}.  While preparing this paper, we became aware of a recent paper \cite{tomonari2021PhysRevB.103.195310} which includes a similar problem and their result for $\phi=\pi/2$ has the same simple mathematical form as ours. 
	
	Figure 2(a) compares the energy dispersions and Figs. 2(b,c) compare the population distributions of the edge modes, as obtained analytically for semi-infinite plane geometries, with those obtained numerically from  eigenvalues of the Hamiltonian $h_\uparrow(k_x)$ in Eq. (\ref{eq:A_spin_up(k_x)}) in the stripe geometry. We can see that the dispersions agree quite well; discrepancies due to finite-size effects are not discernible even when $N_y$ is as low as $N_y=20$.
	
	The edge mode that we have studied so far lies at the bottom edge. There shall be an edge mode at the top, traveling in the opposite direction, according to the bulk-boundary correspondence for a Chern insulator  \cite{hatsugai1993PhysRevLett.71.3697,asboth2016}.  
	An analogous set of results can be obtained for this top edge mode by analyzing a lower-half plane. We will postpone this and other related results until Sec. \ref{sec:edge modes for our model} where we apply them to gain insight into the edge modes of an open system with pairing interactions.
	
	\section{Our model without time-reversal symmetry:topological properties}
	\label{sec:our model without time reversal symmetry}
	Having reviewed the Haldane model in Sec. \ref{sec:Haldane Model}, we now turn our attention to our model in Sec. \ref{sec:Model} which includes pairing interactions and is described by a bosonic BdG Hamiltonian. We study the topological properties of such a system in connection with questions such as what topological symmetry class our system belongs to under the 38-fold way, what topological invariant we use to characterize the bulk, and how pairing interactions affect the edge modes. 
	We pursue these questions for systems without time-reversal symmetry in this section and make a related study for systems with time-reversal symmetry in the next section, where we study the physics behind topological amplifiers.
	
	To avoid confusion with terminology, we stress at the onset that we follow the convention set by Kawabata et al. \cite{kawabata2019PhysRevX.9.041015}, where, for example, $\mathcal{C} H^T_{BdG}(\vb{k})\mathcal{C}^{-1} =-H_{BdG}(-\vb{k})$, rather than the usual $\mathcal{C} H^*_{BdG}(\vb{k})\mathcal{C}^{-1} =-H_{BdG}(-\vb{k})$, is defined as the particle-hole symmetry.  A topological phase describes a group of Hamiltonians that can be continuously deformed into each other without causing their energy bands to close, i.e., to cross a reference, which, for non-Hermitian Hamiltonians, where eigenenergies live in a complex plane, can be either a point or a line \cite{kawabata2019PhysRevX.9.041015}.  In the present study, we always deal with gaps that exist in the real parts of complex energies and topological phases are thus always classified according to a real line gap in which the imaginary axis in the complex-energy plane serves as the reference for an energy-band crossing.   
	
	\subsection{Symmetry and topological classification}
	\label{sec:symmetries and topological classification}
	We begin the topological study of our model by analyzing how the Bloch BdG Hamiltonian (\ref{eq:H_BdG first}) or equivalently,
	\begin{equation}
		\label{eq:H_BdG matrix}
		\begin{split}
			&{H}_{BdG}(\vb{k})  = \mathbb{P}_+\mathbb{P}_\uparrow h_\uparrow(\vb{k}) + \mathbb{P}_{+}\mathbb{P}_\downarrow h_\downarrow(\vb{k})\\ &-\mathbb{P}_-\mathbb{P}_\uparrow h^*_\uparrow(-{\vb{k}}) - \mathbb{P}_- \mathbb{P}_\downarrow h^*_\downarrow(-{\vb{k}}) +i c \tau_y s_x \sigma_0,
		\end{split}
	\end{equation}
	transforms under various symmetry operations, 
	where $\mathbb{P}_{\uparrow, \downarrow}$ and $\mathbb{P}_{+,-}$ are the projection operators in spin and Nambu space defined in Eq. (\ref{eq:projection operators}).  If confusion is unlikely to arise, we do not explicitly write the tensor product symbol $\otimes$ (as well as the subspace identity matrices).
	The bosonic BdG Hamiltonian (\ref{eq:H_BdG matrix}) obeys the pseudo-Hermiticity condition,
	\begin{equation}
		\label{eq:pseudo-Hermiticity}
		\eta H^\dag_{BdG}(\vb{k})\eta^{-1} =H_{BdG}(\vb{k}),
	\end{equation}
	where $\eta =\tau_z(=\Sigma_z$) is the matrix representing pseudo-Hermiticity.  That $\eta$ and $\Sigma_z $ being equivalent follows from the ``first quantized" Hamiltonian $\Sigma_z H_{BdG}(\vb{k})$ being  Hermitian.  In addition, the Bose BdG Hamiltonian obeys particle-hole symmetry,
	\begin{equation}
		\label{eq:particle-hole symmetry 1}
		\mathcal{C} H^T_{BdG}(\vb{k}) \mathcal{C}^{-1}=-H_{BdG}(-\vb{k}),
	\end{equation}
	where $\mathcal{C} = \tau_y$ is the matrix representing particle-hole symmetry. The pseudo-Hermiticity and particle-hole symmetry are intrinsic to the bosonic BdG Hamiltonian; they act more as constraints arising from Bose statistics than as symmetries imposed by additional conditions.  As to how it transforms under time reversal, we note that,  for our pseudo-spin-$1/2$ system, the  time-reversal operator is given by $\mathcal{T} = s_x$, which is a time-reversal even operator, i.e., $\mathcal{T}^2=1$. It can be shown that Eq. (\ref{eq:H_BdG matrix}) does not possess time-reversal symmetry, i.e.,
	\begin{equation}
		\label{eq:time reversal symmetry for the total system}
		\mathcal{T} H^*_{BdG}(\vb{k}) \mathcal{T}^{-1} =  H_{BdG}(-\vb{k}),
	\end{equation}
	unless $\cos\phi = 0$, i.e., $\phi =\pi/2$ (or any half-integral multiple of $\pi$), a situation which we will deal with separately in Sec. \ref{sec:our model with time reversal symmetry}. 
	
	For our model, topological properties are not determined by $ H_{BdG}(\vb{k})$ but rather by Hamiltonians in smaller Hilbert spaces as we now explain.  It comes down to magnetization being conserved under Hamiltonian (\ref{eq:H_half}), i.e., $[\hat{Q},\hat{H}_{1/2}]=0$, where 
	\begin{equation}
		\begin{split}
			\hat{Q} &=\sum_{\vb{i}} \left( \hat{b}^\dag_{\vb{i},\uparrow}\sigma_0\hat{b}_{\vb{i},\uparrow} - \hat{b}^\dag_{\vb{i},\downarrow}\sigma_0\hat{b}_{\vb{i},\downarrow}  \right) \\
		\end{split}
	\end{equation}
	is the magnetization operator (often called the charge operator \cite{blaizot96QuantumTheoryBook}) in position space. In momentum space, conservation of magnetization amounts to 
	\begin{equation}
		[J_z, H_{BdG}(\vb{k})]=0,
	\end{equation}
	so that $J_z$ and $H_{BdG}(\vb{k})$ can be simultaneously diagonalized, where $J_z=\tau_z s_z$ is the matrix representation of the magnetization operator in momentum space (also called the $z$-component of the spin rotation generator represented on Nambu space \cite{altland1997PhysRevB.55.1142}).  In fact, $J_z$ partitions $H_{BdG}(\vb{k})$ to 
	\begin{equation}
		\label{eq:bulk H_BdG(k)}
		H_{BdG}(\vb{k}) = H_\Uparrow(\vb{k}) \oplus H_\Downarrow(\vb{k}),
	\end{equation}
	a direct sum of  $H_\Uparrow(\vb{k})$ on the degenerate subspace of $J_z$ with eigenvalue $+1$ and $H_\Downarrow(\vb{k})$ on the degenerate subspace of $J_z$ with eigenvalue $-1$, where
	\begin{equation}
		\label{eq:H_Uparrow}
		H_{\Uparrow, \Downarrow}(\vb{k}) = 
		\begin{pmatrix}
			h_{\uparrow, \downarrow}(\vb{k}) & c \sigma_0 \\
			-c \sigma_0 & -h^*_{\downarrow, \uparrow}(-\vb{k})
		\end{pmatrix},
	\end{equation}
	in Nambu space.
	As a result, our system inherits the topological classes of and can therefore be classified according to the Hamiltonians in smaller sectors, $H_\Uparrow(\vb{k})$  and $H_\Downarrow(\vb{k})$.
	
	Without loss of generality, we focus on the $J_z$-$\Uparrow$ sector and investigate symmetry as well as energy-band properties of $H_{\Uparrow}(\vb{k})$.  We begin by rewriting $H_{\Uparrow}(\vb{k})$ in terms of the matrices in Nambu space as
	\begin{equation}
		\label{eq:H_Uparrow 1}
		H_{\Uparrow}(\vb{k}) = \mathbb{P}_+ h_\uparrow(\vb{k}) - \mathbb{P}_- h^*_\downarrow(-\vb{k}) + i c \tau_y \sigma_0,
	\end{equation}
	which is perhaps the simplest spinful system built on copies of the Haldane model that involve pairing interactions. 
	
	The Hamiltonian in the $J_z$-$\Uparrow$ sector obeys the pseudo-Hermiticity,
	\begin{equation}
		\label{eq:pseudo-hermiticity H Uparrow}
		\eta H^\dag_\Uparrow(\vb{k})\eta^{-1}=H_\Uparrow(\vb{k}).
	\end{equation}
	However, unlike its parent Hamiltonian $H_{BdG}(\vb{k})$ in Eq. (\ref{eq:H_BdG matrix}), $H_\Uparrow(\vb{k})$ does not preserve particle-hole symmetry,
	\begin{equation}
		\label{eq:not particle-hole symmetric H Uparrow}
		\mathcal{C} H^T_\Uparrow(\vb{k})\mathcal{C}^{-1}\neq -H_\Uparrow(-\vb{k}),
	\end{equation}
	since  $\mathcal{C} H^T_\Uparrow(\vb{k})\mathcal{C}^{-1}$ is found to equal $-H_\Downarrow(-\vb{k})$ instead of $-H_\Uparrow(-\vb{k})$. For this reason, $ H_\Uparrow(\vb{k})$ is not a BdG Hamiltonian. Furthermore, it possesses neither time-reversal symmetry nor chiral symmetry. As such,  $H_\Uparrow(\vb{k})$ belongs to symmetry class  $A + \eta$ whose topological invariant in two dimensions is $\mathbb{Z}\oplus \mathbb{Z}$ when classified by a real line gap. In the next section, we further discuss this topological invariant and the related Chern numbers for systems with stable bulk energy bands.

	\begin{figure*}[t]
		\centering

		\includegraphics[width=1\textwidth]{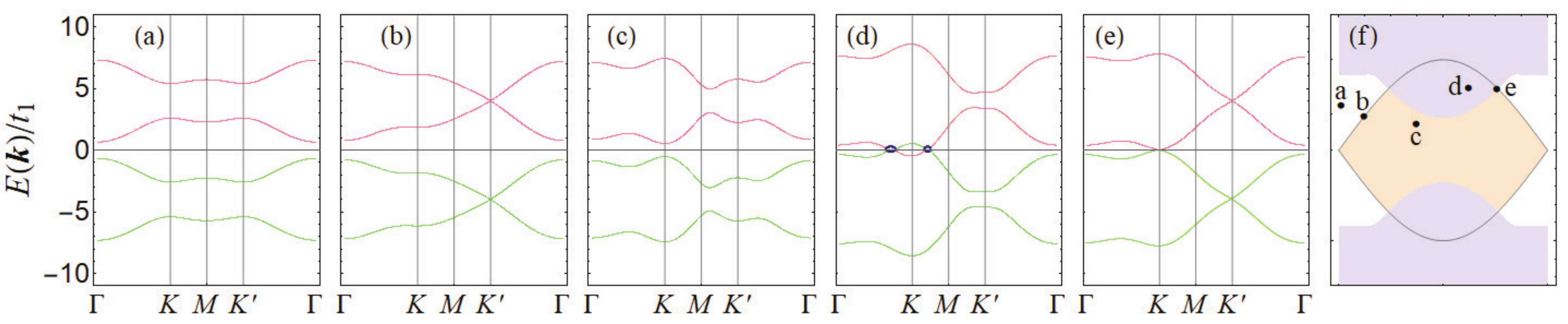}
		\caption{The Bloch energy spectra [relative to $\epsilon(\vb{k})$], sampled at points a, b, c, d, and e as indicated in panel (f),  are shown, respectively, in panels (a)-(e) where  $\phi=0$ and $m = 0.5 \times 3\sqrt{3}t_2$ at point a, $\phi=\pi/8$ and $m = \sin(\pi/8)\times 3\sqrt{3}t_2$ at point b,  $\phi=3\pi/8$ and $m = 0.3 \times 3\sqrt{3}t_2$ at point c, $\phi=\pi/2$ and $m = 0.7 \times 3\sqrt{3}t_2$ at point d, and $\phi=25\pi/32$ and $m = 0.7\times 3\sqrt{3}t_2$ at point e.
			These Bloch energy spectra are calculated along a closed path, $\Gamma\rightarrow K \rightarrow M \rightarrow K' \rightarrow \Gamma$, in the first Brillouin zone as indicated in Fig. \ref{fig:fig1}(b). The red and green curves represent, respectively, the particle and hole energy bands (of real energies) while the dark blue curves [only in panel (d)] depict the imaginary parts of complex energies. (The distinction between the particle and hole bands is clarified in Sec. \ref{sec:edge modes for our model}.)   Other parameters are the same as in Fig. \ref{fig:fig3}(a).}
		\label{fig:fig4}
	\end{figure*}

	\subsection{Topological invariant: The Chern number}
	\label{subsec:Bulk Topology}
	The Chern number traces its root to the geometric phase, which arises naturally in Hermitian Hamiltonians due to the interplay between unitary nature of time evolution and orthornormality property of eigenenergy states.  In 1988, Garrison and Wright \cite{garrison1988PhysLettA.128.177} extended the Berry phase to non-Hermitian systems with the help of the bi-orthonormality relation constructed from right and left eigenvectors, which are defined, within the context of our work, as 
	\begin{equation}
		\begin{split}
			H_\Uparrow(\vb{k})\ket{E} = & \left[E(\vb{k}) + \epsilon(\vb{k})\right] \ket{E},\\
			\left<\left< E\right|\right. H_\Uparrow(\vb{k}) = & \left[E(\vb{k}) + \epsilon(\vb{k}) \right]\left<\left< E\right|\right..
		\end{split}
	\end{equation}
	
	The energy spectrum [relative to $\epsilon(\vb{k})$] consists of four branches given by
	\begin{equation}
		\label{eq:bulk energy bands periodic}
		E_{n,\alpha}(\vb{k})=\alpha E_n(\vb{k}), \quad n =1, 2, \quad \alpha = \pm,
	\end{equation}
	where
	\begin{equation}
		\label{eq:E_n(k)}
		E_n(\vb{k})=\sqrt{\omega^2_n(\vb{k})-c^2}.
	\end{equation}
	Figure \ref{fig:fig4} showcases several bulk (excitation) energy spectra, which we will explain with more detail towards the end of this section,  along a closed path, $\Gamma\rightarrow K \rightarrow M \rightarrow K' \rightarrow \Gamma$, in the first Brillouin zone as shown in Fig. \ref{fig:fig1}(b).  The eigenvectors of energy $E_{n,\alpha}(\vb{k})$ are given by
	\begin{equation}
		\label{eq:eigenenergy states}
		\begin{split}
			\ket{E_{n,\alpha}}= &\ket{n,\alpha} 
			\otimes \ket{n},\\
			\left<\left< E_{n,\alpha} \right|\right. = & \left<\left< n,\alpha \right|\right. \otimes \bra{n},
		\end{split}
	\end{equation}
	where $\ket{n}$ are the eigenvectors [with eigenvalues $\omega_n(\vb{k})$] of $h_\uparrow(\vb{k})$ given by Eq. (\ref{eq:two states}) and
	\begin{equation}
		\label{eq:left and right vectors}
		\begin{split}
			\ket{n,\alpha} =& -u_{n\alpha}(\vb{k})\ket{+} + v_{n\alpha}(\vb{k})\ket{-},\\
			\left<\left< n,\alpha \right|\right.=& \left[-u_{n\alpha}(\vb{k}) \bra{+} + v_{n\alpha}(\vb{k})\bra{-}\right]\tau_z,
		\end{split}
	\end{equation}
	are the right and left eigenvectors in Nambu space $(\ket{+},\ket{-})$,
	where
	\begin{equation}
		\label{eq:uv alpha}
		\begin{split}
			u_{n\alpha}(\vb{k}) =& \frac{c}{\sqrt{2\alpha E_n(\vb{k})\left[\omega_n(\vb{k})-\alpha E_n(\vb{k})\right]}},\\
			v_{n\alpha}(\vb{k}) = &  \frac{\omega_n(\vb{k}) - \alpha E_n(\vb{k})}{\sqrt{2\alpha E_n(\vb{k})\left[\omega_n(\vb{k}) -\alpha E_n(\vb{k})\right]}}.
		\end{split}
	\end{equation}
	This set of four eigenstates are normalized in accordance with the bi-orthonormality condition,
	\begin{equation}
		\label{eq:bi-orthonormality condition}
		\left<\left< E_{n,\alpha} \right|\ket{E_{m,\beta}} \right. = \delta_{n,m}\delta_{\alpha,\beta},
	\end{equation}
	and obey the completeness relation,
	\begin{equation}
		\label{eq:biorthonormality closure identity}
		\sum_{n,\alpha} \ket{E_{n,\alpha}} \left<\left< E_{n,\alpha} \right| \right. = \tau_0\sigma_0,
	\end{equation}
	which, together with the linear superposition principle, forms the two pillars upon which phenomena associated with the Berry phase can be generalized from Hermitian to non-Hermitian systems in a straightforward manner.

	The Berry phase for the $E_{n,\alpha}$ energy band is a (closed) line integral of the Berry potential
	\begin{equation}
		A_{n\alpha} = i \left<\left< E_{n,\alpha} \right| \right. d \ket{E_{n,\alpha}},
	\end{equation}
	which is a 1-form involving both the right and left eigenenergy states, where $d$ is the exterior derivative.  The Berry curvature is then a 2-form obtained by taking the exterior derivative of the Berry potential,
	\begin{equation}
		F_{n\alpha}= dA_{n\alpha}.
	\end{equation}
	Let
	\begin{equation}
		A_{n\alpha, j}(\vb{k}) \equiv i \left<\left< E_{n,\alpha}(\vb{k}) \right| \right. \partial_{j} \ket{E_{n,\alpha}(\vb{k})},
	\end{equation}
	where $\partial_j \equiv \partial/\partial k_j$ with $j = x, y$.
	In momentum space, the Berry potential and the corresponding curvature are given, respectively, by 
	\begin{equation}
		A_{n\alpha}(\vb{k})= A_{n\alpha, x}(\vb{k})  dk_x + A_{n\alpha, y}(\vb{k})  dk_y,
	\end{equation}
	and
	\begin{equation}
		F_{n\alpha}(\vb{k})= \left[ \partial_{x}A_{n\alpha, y}(\vb{k}) - \partial_{y}A_{n\alpha, x}(\vb{k})\right]
		dk_x\wedge dk_y.
	\end{equation}
	The first Chern number is defined as a surface integral over an entire first Brillouin zone, which is a closed manifold (a torus),
	\begin{equation}
		\label{eq:Chern number integral}
		C_{n\alpha} = \frac{1}{2\pi} \int_{BZ}F_{n\alpha}.
	\end{equation}
	
\begin{figure}[ht]
		\centering
		\includegraphics[width=0.45\textwidth]{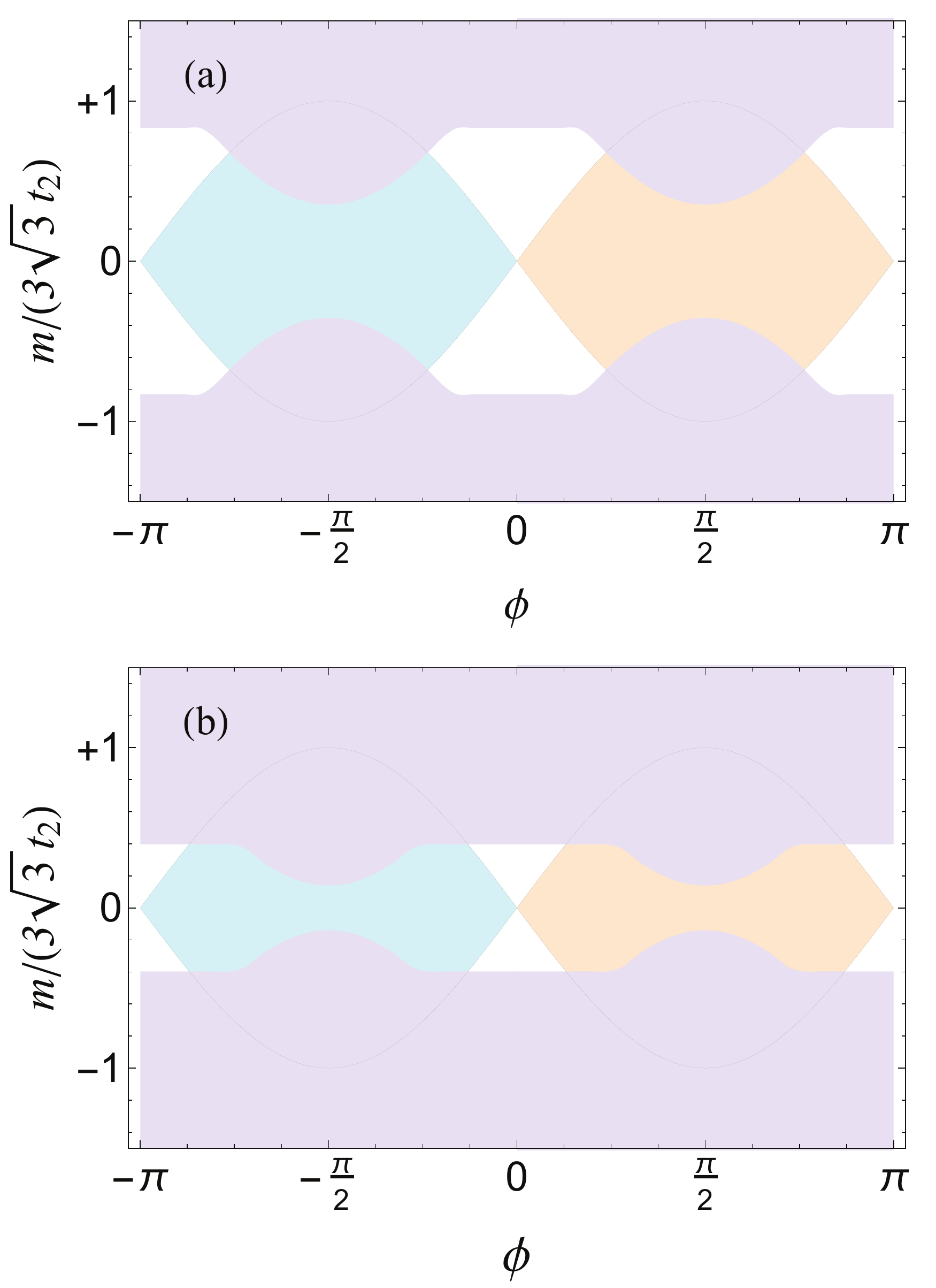}
		\caption{Phase diagrams in the $\phi-m$ space for $q=4t_1$ and (a) $c=0.2t_1$ and (b) $c=0.8t_1$, which are identical to the phase diagram for the Haldane model in the inset of Fig. \ref{fig:fig2} apart from the unstable region with the light-purple color. Other parameters are $t_1=1$ and $t_2=0.54 t_1$.}
		\label{fig:fig3}
	\end{figure}

	Figure \ref{fig:fig3} is a phase diagram in the $\phi-m$ space which divides into stable and unstable regions depending on whether all bulk energy bands in Eq. (\ref{eq:bulk energy bands periodic}) are free of any complex values. In the stable region, we adapt the gauge-invariant algorithm developed by Fukui et al. \cite{fukui2005JPSJ.74.1674} to Eq. (\ref{eq:Chern number integral}) where the Berry connection is formulated in terms of both right and left eigenvectors; we apply it to numerically determine, without loss of generality, the Chern number for the $E_{n=1,\alpha=+}(\vb{k})$ band. We find that the boundary separating the topologically trivial phase from the topologically nontrivial phase is the same as in the Haldane model [see the inset of Fig. \ref{fig:fig2}(a)].
	
	We now work to prove that the Chern number in the stable region should indeed be that of the Haldane model.
	We first recall that the integral for the Chern number [Eq. (\ref{eq:Chern number integral})] has been formulated in momentum $\vb{k}=(k_x,k_y)$ space.  However, states in Eq. (\ref{eq:eigenenergy states}) depend on $\vb{k}$ only via the map $\vb{k}\mapsto 
	\vb{h(\vb{k})}$, and have nothing to do with $\epsilon(\vb{k}) $. As a result, our system can be understood as living in the $\vb{h} = (h_x,h_y,h_z)$ space as far as band topology is concerned. Thus, we consider the Berry curvature 2-form in the $\vb{h} = (h_x,h_y,h_z)$ space,
	\begin{widetext}
		\begin{equation}
			\label{eq:gauge-independent curvature}
			F_{n\alpha}(\vb{h})= i \sum_{m\beta\neq n\alpha} \frac{\left<\left< E_{n,\alpha} \right| \right. \partial_i H_\Uparrow\ket{E_{m,\beta}}\left<\left< E_{m,\beta} \right| \right. \partial_j H_\Uparrow\ket{E_{n,\alpha}}}{(E_{n,\alpha} - E_{m,\beta})^2} dh_i\wedge dh_j,
		\end{equation}
	\end{widetext}
	where repeated indices are summed over and $\partial_j = \partial /\partial h_j$.  Equation (\ref{eq:gauge-independent curvature}) generalizes the computation of the Berry curvature (in a manifestly gauge-independent manner) from Hermitian systems where the completeness relation is founded on the usual orthonormality condition \cite{berry1984BerryPhase} to non-Hermitian system where the completeness relation (\ref{eq:biorthonormality closure identity}) is founded on the bi-orthonormality condition (\ref{eq:bi-orthonormality condition}).  Replacing $\partial_i H_\Uparrow$ in Eq. (\ref{eq:gauge-independent curvature}) with  a separable form
	\begin{equation}
		\partial_i H_\Uparrow = \tau_z \sigma_i,
	\end{equation}
	which we obtain from Eqs (\ref{eq:H_Uparrow 1}), (\ref{eq:hadane model for spin up}), and (\ref{eq:hadane model for spin down}), we simplify Eq. (\ref{eq:gauge-independent curvature}) to
	\begin{equation}
		\label{eq:simplified F_n alpha}
		F_{n\alpha} = i  D_{n\alpha} \bra{n}\sigma_i \ket{m}\bra{m}\sigma_j\ket{n} dh_i \wedge dh_j,
	\end{equation}
	where
	\begin{equation}
		\label{eq:D_(n alpha)}
		D_{n\alpha} = \sum_{\beta}\frac{\left<\left< n,\alpha \right| \right. \tau_z\ket{m,\beta}\left<\left<m,\beta \right| \right. \tau_z \ket{n,\alpha}}{(\alpha E_{n} -\beta E_{m})^2},
	\end{equation}
	with $n\neq m$.  In Appendix \ref{sec:simplification}, we show that the summation in Eq. (\ref{eq:D_(n alpha)}) can be reduced to 
	\begin{equation}
		\label{eq: D n alpha}
		D_{n\alpha} =\frac{1}{(\omega_n - \omega_m)^2},
	\end{equation}
	so that 
	\begin{equation}
		\label{eq:Berry curvature of the Haldane model}
		F_{n\alpha}(\vb{h}) = i \frac{\bra{n}\partial_i h_\uparrow \ket{m}\bra{m}\partial_j h_\uparrow \ket{n}}{(\omega_n - \omega_m)^2} dh_i \wedge dh_j,
	\end{equation}
	where $\partial_i h_\uparrow=\sigma_i$ has been used. Equation (\ref{eq:Berry curvature of the Haldane model}) is the Berry curvature of the Haldane model described by the two-band Hamiltonian (\ref{eq:hadane model for spin up}) and therefore has the well-known result \cite{berry1984BerryPhase},
	\begin{equation}
		\label{eq:berry curvature 2-form in h space}
		F_{1\alpha}(\vb{h}) =-F_{2\alpha}(\vb{h}) =\epsilon_{ijk}\frac{h_k(\vb{k})}{4h^3(\vb{k})} dh_i(\vb{k}) \wedge dh_j(\vb{k})
	\end{equation}
	where $\epsilon_{ijk}$ is the Levi-Civita symbol. Following standard practice, we pull the Berry curvature 2-form in Eq. (\ref{eq:berry curvature 2-form in h space}) back, via the map $\vb{k}\mapsto 
	\vb{h}(\vb{k})$,  to momentum space where the first Brillouin zone is well defined so that the Chern number integral (\ref{eq:Chern number integral}) becomes
	\begin{equation}
		\label{eq:chern number integral in the k space}
		C_{1\alpha} = - C_{2\alpha} =  \int_{BZ} \frac{ \vb{h}(\vb{k}) \cdot \left[ \partial_x \vb{h}(\vb{k}) \times \partial_y \vb{h}(\vb{k}) \right]}{4\pi h^3(\vb{k})} d^2\vb{k}. 
	\end{equation}
	Expressed in this way, the Chern number is recognized as the winding number --- the number of times that the image of the first Brillouin zone in the $\vb{h}$ space wraps around the origin.  This can be evaluated analytically (see, for example, Ref. \cite{fruchart2013ComptesRendusPhysique.14.779}) to give Haldane's result \cite{haldane88PhysRevLett.61.2015} in Eq. (\ref{eq:haldane's Chern number}).  In conclusion, we have
	\begin{equation}
		\label{eq:single churn number}
		\begin{split}
			C_{1+} & = -C_{2+}=C,\\
			C_{1-} & = -C_{2-} = C.
		\end{split}
	\end{equation}
	
	A comment is in order. In the previous section, we have classified $H_{\Uparrow}(\vb{k})$, within the 38-fold way, as symmetry class A+$\eta$ with the $\mathbb{Z}\oplus \mathbb{Z}$ topological invariant. However, as one may verify, for stable bands, states $\ket{E_{n,-}(\vb{k})}$ are connected to states $\ket{E_{n,+}(\vb{k})}$ via a unitary transformation, i.e., $\ket{E_{n,-}(\vb{k})} = i\tau_x \ket{E_{n,+}(\vb{k})}$. 
	As a result, $\mathbb{Z}\oplus \mathbb{Z}$ is reduced to $\mathbb{Z}$. In arriving at Eq. (\ref{eq:single churn number}), we have shown, in essence, that this single $\mathbb{Z}$ corresponds to $C$, the Chern number for the Haldane model in Eq. (\ref{eq:haldane's Chern number}). 

	We now return to Fig. \ref{fig:fig3} where pairing interactions increase from Figs. \ref{fig:fig3}(a) to \ref{fig:fig3}(b).  We can see that an increased pairing interaction leads to an increased region of instability, but also that, in the stable region, the Chern number remain unchanged.  There are two simple reasons that account for this.  First, pairing terms, i.e., the last line of Eq. (\ref{eq:H_half}), do not affect the $C_3$ point-group symmetry underlying a honeycomb lattice, which our model inherits from the Haldane model, since it is built upon two copies of it. Thus, even in the presence of pairing interations, a gap closing and reopening transition still occurs at Dirac points as guaranteed by $C_3$ point group symmetry \cite{bernevig2013}.  Second, pairing in our model takes the form $ic\tau_y \sigma_0$ [the last term in Eq. (\ref{eq:H_Uparrow 1})] and therefore breaks neither time-reversal symmetry nor inversion symmetry. Thus, as far as closing and reopening a gap at a Dirac point is concerned, such a pairing remains a spectator, having no effect on the topologically trivial-nontrivial phase transition. This is to be contrasted to the Bose-Hubbard extension of the Haldane model, where pairing may break inversion symmetry and therefore may affect  the topologically trivial-nontrivial phase transition \cite{furukawa2017NewJournalOfPhysics.17.115014}.

	We conclude this section by taking a look at Fig. \ref{fig:fig4}(a)-\ref{fig:fig4}(e), which show Bloch energy spectra sampled from states marked, respectively, as points a-e in Fig. \ref{fig:fig4}(f) [where \ref{fig:fig4}(f) a sketch representing the right-half of Fig. \ref{fig:fig3}(a)].  Figures \ref{fig:fig4}(a)-\ref{fig:fig4}(c) depict a gap-closing and -reopening transition from the trivial gapped state at point a to the nontrivial gapped state at point c via the gapless state at point b where two particle energy bands touch at Dirac point $K'$ [and two hole energy bands are guaranteed to touch also at the same $K'$ by pseudo-Hermiticity of $H_\Uparrow(\vb{k})$].  Figure \ref{fig:fig4}(d) shows a typical unstable bulk state at point d where energies become complex near the regions where there are level crossings between particle and hole energy bands.  Finally, Fig. \ref{fig:fig4}(e) displays the energy spectrum at point e where three phases, i.e., topologically trivial, topologically nontrivial, and unstable states, meet; this spectrum is characterized by two distinct types of gapless points, one between two particle energy bands (and two hole energy bands) as in Fig. \ref{fig:fig4}(b) and the other between a particle and hole energy band, which signals the onset of dynamical instability.


	\subsection{Edge modes}
	\label{sec:edge modes for our model}
	We now investigate our system with zigzag boundaries, as first presented in Sec. \ref{sec:The Model with Open Boundaries}. We begin by noting that $J_z$ is still a conserved quantity under $H_{BdG}(k_x)$ in Eq. (\ref{eq:HBdG(k_x)}) and partitions ${H}_{BdG}(k_x)$ into the direct sum,
	\begin{equation}
		{H}_{BdG}(k_x) = {H}_{\Uparrow}(k_x) \oplus {H}_{\Downarrow}(k_x),
	\end{equation}
	where
	\begin{equation}
		\label{eq:H_Uparrow kx}
		{H}_{\Uparrow,\Downarrow}(k_x) = \begin{pmatrix}
			{h}_{\uparrow,\downarrow}(k_x) & c I_0 \sigma_0\\
			-c I_0 \sigma_0 & -{h}_{\downarrow,\uparrow}(-k_x)
		\end{pmatrix}.
	\end{equation}
	with ${h}_{s=\uparrow, \downarrow}(k_x)$ being the matrices defined earlier in Eq. (\ref{eq:A_spin_up(k_x)}).
	As in the previous section, we focus on the sub-Hamiltonian in the $J_z-\Uparrow$ sector and investigate the band structure of ${H}_{\Uparrow}(k_x)$ in  Eq. (\ref{eq:H_Uparrow kx}) or equivalently
	\begin{equation}
		\label{eq:H_Uparrow not a matrix}
		{H}_\Uparrow(k_x) = P_+ {h}_\uparrow(k_x) -P_-{h}_\downarrow(-k_x) + ic \tau_yI_0 \sigma_0
	\end{equation} by analyzing the Schr\"{o}dinger equation,
	\begin{equation}
		{H}_{\Uparrow}(k_x)\ket{E_i}= E_i(k_x)\ket{E_i}.
	\end{equation}

	A comment is in order concerning the bi-orthonormalization scheme. In the previous subsection, state $\ket{E_i}$ is normalized according to $\left<\left< E_{i} \right|\ket{E_{j}} \right. =\delta_{i,j}$, which is traditionally used in the study of non-Hermitian systems with gain and loss \cite{garrison1988PhysLettA.128.177}. It is a convenient normalization convention for problems where there is no need to distinguish between the particle and hole band, as we demonstrated in the previous section when we used this normalization to prove the result in Eq. (\ref{eq:single churn number}).  In the present section, we use instead the bi-orthonormalization scheme, which is traditionally used in the study of the bosonic BdG systems \cite{blaizot96QuantumTheoryBook}. A stable state in this scheme is classified either as a particle state or as a hole state, depending on whether its norm with metric $\eta$ can be scaled to $+1$ or $-1$,
	so that particle and hole states obey different orthonormality relations given, respectively, by
	$\bra{E_i}\eta\ket{E_j} = + \delta_{i,j}$
	and $\bra{E_i}\eta\ket{E_j} = -\delta_{i,j}$.
	
	\begin{figure}[ht]
		\centering
		\includegraphics[width=0.45\textwidth]{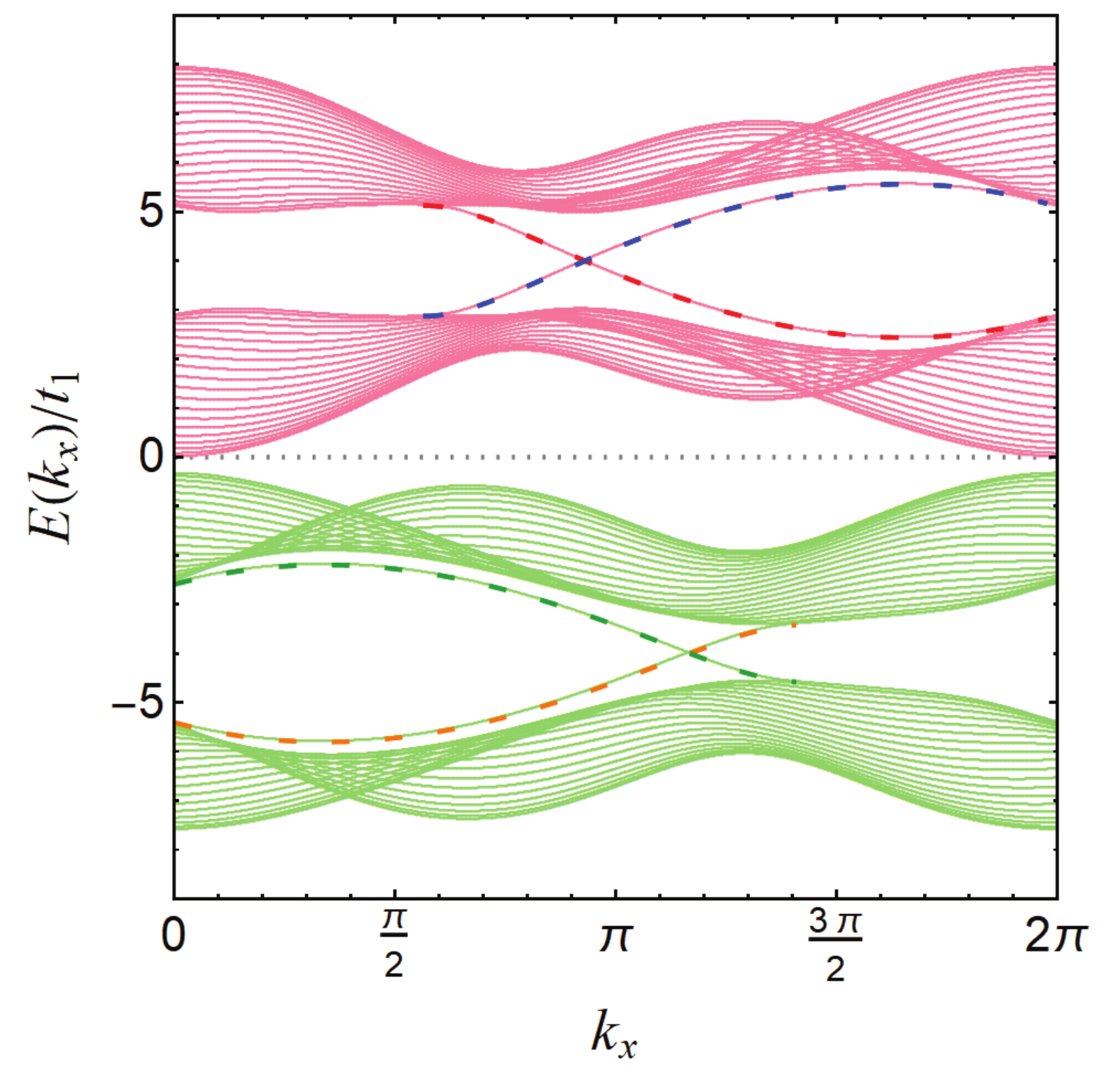}
		\caption{A typical (open-boundary) energy spectrum in a stable region with $\phi=\pi/4$ and $m=0.1\times 3\sqrt{3}t_2$. The analytical mode dispersions, $\epsilon_p^b(k_x)$ at the bottom edge (dashed red line) and  $\epsilon_p^t(k_x)$ at the top edge (dashed blue line), that lie inside the gap of the particle bands (solid red lines), are determined from Eqs. (\ref{eq:edge mode dispersions for A up}) - (\ref{eq:condition p t}). The analytical mode dispersions, $\epsilon_h^b(k_x)$ at the bottom edge (dashed orange line) and $\epsilon_h^t(k_x)$ at the top edge  (dashed green line), that lie inside the gap of the hole bands (solid green lines), are determined from Eqs. (\ref{eq:edge mode dispersions for A up hole})  -(\ref{eq:condition h t}).  Other parameters are the same as in Fig. \ref{fig:fig3}(a).}
		\label{fig:fig5}
	\end{figure}
	
	We stress that ${H}_\Uparrow(k_x) $ in Eq. (\ref{eq:H_Uparrow not a matrix}) for an open-boundary system is not a BdG Hamiltonian since it does not obey  particle-hole symmetry.   However, as we show in Appendix \ref{sec:a generalized theorem}, the aforementioned normalization involving metric $\eta$ holds for any Hamiltonian that is pseudo-Hermitian, irrespective of whether it obeys particle-hole symmetry; classification of eigenenergy states into hole and particle types is solely determined by pseudo-Hermiticity.
	
	Figure \ref{fig:fig5} displays a typical open-boundary energy spectrum in the topologically nontrivial region.  There are two particle edge modes inside the gap between two particle bulk bands (solid red lines) and two hole edge modes inside the gap between two hole bulk bands (solid green lines).  Also plotted in Fig. \ref{fig:fig5} are the edge mode dispersions of systems with semi-infinite plane geometries in the limit of weak pairing where $c\rightarrow 0$ (dashed lines).  In this limit, the particle spectrum $\epsilon_p(k_x)$ is the eigenenergy of $h_\uparrow(k_x)$
	\begin{equation}
		\label{eq:particle edge modes without pairing}
		h_\uparrow(k_x)\ket{\psi_p} = \epsilon_p(k_x)\ket{\psi_p}.
	\end{equation}
	while the hole spectrum $\epsilon_h(k_x)$ is the eigenenergy of  $-h_\uparrow(-k_x)$,  
	\begin{equation}
		\label{eq:hole edge modes without pairing}
		-h_\uparrow(-k_x)\ket{\psi_h} = \epsilon_h(k_x)\ket{\psi_h}.
	\end{equation}
	What we obtained in Sec. \ref{eq:Edge Modes Haldane Model} is the particle edge mode of an upper-half lane, a solution to Eq. (\ref{eq:particle edge modes without pairing}). The method used there can be generalized to solve, from Eqs. (\ref{eq:particle edge modes without pairing}) and (\ref{eq:hole edge modes without pairing}), all types of edge modes in semi-infinite planes. We summarize all analytical results (including, for convenience, those from in Sec. \ref{eq:Edge Modes Haldane Model}) in terms of three functions of $k_x$,
	\begin{equation}
		\begin{split}
			d(k_x) &=  \cos^{2}\phi_+(k_x) \label{eq:d(kx)}\\
			p(k_x) &= \cos\phi_+(k_x) \cos\phi_-(k_x),\\
			r(k_x) &= \cos\phi_-(k_x)- 4\cos(k_x/2)\cos\phi_+(k_x).
		\end{split}
	\end{equation}
	
	Let $i = b$ or $t$ and define $\text{sgn}(i)$ as 
	\begin{equation}
		\text{sgn}(i) = \begin{cases}
			+1 & \mbox{ if } i = b,\\
			-1 & \mbox{ if } i = t,
		\end{cases}
	\end{equation}
	where ``$b$" and ``$t$" are abbreviations representing,  respectively, the bottom and top edge of a half plane. Then, we find from Eq. (\ref{eq:particle edge modes without pairing}),
	\begin{equation}
		\label{eq:edge mode dispersions for A up}
		\epsilon_p^{i}(k_x) = q + \text{sgn}(i) \frac{ \abs{t_1}\left[m+ 2t_2r(k_x)\right]}{\sqrt{t_1^2 + 16 t_2^2 \cos^2\phi_+(k_x)}}
	\end{equation}
	gives the particle edge mode dispersion at the bottom (dashed red line), $\epsilon_p^{b}(k_x)$, under the condition,
	\begin{equation}
		\label{eq:condition p b}
		\abs{\lambda_{p,+}^b(k_x)}<1 \mbox{ and } \abs{\lambda_{p,-}^b(k_x)}<1,
	\end{equation}
	and gives the particle edge mode dispersion at the top (dashed blue line), $\epsilon_p^{t}(k_x)$, under the condition, 
	\begin{equation}
		\label{eq:condition p t}
		\abs{\lambda_{p,+}^t(k_x)}>1 \mbox{ and } \abs{\lambda_{p,-}^t(k_x)}>1,
	\end{equation}
	where 
	\begin{equation}
		\label{eq:lambda-i}
		\lambda^{i}_{p,\pm}(k_x) =\frac{a_p^{i} a^0_p \pm \sqrt{(a_p^{i}a^{0}_p)^2 - 4(1+a_p^{i})}}{2}
	\end{equation}
	are functions of 
	\begin{equation}
		\label{eq:ap-i}
		a_p^{i} (k_x) = \frac{(t_1/t_2)^2}{8d(k_x)} -\text{sgn}(i) \sqrt{\left[1 +\frac{(t_1/t_2)^2}{8d(k_x)} \right]^2 -1},
	\end{equation}
	and
	\begin{equation}
		a_p^0(k_x) =\frac{t_1^2 \cos\frac{k_x}{2} +4t_2^2p(k_x) +2t_2 m \cos\phi_+(k_x)}{ 0.5 t_1\sqrt{t_1^2 + 16t_2^2\cos^2\phi_+(k_x)}}.
	\end{equation}
	
	Similarly, we find from Eq. (\ref{eq:hole edge modes without pairing}), 
	\begin{equation}
		\label{eq:edge mode dispersions for A up hole}
		\epsilon_h^{i}(k_x) = -q -\text{sgn}(i) \frac{t_1 \left[m  - 2t_2 r(-k_x)\right]}{\sqrt{t_1^2 + 16 t_2^2 \cos^2\phi_+(-k_x)}},
	\end{equation}
	gives the hole edge mode dispersion at the bottom (dashed orange line), $\epsilon_h^{b}(k_x)$, under the condition,
	\begin{equation}
		\label{eq:condition h b}
		\abs{\lambda_{h,+}^b(k_x)}<1 \mbox{ and } \abs{\lambda_{h,-}^b(k_x)}<1,
	\end{equation}
	and gives the hole edge mode dispersion at the top (dashed green line), $\epsilon_h^{t}(k_x)$, under the condition,
	\begin{equation}
		\label{eq:condition h t}
		\abs{\lambda_{h,+}^t(k_x)}>1 \mbox{ and } \abs{\lambda_{h,-}^t(k_x)}>1,
	\end{equation}
	where 
	\begin{equation}
		\lambda_{h,\pm}^{i}(k_x) =\frac{a_h^{i} a^0_h \pm \sqrt{(a_h^{i}a^{0}_h)^2 - 4(1+a_h^{i})}}{2},
	\end{equation}
	are functions of
	\begin{equation}
		a_h^{i}(k_x) = a_p^{i}(-k_x),
	\end{equation}
	and 
	\begin{equation}
		\label{eq:a_h0(kx)}
		a_h^0(k_x) =\frac{t_1^2 \cos\frac{k_x}{2} + 4t_2^2p(-k_x) -2t_2 m \cos\phi_+(-k_x)}{ 0.5 t_1\sqrt{t_1^2 + 16t_2^2\cos^2\phi_+(-k_x)}}.
	\end{equation}

	\section{Our model with time reversal symmetry:topological properties}
	
	\label{sec:our model with time reversal symmetry}
	A bosonic system, to which the Pauli exclusion principle is not applicable, has a fascinating prospect of being made to operate as active topological matter (``lasers") where population amplification occurs in modes on boundaries \cite{schomerus2013OptLett.38.1912,poli2015NatCommun.6.6710,st-jean2017NatPhoton.11.651,zhao2018NatCommun.9.981,harari2018Science.359.eaar4003,bandres2018Science.359.eaar4005,bahari2017Science.358.636}.
	A topological amplifier in a bosonic BdG system has unstable edge modes and a stable bulk, and is founded on the ability to make dynamical instability occur preferentially in edge modes, which allows edge modes to be populated at an exponentially fast rate \cite{galilo15PhysRevLett.115.245302,ling2021PhysRevA.104.013305}.
	In Ref. \cite{ling2021PhysRevA.104.013305}, we showed that a time-reversal-symmetric system ($\phi=\pi/2$) can be made to behave like a topological amplifier, irrespective of whether the system is inversion symmetric ($m=0$) \cite{galilo15PhysRevLett.115.245302} or inversion asymmetric ($m \neq 0$). In this section,  we study our model with time-reversal symmetry and use it as an opportunity to shed more light on the physics underlying topological amplifiers.
	
	An instability in a bosonic BdG model occurs because of a level crossing between particle and hole states \cite{wu2001PhysRevA.64.061603,kawaguchi2004PhysRevA.70.043610,nakamura2008PhysRevA.77.043601}.  This insight led us to the theorem in Ref. \cite{ling2021PhysRevA.104.013305} which expresses the (purely imaginary) energy splitting of a pair of degenerate states with energy sufficiently far away from zero in terms of an unconventional commutator between the normal and abnormal parts of a BdG Hamiltonian, e.g., $H_{BdG}(k_x)$ in Eq. (\ref{eq:HBdG(k_x)}). The vanishing of this ``commutator" acts as a general and straightforward to use criterion for creating a topological amplifier from systems described by BdG Hamiltonians. In Appendix \ref{sec:a generalized theorem}, we adapt this criterion so that it can be directly applied to sub-Hamiltonians like $H_{\Uparrow}(k_x)$ in Eq. (\ref{eq:H_Uparrow kx})  which is pseudo-Hermitian but not necessarily of a BdG type.
	
	We begin by applying Eq. (\ref{eq:generalized unconventional commutator}) to determine the unconventional commutator for $\mathcal{H}_\Uparrow(k_x)$ in Eq. (\ref{eq:H_Uparrow kx}), which is found to follow
	\begin{equation}
		\label{eq:[A,B] for subhamiltonian}
		\begin{split}
			\left \lceil A,B \right\rfloor
			&= cI_-[\alpha_\uparrow(k_x)  -\alpha_\downarrow(-k_x)] \\
			& + cI_0[\beta_\uparrow(k_x) -\beta_\downarrow(-k_x)] \\
			& + cI_+[\gamma_\uparrow(k_x) -\gamma_\downarrow(-k_x)].
		\end{split}
	\end{equation}
	We next use Eqs. (\ref{eq: alpha beta gamma stripe up}) and (\ref{eq: alpha beta gamma stripe down}) to simplify Eq. (\ref{eq:[A,B] for subhamiltonian}) to
	\begin{equation}
		\label{eq:unconventional commutator for our model}
		\begin{split}
			\left \lceil A,B \right\rfloor = & 4ct_2\cos\phi\\
			&\times\left[\cos k_x I_0 + \cos\frac{k_x}{2}(I_-  + I_+)\right] \sigma_z,
		\end{split}  
	\end{equation}
	which vanishes when $\phi=\frac{\pi}{2}$.  Thus, the analysis of the sub-Hamiltonian (\ref{eq:H_Uparrow}) arrives at the same conclusion as the analysis of the total Hamiltonian (\ref{eq:H_BdG first}) in  \cite{ling2021PhysRevA.104.013305}: As long as time reversal symmetry is preserved, our system may act as a topological amplifier for which the bulk is free of dynamical instabilities.
	
	How does the existence of time-reversal symmetry for the total system affect the topological classification of the sub-Hamiltonians?  We have already shown in Sec. \ref{sec:symmetries and topological classification} that $ H_\Uparrow(\vb{k})$ is pseudo-Hermitian [Eq. \ref{eq:pseudo-hermiticity H Uparrow})] but not invariant under particle-hole symmetry [Eq. \ref{eq:not particle-hole symmetric H Uparrow}]. Furthermore, $ H_\Uparrow(\vb{k})$ is not time-reversal-symmetric even though the total system is. However, it can be shown that when $\phi=\pi/2$, $ H_\Uparrow(\vb{k})$ is invariant under chiral symmetry,
	\begin{equation}
		\Gamma H^\dag_\Uparrow(\vb{k}) \Gamma^{-1} = -H_\Uparrow(\vb{k})
	\end{equation}
	where $\Gamma=\tau_y$ is a matrix representing chiral symmetry.  As a consequence, the symmetry class of this sub-Hamiltonian must be built upon AZ symmetry class AIII. (As a side note, we mention that the AZ symmertry class AIII was known to occur in non-Hermitian systems without pseudo-Hermiticity, examples of which include a passive dimerized photonic crystal in one dimension \cite{weimann2017NatureMaterials.16.433} and an active array of microring resonators \cite{parto2018PhysRevLett.120.113901}. Both Refs. \cite{weimann2017NatureMaterials.16.433} and \cite{parto2018PhysRevLett.120.113901} use variants of the Su-Schrieffer-Heeger model \cite{su1979PhysRevLett.42.1698} with loss and gain, whose topological phases, when classified by a real line gap, are characterized, according to the 38-fold way \cite{kawabata2019PhysRevX.9.041015},  by a $\mathbb{Z}$ topological invariant identified as the winding number.)
	
	For pseudo-Hermitian Hamiltonians,  whether $\eta$ commutes and anticommutes with other symmetry operators plays an important role in the topological classification. In our model, $\eta=\tau_z$ anticommutes with the chiral symmetry operator $\Gamma=\tau_y$, i.e., $\{\eta,\Gamma\}=0$. Thus, $H_\Uparrow(\vb{k})$  belongs to class AIII + $\eta_-$, whose topological invariant, when classified by a real line gap, is $\mathbb{Z}$ in two dimensions,  which is again identified as the Chern number of the Haldane model in Eq. (\ref{eq:haldane's Chern number}), regardless of pairing interactions.  In a nutshell, the criterion that the unconventional commutator be made to vanish may impose additional symmetries on a topological system, thereby altering the classification of the topological system, e.g., it changes our model from the A$+\eta$ class without time-reversal symmetry  to the AIII$+\eta_-$ class with time-reversal symmetry.
	
\begin{figure}[ht]
		\centering
		 \includegraphics[width=0.45\textwidth]{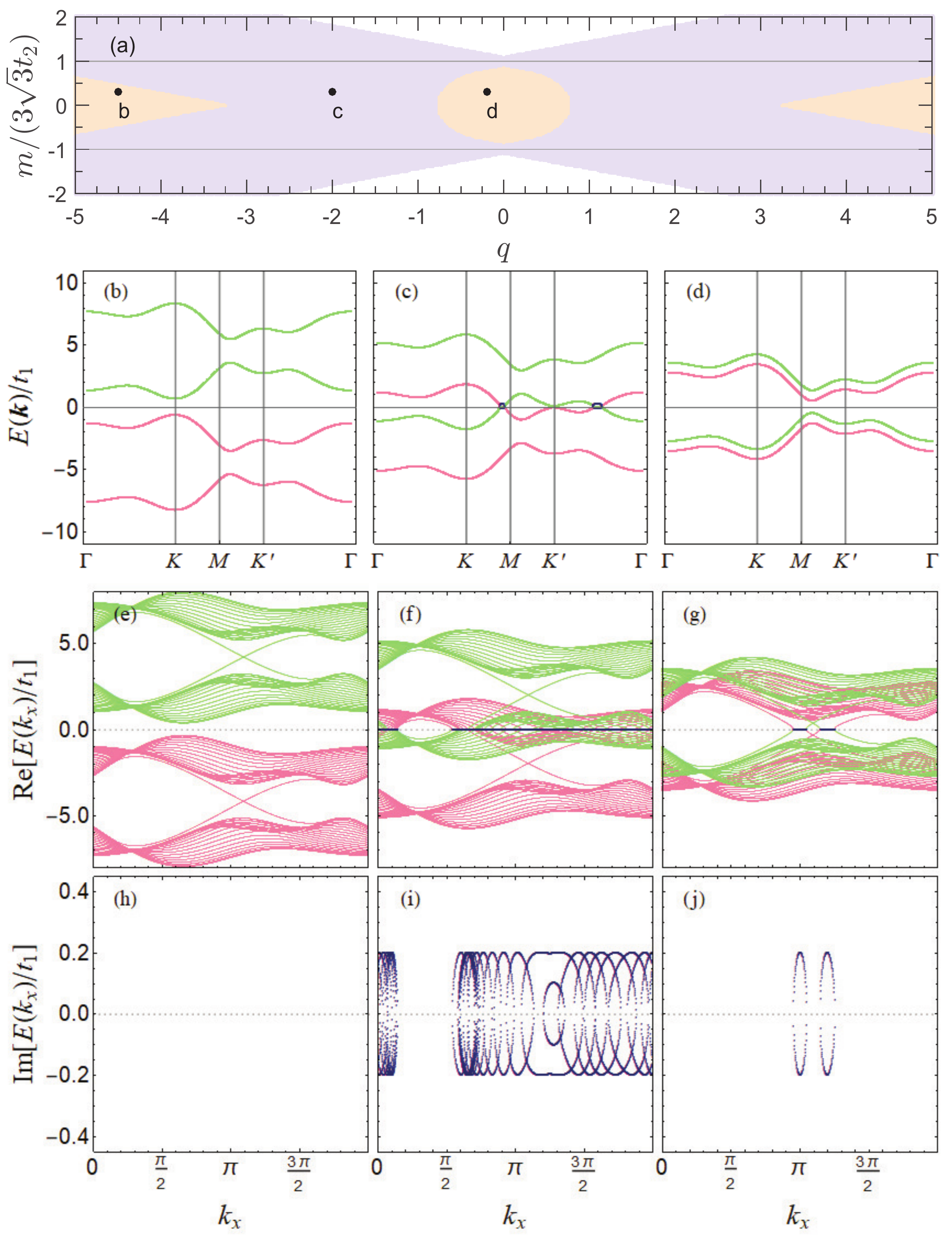}
		\caption{Panel (a) is a phase diagram in the $q-m$ space for $\phi=\pi/2$ and $c=0.2t_1$.  Panels (b)-(d) display the Bloch energy spectra, sampled, respectively, at points b, c and d as indicated in panel (a), where $m = 0.36\times 3\sqrt{3} t_2$ and $q=-4.2t_1$ at point b, $q=-2t_1$ at point c, and $q=-0.4t_1$ at point d.  
			Panels (e)-(g)[(h)-(j)] are the real (imaginary) parts of the corresponding open-boundary energy spectra at points b, c and d, respectively. For panels (e)-(g), red and green indicate, respectively, hole and particle states just as in Fig. \ref{fig:fig5} while dark blue indicates unstable states.
			The other parameters, $t_1$ and $t_2$, are the same as in Fig. \ref{fig:fig3}(a).}
		\label{fig:fig6}
	\end{figure}

	Figure \ref{fig:fig6}(a) displays the phase diagram in the $q-m$ space for the $J_z-\Uparrow$ subsystem with $\phi=\pi/2$.  The topologically nontrivial (trivial) phase would lie inside (outside) the region between the two horizontal lines at $m=-3\sqrt{3}t_2$ and $m=+3\sqrt{3}t_2$, irrespective of $q$, if pairing interactions were absent.  As expected, in the presence of pairing interactions, an unstable region (light purple) emerges, thereby reducing the region with a topologically nontrivial phase (orange) as well as the region with a topologically trivial phase (white).

	The bulk energy spectra for a periodic system can be computed analytically from Eq. (\ref{eq:bulk energy bands periodic}) with $\phi=\pi/2$. For models with open boundaries, we note that for a time-reversal invariant total system, $ h_\downarrow(-k_x) = h_\uparrow(k_x)$
	according to Eq. (\ref{eq:A_spin_up(k_x)}), so that  ${H}_\Uparrow(k_x)$ in Eq. (\ref{eq:H_Uparrow not a matrix}) is invariant under $h_\uparrow(k_x)$,
	\begin{equation}
		[h_\uparrow(k_x),{H}_\Uparrow(k_x)]=0.
	\end{equation}
	This indicates that $\omega_n$, which is the eigenenergy of $h_\uparrow(k_x)$,
	\begin{equation}
		\label{eq:eigenvalue of h uparrow (kx)}
		h_\uparrow(k_x) \ket{\omega_n(k_x)} = \omega_n(k_x) \ket{\omega_n(k_x)},
	\end{equation}
	is a good quantum number.  As such, in the space $\{\ket{\omega_n(k_x)} \}$ where $ h_\uparrow(k_x)$ is diagonal, ${H}_\Uparrow(k_x)$ in Eq. (\ref{eq:H_Uparrow not a matrix}) is block diagonal,
	\begin{equation}
		{H}_\Uparrow(k_x) = \sum_n \left[ \omega_n(k_x)\tau_z +ic\tau_y \right] \ket{\omega_n(k_x)}\bra{\omega_n(k_x)},
	\end{equation}
	where
	\begin{equation}
		\omega_n(k_x)\tau_z +ic\tau_y 
		= \begin{pmatrix}
			\omega_n(k_x) & c\\
			-c & -\omega_n(k_x)
		\end{pmatrix},
	\end{equation}
	is a two state Hamiltonian in Nambu space. The energy spectra are then given by  
	\begin{equation}
		E_{n,\pm}(k_x) = \pm \sqrt{\omega^2_n(k_x) - c^2},
	\end{equation}
	which appear in pairs of ($E_{n,+}(k_x)$, $E_{n,-}(k_x) = -E_{n,+}(k_x)$) as anticipated from an underlying sublattice symmetry: $\sigma_x {H}_\Uparrow(k_x) \sigma_x^{-1} = -{H}_\Uparrow(k_x) $.
	
	Figures \ref{fig:fig6}(b)-\ref{fig:fig6}(d) display Bloch spectra (second column), real open-boundary spectra in Figs. \ref{fig:fig6}(e)-\ref{fig:fig6}(g) (third row), and the corresponding imaginary open-boundary spectra in Figs. \ref{fig:fig6}(h)-\ref{fig:fig6}(j) (fourth row). 
	Figures \ref{fig:fig6}(b), \ref{fig:fig6}(e), and\ref{fig:fig6}(h), made for a stable state at point b in Fig. \ref{fig:fig6}(a) with a relative large $\abs{q}$, are free of any complex energies. Figures \ref{fig:fig6}(c),\ref{fig:fig6}(f), and \ref{fig:fig6}(i), made for an unstable state at point c in Fig. \ref{fig:fig6}(a) with a reduced $\abs{q}$, contain complex energies in its bulk bands, and Figs. \ref{fig:fig6}(d),\ref{fig:fig6}(g), and \ref{fig:fig6}(j), made for a topological amplifier state at point d in Fig. \ref{fig:fig6}(a) with a very small $\abs{q}$, are all real with the exception of some energies of the edge modes, which are complex.
	
	The edge mode dispersions in Fig. \ref{fig:fig6} are found in excellent agreement (not shown) with those on half-plane geometries, which, when pairing terms are taken into consideration, are given by
	\begin{equation}
		E_{edge}^i(k_x) = \pm \sqrt{\epsilon_p^i - c^2}, \quad  i = b, t,
	\end{equation}
	where
	\begin{equation}
		\epsilon^{i}_p(k_x)= q +\text{sgn}(i) \abs{t_1}\frac{6t_2\sin k_x + m }{\sqrt{t_1^2 + 16 t_2^2 \sin^2 \frac{k_x}{2}}}
	\end{equation}
	are the particle edge mode dispersions in the absence of pairing interactions,  i.e., Eq. (\ref{eq:edge mode dispersions for A up}) when $\phi=\pi/2$.  Here, $E_{edge}^i(k_x)$ become the particle edge mode dispersions, $E_{p}^i(k_x)$, when $\omega_n(k_x) + E_{edge}^i(k_x)>0$ and the hole edge mode dispersions, $E_{h}^i(k_x)$, when $\omega_n(k_x) + E_{edge}^i(k_x)< 0$.
	
	Note that complex energies appear only when particle and hole spectra cross near zero energy, irrespective of whether such states belong to bulk bands [Fig. \ref{fig:fig6}(f) for point $c$ in Fig. \ref{fig:fig6}(c)] or to edge modes [Fig. \ref{fig:fig6}(g) for point d in Fig. \ref{fig:fig6}(a)], confirming that the condition $\left \lceil A,B \right\rfloor = 0$ in Appendix \ref{sec:a generalized theorem} \cite{ling2021PhysRevA.104.013305} cannot prevent any degenerate states, other than those with energies away from zero, from undergoing dynamical instability. We are thus led to conclude that a necessary condition for our model to act as a topological amplifier would be inversion (or crossing) between particle and hole edge modes near zero energy. 
	
	It is worth noting that it is not uncommon for a system to enter an unconventional phase with novel properties under some forms of inversion.  An optical cavity emits  highly coherent laser light when there is a population inversion between upper and lower energy levels of the atoms inside the cavity \cite{lamb1974laserPhysics}. An insulator becomes a topological insulator with topologically protected gapless edge modes when there is an inversion between conduction and valance bands of the insulator \cite{bernevig2006Science.314.1757}. In the present system, we can reduce $\abs{q}$ and cause particle and hole edge modes to cross, thereby exchanging/reversing the particle and hole roles, near the zero energy level, leading to amplification of the edge modes.

\begin{figure}
		\centering
		\includegraphics[width=0.45\textwidth]{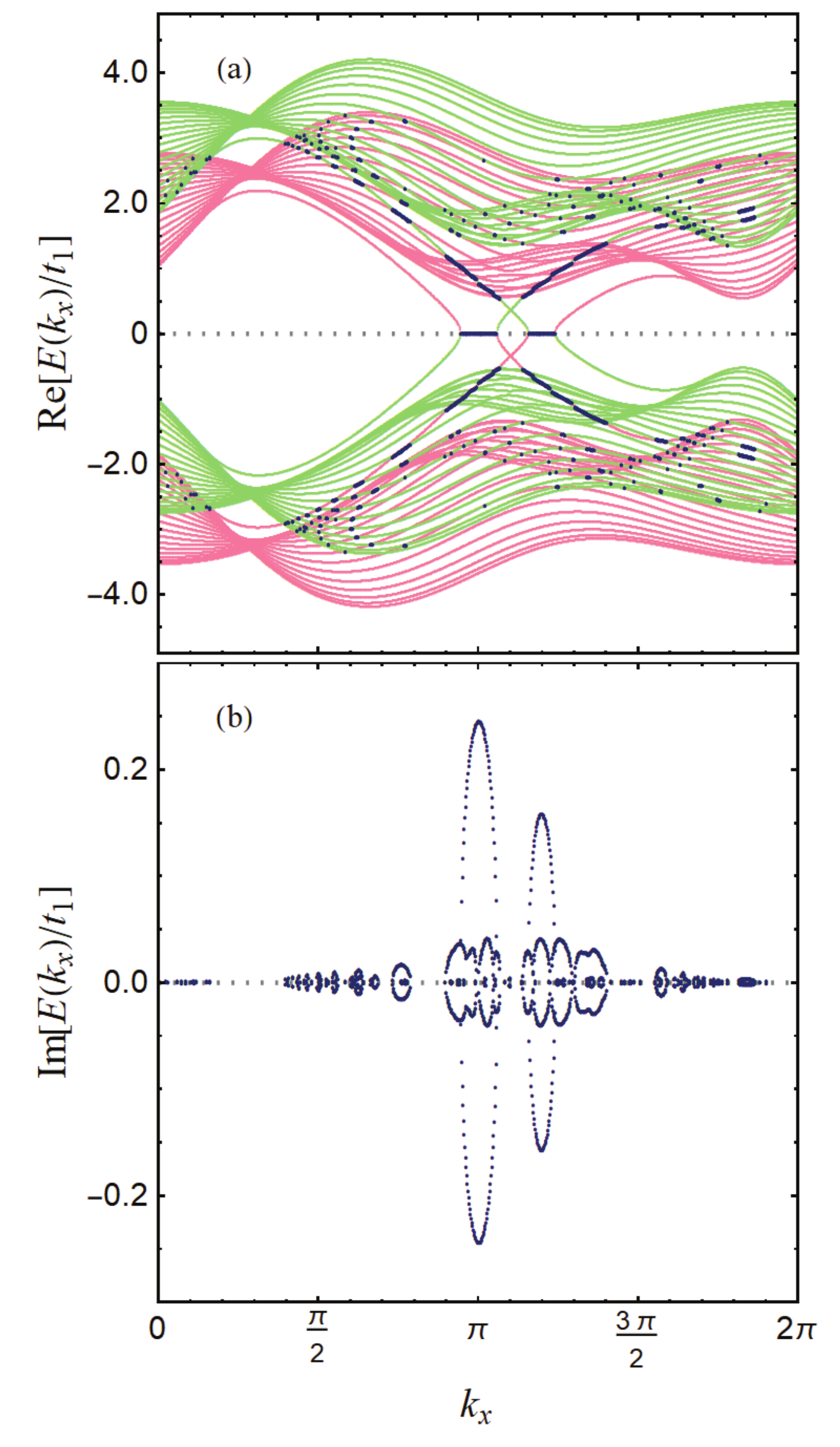}
		\caption{The open-boundary energy spectrum if the pairing term in $H_\Uparrow(k_x)$ [the last term in Eq. (\ref{eq:H_Uparrow 1})] were to change from $i c\tau_y \sigma_0$ to $i c\tau_y \sigma_0 + i c' \tau_y \sigma_z$ (instead of just $i c\tau_y \sigma_0$) with $c'=0.1$. Other parameters are the same as in Fig. \ref{fig:fig6}(g). }
		\label{fig:fig7} 
	\end{figure}

	Before concluding this section, we consider how the spectra of $H_\Uparrow(k_x)$ might modify if the pairing term were to change to $i c \tau_y \sigma_0 + i c' \tau_y \sigma_z$, which contains a component proportional to $\sigma_z$.  We compute the ``commutator"  using Eq. (\ref{eq:generalized unconventional commutator}) and find 
	\begin{equation}
		\begin{split}
			& \left \lceil A,B \right\rfloor  = (\cdots) \\
			& + 4 c't_2\cos\phi [I_0 \cos k_x+ I_-  \cos\frac{k_x}{2}  + I_+ \cos\frac{k_x}{2}]\sigma_0 \\
			&+ i4c't_1 \cos\frac{k_x}{2} I_0 \sigma_y
			+ 2t_1 c'\left( I_-\sigma_+ - I_+\sigma_-\right).
		\end{split}    
	\end{equation}
	where $(\cdots)$ is the right-hand side of Eq. (\ref{eq:unconventional commutator for our model}). Thus, even when the total system is time-reversal symmetric, where both $(\cdots)$ and the second line vanish, the unconventional commutator is nonzero because of the last line.  For a system with such pairing interactions, not only do the boundaries separating trivial from nontrivial topological phases depend on pairing interactions, as in Ref. \cite{furukawa2017NewJournalOfPhysics.17.115014}, since $ic'\tau_y \sigma_z$ breaks inversion symmetry, but also the bulk bands are unstable even when the total system is time-reversal-symmetric, as illustrated in Fig. \ref{fig:fig7}.  We use this example to stress a point in Ref. \cite{ling2021PhysRevA.104.013305}: It is not so much time-reversal symmetry as the vanishing of the unconventional commutator that indicates if a system has the potential to be made into a topological amplifier.
	
	\section{Conclusion}
	\label{sec:conclusion}

	In this work, we explored the topological properties of quadratic Hamiltonians for spin-$1/2$ bosons with conserved magnetization in a honeycomb lattice with both periodic and open boundaries. 
	
	We used this study as an opportunity to construct a relatively simple analytical description for the edge modes of the Haldane model in semi-infinite planes, which not only lays the foundation for an analytical understanding of the edge modes in our model, but also will be useful for many other models across a broad array of disciplines, that are built upon copies of the Haldane model. 
	
	In the spirit of the 38-fold way for topological classification of non-Hermitian Hamiltonians \cite{kawabata2019PhysRevX.9.041015,zhou2019PhysRevB.99.235112}, we carried out an in-depth analysis of the symmetry properties of the Bloch Hamiltonian and identified our pseudo-Hermitian system to be made up of either two copies of  the AIII  $+ \eta_-$ or two copies of the A $+\eta$ symmetry classes depending on whether the total system is time-reversal symmetric.  An important goal of the 38-fold way \cite{kawabata2019PhysRevX.9.041015,zhou2019PhysRevB.99.235112} is to help find and design novel symmetry-protected topological lasers. Indeed, the 38-fold way helped us find that the edge mode amplifier,  which Galilo et al. \cite{galilo15PhysRevLett.115.245302} proposed before the development of the 38-fold way, is classified as a topological ``laser" consisting of two copies of the AIII $+\eta_-$ symmetry class. Furthermore, we verified rigorously, starting from a surface integral of  Berry curvature over the first Brillouin zone, where the Berry curvature is constructed from right and left eigenvectors within the context of non-Hermitian physics, that a system with a dynamical stable bulk is characterized by a single Chern number, independent of pairing interactions, as predicted from symmetry analysis. 
	
	We generalized a theorem presented in Ref.  \cite{ling2021PhysRevA.104.013305} for a BdG Hamiltonian so that the theorem is now expressed in terms of an unconventional commutator between the number-conserving and number-nonconserving part of pseudo-Hermitian Hamiltonians which are similar to but less restrictive than the BdG Hamiltonian.  We applied this generalized theorem to learn directly from the sub-Hamiltonians that it is not so much time-reversal symmetry but rather the vanishing of the unconventional commutator which foretells whether a system can be made to act as a topological amplifier.

	\appendix
	\section{An implementation of Hamiltonian (\ref{eq:H_half}) in a spin-1 cold atom system}
	\label{sec: detailed description of our model}

	A possible realization of the model we explore in this paper is in a honeycomb optical lattice with interacting spin-1 bosons in the context of ultracold atoms \cite{ho1998PhysRevLett.81.742,ohmi1998doi:10.1143/JPSJ.67.1822,law1998PhysRevLett.81.5257}. (Part of what we present in this Appendix can be found in Refs. \cite{galilo15PhysRevLett.115.245302,ling2021PhysRevA.104.013305}.)
	We describe the bosons by a field vector $\hat{b}'_{\vb{i}}=(\hat{b}'_{\vb{i},+1},\hat{b}'_{\vb{i},0}, \hat{b}'_{\vb{i},-1})$ with $\hat{b}'_{\vb{i},\alpha}$ the annihilation operator for a boson with the spin-$\alpha$ component at site $\vb{i}$ and by a spin vector $\vb{S} = S_x \hat{i} + S_y \hat{j} + S_z \hat{k}$ with each of its components being a $3\times 3$ matrix in the spin-1 internal space. Furthermore, $S_0$ is used to denote the identity matrix in this same spin space.
	We model such a system with a grand canonical Hamiltonian 
	\begin{equation}
		\label{eq:H first}
		\hat{H} = \hat{H}_{hop} +\hat{H}_{pot} + \hat{H}_{col} - \mu\sum_{\vb{i}} \hat{b}'^\dag_{\vb{i}}S_0\hat{b}'_{\vb{i}}
	\end{equation}
	which, in addition to $\mu\sum_{\vb{i}} \hat{b}'^\dag_{\vb{i}}S_0\hat{b}'_{\vb{i}}$ with $\mu$ being the chemical potential introduced to conserve the total atom number, is made up of three pieces, which we now explain in some details.  
	
	The first piece in Eq. (\ref{eq:H first}) is given by
	\begin{equation}
		\label{eq:HaldaneModel}
		\hat{H}_{hop} = -t_1 \sum_{\left< \vb{ij} \right>} \hat{b}'^\dag_{\vb{i}} S_0 \hat{b}'_{\vb{j}} + t_2\sum_{\left<\left< \vb{ij} \right> \right>} e^{-i\nu_{\vb{ij}} \phi} \hat{b}'^\dag_{\vb{i}} S_z \hat{b}'_{\vb{j}}
	\end{equation}
	where the first term describes hopping of an atom between nearest neighbors (NN) with real amplitude $t_1$ and the second one describes hopping of an atom between the next-nearest neighbors (NNN) with complex amplitude $t_2e^{-i\nu_{\vb{ij}}\phi}$ where the phase changes sign, $\nu_{\vb{ij}}=\pm$, periodically in the manner of Kane and Mele \cite{kane2005PhysRevLett.95.146802}. The latter hopping is the generalization of the hopping  first introduced by Haldane for spinless fermions in graphene \cite{haldane88PhysRevLett.61.2015}.
	
	The second piece in Eq. (\ref{eq:H first}) takes the form 
	\begin{equation}
		\label{eq:H-pot}
		\hat{H}_{pot} = 
		q' \sum_{\vb{i}} \hat{b}'^\dag_{\vb{i}} S_z^2 \hat{b}'_{\vb{i}} 
		+ m \sum_{\vb{i}} \xi_{\vb{i}}  \hat{b}'^\dag_{\vb{i}} S_z^2 \hat{b}'_{\vb{i}},
	\end{equation}
	which summarizes the onsite potential (with staggering), which can be generated by a magnetic field through the quadratic Zeeman shift for a spinor condensate \cite{stamperKurn2013RevModPhys.85.1191} or by eletromagnetic waves (such as the microwaves in Ref. \cite{gerbier2006PhysRevA.73.041602}) through the ac-Stark shift. Here, $\xi_{\vb{i}}=+1$ for sites on sublatice $A$ and $-1$ for sites on sublatice $B$. 
	
	The third piece in Eq. (\ref{eq:H first}) accounts for the on-site two-body interaction that preserves spin rotation invariance \cite{ho1998PhysRevLett.81.742}
	\begin{equation}
		\hat{H}_{col} = \frac{c_0}{2}\sum_{\vb{i}}:\left(\hat{b}'^\dag_{\vb{i}} S_0 \hat{b}'_{\vb{i}}\right)^2 :+\frac{c_2}{2}\sum_{\vb{i}}:\left( \hat{b}'^\dag_{\vb{i}} \vb{S} \hat{b}'_{\vb{i}} \right)^2:
	\end{equation}
	where $c_0$ and $c_2$ describe, respectively, the magnitude of the density and spin interactions, and the terms between $:$  and $:$ are to follow normal orderings.
	
	This model, for a purely imaginary NNN hopping amplitude ($\phi = \pi/2$) and in the absence of the staggered potential ($m = 0$), reduces to the one proposed by Galilo et al. \cite{galilo15PhysRevLett.115.245302} for realizing topological atom amplifiers characterized with nonreciprocal amplification of matter fields in the chiral modes that are hosted in the system boundaries.

	Cold atom systems provide an excellent platform for the exploration of the dynamics of a system driven out of equilibrium, thanks to the ability to tune many of its key parameters with high precision.  In the present work,  we follow \cite{galilo15PhysRevLett.115.245302} and consider a quench process in which we abruptly change the onsite potential away from those that support an initial spinor condensate in a polar state $b'_0(0,1,0)$, where we assume that the spin-0 component be in a spatially uniform coherent state $\ket{b'_0}$, i.e., $\hat{b}'_{i,0}\ket{b'_0} = b'_0\ket{b'_0}$ (where $b'_0$ will be taken as a real number without loss of generality).  The polar condensate is then characterized by a chemical potential
	\begin{equation}
		\mu = -3t_1 + c_0 n_B,
	\end{equation}
	which we obtain by minimizing the energy per lattice site: $ (-3t_1 -\mu) b'^2_0 + c_0 b'^4_0/2$, where $n_B\equiv b'^2_0$ is the filling factor for the condensed bosons.  Such a polar state is energetically favored as long as $q'$ is set to a sufficiently large positive value.  
	
	Following the usual practice (see, for example, Ref. \cite{kain2014PhysRevA.90.063626}), we apply the Bogoliubov perturbation ansatz, $\hat{b}'_{\vb{i}} \approx (0,\sqrt{n_B},0) + \hat{b}_{\vb{i}}$, to the postquench Hamiltonian where the system parameters are fixed at their quenched values, with $\hat{b}_{\vb{i}}$  the bosonic operator describing the quantum fluctuation around the initial state. The treatment of the two-body collision, $\hat{H}_{col} $, which is by far the most complicated term, is significantly simplified thanks to the average spin vector of a spinor condensate being zero in a polar state \cite{ho1998PhysRevLett.81.742}.  This, along with the fact that the remaining terms in Eq. (\ref{eq:H first}) preserve the $S_z$-rotation invariance,  makes the (postquench) Hamiltonian under the Bogoliubov approximation, where the Hamiltonian is expanded in the second order in $\hat{b}_{\vb{i}}$, block diagonal,
	\begin{equation}
		\hat{H} = \hat{H}_{0} \oplus \hat{H}_{1/2},
	\end{equation}
	where
	\begin{equation}
		\begin{split}
			\hat{H}_{0} =& - t_1 \sum_{\left< \vb{ij} \right>} \hat{b}_{\vb{i},0}^\dag \hat{b}_{\vb{j},0} + (2c_0n_B-\mu)\sum_{\vb{i}} \hat{b}^\dag_{i,0}\hat{b}_{i,0} \\ &+\frac{c_0n_B}{2}\sum_{\vb{i}}(\hat{b}_{\vb{i},0}\hat{b}_{\vb{i},0}+H.c.)
		\end{split} 
	\end{equation}
	is the Hamiltonian in the $\ket{S=1,M=0}$ spin space and $\hat{H}_{1/2}$ is the Hamiltonian in the (pseudo) spin-$1/2$ space $\mathcal{H}_s$ defined in Eq. (\ref{eq:spin space}) with $\ket{\uparrow}\equiv \ket{S=1,M=+1}$ and $ \ket{\downarrow} \equiv \ket{S=1,M=-1}) $.  Explicitly, $\hat{H}_{1/2}$ is given by Eq. (\ref{eq:H_half}), in which
	\begin{equation}
		c = n_Bc_2,
	\end{equation}
	and 
	\begin{equation}
		q = q'+3t_1+c.
	\end{equation}
	
	As can be seen, $\hat{H}_0$ is the Hamiltonian for a single-component Bose-Hubbard model in the Bogoliubov approximation, which is known to support a superfluid state as long as the two-body onsite interaction is repulsive ($c_0>0$) and relatively weak. Indeed, a simple analysis shows that its excitation spectrum consists of two stable positive branches (and their negative counterparts) with the lower branch 
	\begin{equation}
		\sqrt{ (3t_1 -t_1\abs{\Gamma_{\vb{k}}}) (3t_1 -t_1\abs{\Gamma_{\vb{k}}} + 2n_Bc_0) }
	\end{equation}
	featuring, as expected, a Goldstone mode at $\vb{k}=0$, and a phonon dispersion in the small momentum region near the Goldstone mode, where $\Gamma_{\vb{k}}(\leq 3)$ is defined in Eq. (\ref{eq:Gamma(k)}).
	
	Thus, within the Bogoliubov approximation, the dynamics of the spin-0 component is decoupled from that of the spin-1/2 Bose system. Moreover, $\hat{H}_0$ does not possess any quench parameters and the spin-0 component remains a spectator for a quench process involving a sudden change of the onsite potential.  Nevertheless, it is the superfluid in the spin-0 component that induces the superfluid pairing [i.e., the last line in Eq. (\ref{eq:H_half})] which is an essential ingredient in this spin-$1/2$ bosonic BdG Hamiltonian. In this work, we confine our interest to topological properties (i.e., bulk invariants, edge modes, and symmetry classifications).  We refer the readers, who are interested in how topology affects quench dynamics, to Ref. \cite{galilo15PhysRevLett.115.245302} for details.
	
	\section{The steps from Eq. (\ref{eq:D_(n alpha)}) to Eq. (\ref{eq: D n alpha})}
	\label{sec:simplification}
	In this appendix, we show that the summation in Eq. (\ref{eq:D_(n alpha)}) in the main text,
	\begin{equation}
		D_{n\alpha}=\sum_{\beta}\frac{\left<\left< n,\alpha \right| \right. \tau_z\ket{m,\beta}\left<\left<m,\beta \right| \right. \tau_z \ket{n,\alpha}}{(\alpha E_{n} -\beta E_{m})^2} 
	\end{equation}
	can be simplified to $1/(\omega_n - \omega_m)^2$, where $n\neq m$, $n, m = 1, 2$, and $\alpha, \beta = +, -$. We begin by noting that, for a given $\alpha$, $\beta$ can equal either $\alpha$ or $-\alpha$ so that upon expansion, $D_{n\alpha}$ becomes
	\begin{equation}
		\begin{split}
			D_{n\alpha}= & \frac{\left<\left< n,\alpha \right| \right. \tau_z\ket{m,\alpha}\left<\left<m,\alpha \right| \right. \tau_z \ket{n,\alpha}}{( E_{n} -E_{m})^2} \\
			+ & \frac{\left<\left< n,\alpha \right| \right. \tau_z\ket{m,-\alpha}\left<\left<m,-\alpha \right| \right. \tau_z \ket{n,\alpha}}{(E_{n} +E_{m})^2}.
		\end{split}
	\end{equation}
	We replace the left and right eigenvectors with Eq. (\ref{eq:left and right vectors}) and express $D_{n\alpha}$ in terms of the Bogoliubov parameters 
	\begin{equation}
		\begin{split}
			\label{eq:D_(n alpha) uv appendix}
			D_{n\alpha}= & \frac{(u_{n\alpha}u_{m\alpha} + v_{n\alpha}v_{m\alpha})^2}{( E_{n} -E_{m})^2} \\
			+ & \frac{(u_{n\alpha}u_{m\bar{\alpha}} + v_{n\alpha}v_{m\bar{\alpha}})^2}{(E_{n} +E_{m})^2}.
		\end{split}
	\end{equation}
	where $u_{m\bar{\alpha}}\equiv u_{m(-\alpha)}$ and $v_{m\bar{\alpha}}\equiv v_{m(-\alpha)}$. 
	In Eq. (\ref{eq:D_(n alpha) uv appendix}), we find, with the help of Eq. (\ref{eq:uv alpha}), that
	\begin{equation}
		\begin{split}
			& (u_{n\alpha}u_{m\alpha} + v_{n\alpha}v_{m\alpha})^2 \\
			=&\frac{\left[ c^2 - (\omega_n - \alpha E_n)(\omega_m - \alpha E_m)\right]^2}{4 E_n E_m (\omega_n - \alpha E_n) (\omega_m - \alpha E_m)} \\
			& (u_{n\alpha}u_{m\bar{\alpha}} + v_{n\alpha}v_{m\bar{\alpha}})^2\\
			=&\frac{\left[ c^2 - (\omega_n - \alpha E_n)(\omega_m + \alpha E_m)\right]^2}{-4 E_n E_m (\omega_n - \alpha E_n) (\omega_m + \alpha E_m)}
		\end{split}
	\end{equation}
	which, with the help of the identity
	\begin{equation}
		c^2 = \sqrt{(\omega_n^2 - \alpha^2 E_n^2)(\omega_m^2 - \alpha^2 E_m^2)}
	\end{equation}
	simplifies to
	\begin{equation}
		\label{eq: (uu+vv)2 alpha - alpha}
		\begin{split}
			(u_{n\alpha}u_{m\alpha} + v_{n\alpha}v_{m\alpha})^2 =& \frac{1}{2} + \frac{\omega_n\omega_m + c^2}{2E_nE_m}\\
			(u_{n\alpha}u_{m\bar{\alpha}} + v_{n\alpha}v_{m\bar{\alpha}})^2 = &  \frac{1}{2} -
			\frac{\omega_n\omega_m + c^2}{2E_nE_m}
		\end{split}
	\end{equation}
	which are independent of $\alpha$. Inserting Eq. (\ref{eq: (uu+vv)2 alpha - alpha}) into Eq. (\ref{eq:D_(n alpha) uv appendix}), we have 
	\begin{equation}
		\begin{split}
			D_{n\alpha} = &  \frac{\omega_n\omega_m + c^2}{2E_nE_m} \times \\
			&  \left[ \frac{1}{(E_n-E_m)^2} -\frac{1}{(E_n + E_m)^2}\right]\\
			& +\frac{1}{2}
			\left[ \frac{1}{(E_n-E_m)^2} + \frac{1}{(E_n + E_m)^2} \right]
		\end{split}
	\end{equation}
	which simplifies to
	\begin{equation}
		D_{n\alpha}=   \frac{E^2_n+E^2_m + 2(\omega_n\omega_m +c^2)}{(E_n^2-E_m^2)^2}
	\end{equation}
	Finally, when use of Eq. (\ref{eq:E_n(k)}) is made, we arrive at Eq. (\ref{eq: D n alpha}) in the main text.
	
	\section{Pseudo-Hermitian Hamiltonians and particle and hole states}
	\label{sec:pseudo-Hermitian Hamiltonian}
	Pseudo-Hermiticity of a Hamiltonian $H$ in non-Hermitian physics is defined as
	\begin{equation}
		\eta H^\dag \eta^{-1} = H, \quad \eta^2=1,
	\end{equation}
	which is a similarity transformation characterized by a unitary and Hermitian matrix $\eta$, i.e.,
	\begin{equation}
		\eta\eta^\dag = \eta^\dag\eta =1,\quad \eta = \eta^\dag.
	\end{equation}
	Note that because $ \eta^2=1$, the same pseudo-Hermiticity condition can also be written as
	\begin{equation}
		\label{eq:pseudo Hermiticity eta}
		H^\dag = \eta H \eta^{-1}, \quad \eta^2=1.
	\end{equation}
	
	In what follows, we show that states in any pseudo-Hermitian Hamiltonian can be divided into either particle states or hole states according to whether their norm with metric $\eta$ is $+1$ or $-1$.  Let $\ket{\phi_n}$ be the eigenstate of $H$ with eigenvalue $\mathcal{E}_n$,
	\begin{equation}
		H\ket{\phi_n} =\mathcal{E}_n \ket{\phi_n},\label{eq:H(k) generalized}
	\end{equation}
	or 
	\begin{equation}
		\bra{\phi_m} H^\dag =\mathcal{E}^*_m \bra{\phi_m} \label{eq:H(k)* generalized}.
	\end{equation}
	On one hand, multiplying Eq. (\ref{eq:H(k) generalized}) from the left with $\bra{\phi_m}\eta$, we have
	\begin{equation}
		\bra{\phi_m}\eta H(k)\eta^{-1} \eta \ket{\phi_n}=\mathcal{E}_n \bra{\phi_m}\eta\ket{\phi_n},
	\end{equation}
	which, when use of the pseudo-Hermiticity condition (\ref{eq:pseudo Hermiticity eta}) is made, is changed to
	\begin{equation}
		\label{eq:energy n}
		\bra{\phi_m} H^\dag(k)\eta \ket{\phi_n}=\mathcal{E}_n \bra{\phi_m}\eta\ket{\phi_n}.
	\end{equation}
	On the other hand, multiplying Eq. (\ref{eq:H(k)* generalized}) from the right with $\eta\ket{\phi_n}$, we have
	\begin{equation}
		\bra{\phi_m} H^\dag(k)\eta\ket{\phi_n} =\mathcal{E}^*_m \bra{\phi_m}\eta\ket{\phi_n}
	\end{equation}
	which, when combined with Eq. (\ref{eq:energy n}), leads to 
	\begin{equation}
		\label{eq:prelude to bi-orthonormality generalized}
		\left(\mathcal{E}^*_m - \mathcal{E}_n\right) \bra{\phi_m}\eta \ket{\phi_n} = 0.
	\end{equation}
	
	From Eq. (\ref{eq:prelude to bi-orthonormality generalized}), we conclude that $\bra{\phi_n}\eta\ket{\phi_n}$, the norm with metric $\eta$, is zero for an unstable eigenstate (i.e., a state with a complex eigenvalue) and is either a positive or a negative real number for a stable state (i.e., a state with a real eigenvalue).  In arriving at the latter conclusion, we have considered the fact that (a) the norm with metric $\eta$ is an average of a Hermitian matrix, $\eta$, and is therefore always real, and (b) in contrast to the usual norm, which is always positive, the norm with metric $\eta$ is not guaranteed to be a positive number. 
	
	In summary, Eq. (\ref{eq:prelude to bi-orthonormality generalized}) means that the eigenstates of a pseudo-Hermitian Hamiltonian, $H(k)$ in Eq. (\ref{eq:generalized Hamiltonian}), still obey the same bi-orthonormality relation as that for a BdG Hamiltonian, in which a stable state is classified either as a particle state $\ket{p}$ where $\bra{p}\eta\ket{p}=1$ or as a hole state $\ket{h}$, where $\bra{h}\eta\ket{h}=-1$.

	\section{Generalized theorem}
	\label{sec:a generalized theorem}
	In this appendix, we consider a Hamiltonian
	\begin{equation}
		\label{eq:generalized Hamiltonian}
		H(k) = 
		\begin{pmatrix}
			A_1(k) & B_1(k)\\
			-B^*_2(-k) & -A^*_2(-k)
		\end{pmatrix}
	\end{equation}
	where  $k$ stands for the crystal momentum (which may contain several components) within the first Brillouin zone. 
	The submatrices in Eq. (\ref{eq:generalized Hamiltonian}) satisfy the conditions,
	\begin{equation}
		\label{eq:generalized H(k)}
		A_{1,2}(k)=A_{1,2}^\dag(k),\quad B_2(k) = B_1^T(-k),
	\end{equation}
	which guarantees that $ H(k)$ is pseudo-Hermitian,
	\begin{equation}
		\label{eq:pseudo-Hermiticity generalized}
		\eta H^\dag(k) \eta^{-1} = H(k),
	\end{equation}
	which is equivalent to $\eta H(k)$ being Hermitian where $\eta = \tau_z$. $H(k)$ in Eq. (\ref{eq:generalized Hamiltonian}) generalizes  a bosonic BdG Hamiltonian for which
	\begin{equation}
		\label{eq:BdG Condition}
		A_1(k) = A_2(k) \mbox{ and } B_1(k)= B_2(k).
	\end{equation}  
	Here, we consider situations where condition (\ref{eq:BdG Condition}) does not hold. Then, $H(k)$ in Eq. (\ref{eq:generalized Hamiltonian}) is not invariant under particle-hole symmetry,
	\begin{equation}
		\mathcal{C} H^T(k)\mathcal{C}^{-1} \neq -H(-k),
	\end{equation}
	where $\mathcal{C}= \tau_y$ and is therefore not a BdG Hamiltonian. 
	
	We showed in Appendix \ref{sec:pseudo-Hermitian Hamiltonian} that a stable state, of a pseudo-Hermitian Hamiltonian in Eq. (\ref{eq:generalized Hamiltonian}), is classified either as a particle state or as a hole state, independent of any other symmetries. This means that the  bi-orthonormality relation built upon particle and hole states, which forms the basis of our study in Ref. \cite{ling2021PhysRevA.104.013305}, still holds true for a system described by Hamiltonian (\ref{eq:generalized Hamiltonian}).  We can thus expect to generalize our theorem in \cite{ling2021PhysRevA.104.013305} straightforwardly  from a BdG Hamiltonian to its non-BdG generalization (\ref{eq:generalized Hamiltonian}), beginning with the definition of a pair of a particle state, $\ket{\psi_{p_0}}$, and a hole state, $\ket{\psi_{h_0}}$, that are degenerate with energy $E_0$ in the absence of $B_1(k)$ and $B_2(k)$,
	\begin{equation}
		\begin{split}
			A_1(k)\ket{\psi_{p_0}}& = E_0 \ket{\psi_{p_0}},\\ -A^*_2(-k)\ket{\psi_{h_0}} & = E_0 \ket{\psi_{h_0}}.\\
		\end{split}
	\end{equation}
	We do not present the derivation here, but it can be shown that weak pairing lifts the degeneracy, splitting $E_0$ into a  pair of complex-conjugate energies, $E_0 + i \abs{E^{(1)}}$ and $E_0 - i \abs{E^{(1)}}$, where 
	\begin{equation}
		|E^{(1)}| = \abs{\bra{\psi_{p_0}} B_1(k) \ket{\psi_{h_0}}} 
	\end{equation}
	or
	\begin{equation}
		|E^{(1)}| = \abs{\bra{\psi_{p_0}}\left \lceil A,B \right\rfloor  \ket{\psi_{h_0}}}/{2\abs{E_0}}
	\end{equation}
	with
	\begin{equation}
		\label{eq:generalized unconventional commutator}
		\left \lceil A,B \right\rfloor \equiv A_1(k)B_1(k) - B_1(-k)A^*_2(-k)
	\end{equation}
	being defined symbolically as an unconventional commutator between $A$ and $B$ matrices in (\ref{eq:generalized Hamiltonian}). A sufficient condition for creating stable high-$\abs{E_0}$ states is that the unconventional commutator vanishes, i.e., $ \left \lceil A,B \right\rfloor =0$.

	%


\begin{thebibliography}{99}%
		\makeatletter
		\providecommand \@ifxundefined [1]{%
			\@ifx{#1\undefined}
		}%
		\providecommand \@ifnum [1]{%
			\ifnum #1\expandafter \@firstoftwo
			\else \expandafter \@secondoftwo
			\fi
		}%
		\providecommand \@ifx [1]{%
			\ifx #1\expandafter \@firstoftwo
			\else \expandafter \@secondoftwo
			\fi
		}%
		\providecommand \natexlab [1]{#1}%
		\providecommand \enquote  [1]{``#1''}%
		\providecommand \bibnamefont  [1]{#1}%
		\providecommand \bibfnamefont [1]{#1}%
		\providecommand \citenamefont [1]{#1}%
		\providecommand \href@noop [0]{\@secondoftwo}%
		\providecommand \href [0]{\begingroup \@sanitize@url \@href}%
		\providecommand \@href[1]{\@@startlink{#1}\@@href}%
		\providecommand \@@href[1]{\endgroup#1\@@endlink}%
		\providecommand \@sanitize@url [0]{\catcode `\\12\catcode `\$12\catcode
			`\&12\catcode `\#12\catcode `\^12\catcode `\_12\catcode `\%12\relax}%
		\providecommand \@@startlink[1]{}%
		\providecommand \@@endlink[0]{}%
		\providecommand \url  [0]{\begingroup\@sanitize@url \@url }%
		\providecommand \@url [1]{\endgroup\@href {#1}{\urlprefix }}%
		\providecommand \urlprefix  [0]{URL }%
		\providecommand \Eprint [0]{\href }%
		\providecommand \doibase [0]{https://doi.org/}%
		\providecommand \selectlanguage [0]{\@gobble}%
		\providecommand \bibinfo  [0]{\@secondoftwo}%
		\providecommand \bibfield  [0]{\@secondoftwo}%
		\providecommand \translation [1]{[#1]}%
		\providecommand \BibitemOpen [0]{}%
		\providecommand \bibitemStop [0]{}%
		\providecommand \bibitemNoStop [0]{.\EOS\space}%
		\providecommand \EOS [0]{\spacefactor3000\relax}%
		\providecommand \BibitemShut  [1]{\csname bibitem#1\endcsname}%
		\let\auto@bib@innerbib\@empty
		\bibitem [{\citenamefont {Hasan}\ and\ \citenamefont
			{Kane}(2010)}]{hasan10RevModPhys.82.3045}%
		\BibitemOpen
		\bibfield  {author} {\bibinfo {author} {\bibfnamefont {M.~Z.}\ \bibnamefont
				{Hasan}}\ and\ \bibinfo {author} {\bibfnamefont {C.~L.}\ \bibnamefont
				{Kane}},\ }\bibfield  {title} {\bibinfo {title} {\textit{Colloquium} :
				Topological insulators},\ }\href {https://doi.org/10.1103/RevModPhys.82.3045}
		{\bibfield  {journal} {\bibinfo  {journal} {Rev. Mod. Phys.}\ }\textbf
			{\bibinfo {volume} {82}},\ \bibinfo {pages} {3045} (\bibinfo {year}
			{2010})}\BibitemShut {NoStop}%
		\bibitem [{\citenamefont {Qi}\ and\ \citenamefont
			{Zhang}(2011)}]{qi2011RevModPhys.83.1057}%
		\BibitemOpen
		\bibfield  {author} {\bibinfo {author} {\bibfnamefont {X.-L.}\ \bibnamefont
				{Qi}}\ and\ \bibinfo {author} {\bibfnamefont {S.-C.}\ \bibnamefont {Zhang}},\
		}\bibfield  {title} {\bibinfo {title} {Topological insulators and
				superconductors},\ }\href {https://doi.org/10.1103/RevModPhys.83.1057}
		{\bibfield  {journal} {\bibinfo  {journal} {Rev. Mod. Phys.}\ }\textbf
			{\bibinfo {volume} {83}},\ \bibinfo {pages} {1057} (\bibinfo {year}
			{2011})}\BibitemShut {NoStop}%
		\bibitem [{\citenamefont {Haldane}(1988)}]{haldane88PhysRevLett.61.2015}%
		\BibitemOpen
		\bibfield  {author} {\bibinfo {author} {\bibfnamefont {F.~D.~M.}\
				\bibnamefont {Haldane}},\ }\bibfield  {title} {\bibinfo {title} {Model for a
				quantum hall effect without landau levels: Condensed-matter realization of
				the "parity anomaly"},\ }\href {https://doi.org/10.1103/PhysRevLett.61.2015}
		{\bibfield  {journal} {\bibinfo  {journal} {Phys. Rev. Lett.}\ }\textbf
			{\bibinfo {volume} {61}},\ \bibinfo {pages} {2015} (\bibinfo {year}
			{1988})}\BibitemShut {NoStop}%
		\bibitem [{\citenamefont {Klitzing}\ \emph {et~al.}(1980)\citenamefont
			{Klitzing}, \citenamefont {Dorda},\ and\ \citenamefont
			{Pepper}}]{klitzing1980PhysRevLett.45.494}%
		\BibitemOpen
		\bibfield  {author} {\bibinfo {author} {\bibfnamefont {K.~v.}\ \bibnamefont
				{Klitzing}}, \bibinfo {author} {\bibfnamefont {G.}~\bibnamefont {Dorda}},\
			and\ \bibinfo {author} {\bibfnamefont {M.}~\bibnamefont {Pepper}},\
		}\bibfield  {title} {\bibinfo {title} {New method for high-accuracy
				determination of the fine-structure constant based on quantized hall
				resistance},\ }\href {https://doi.org/10.1103/PhysRevLett.45.494} {\bibfield
			{journal} {\bibinfo  {journal} {Phys. Rev. Lett.}\ }\textbf {\bibinfo
				{volume} {45}},\ \bibinfo {pages} {494} (\bibinfo {year} {1980})}\BibitemShut
		{NoStop}%
		\bibitem [{\citenamefont {Thouless}\ \emph {et~al.}(1982)\citenamefont
			{Thouless}, \citenamefont {Kohmoto}, \citenamefont {Nightingale},\ and\
			\citenamefont {den Nijs}}]{thouless1982PhysRevLett.49.405}%
		\BibitemOpen
		\bibfield  {author} {\bibinfo {author} {\bibfnamefont {D.~J.}\ \bibnamefont
				{Thouless}}, \bibinfo {author} {\bibfnamefont {M.}~\bibnamefont {Kohmoto}},
			\bibinfo {author} {\bibfnamefont {M.~P.}\ \bibnamefont {Nightingale}},\ and\
			\bibinfo {author} {\bibfnamefont {M.}~\bibnamefont {den Nijs}},\ }\bibfield
		{title} {\bibinfo {title} {Quantized hall conductance in a two-dimensional
				periodic potential},\ }\href {https://doi.org/10.1103/PhysRevLett.49.405}
		{\bibfield  {journal} {\bibinfo  {journal} {Phys. Rev. Lett.}\ }\textbf
			{\bibinfo {volume} {49}},\ \bibinfo {pages} {405} (\bibinfo {year}
			{1982})}\BibitemShut {NoStop}%
		\bibitem [{\citenamefont {Kane}\ and\ \citenamefont
			{Mele}(2005{\natexlab{a}})}]{kane2005PhysRevLett.95.146802}%
		\BibitemOpen
		\bibfield  {author} {\bibinfo {author} {\bibfnamefont {C.~L.}\ \bibnamefont
				{Kane}}\ and\ \bibinfo {author} {\bibfnamefont {E.~J.}\ \bibnamefont
				{Mele}},\ }\bibfield  {title} {\bibinfo {title} {${Z}_{2}$ topological order
				and the quantum spin hall effect},\ }\href
		{https://doi.org/10.1103/PhysRevLett.95.146802} {\bibfield  {journal}
			{\bibinfo  {journal} {Phys. Rev. Lett.}\ }\textbf {\bibinfo {volume} {95}},\
			\bibinfo {pages} {146802} (\bibinfo {year} {2005}{\natexlab{a}})}\BibitemShut
		{NoStop}%
		\bibitem [{\citenamefont {Kane}\ and\ \citenamefont
			{Mele}(2005{\natexlab{b}})}]{kane2005PhysRevLett.95.226801}%
		\BibitemOpen
		\bibfield  {author} {\bibinfo {author} {\bibfnamefont {C.~L.}\ \bibnamefont
				{Kane}}\ and\ \bibinfo {author} {\bibfnamefont {E.~J.}\ \bibnamefont
				{Mele}},\ }\bibfield  {title} {\bibinfo {title} {Quantum spin hall effect in
				graphene},\ }\href {https://doi.org/10.1103/PhysRevLett.95.226801} {\bibfield
			{journal} {\bibinfo  {journal} {Phys. Rev. Lett.}\ }\textbf {\bibinfo
				{volume} {95}},\ \bibinfo {pages} {226801} (\bibinfo {year}
			{2005}{\natexlab{b}})}\BibitemShut {NoStop}%
		\bibitem [{\citenamefont {Murakami}\ \emph {et~al.}(2004)\citenamefont
			{Murakami}, \citenamefont {Nagaosa},\ and\ \citenamefont
			{Zhang}}]{murakami04PhysRevLett.93.156804}%
		\BibitemOpen
		\bibfield  {author} {\bibinfo {author} {\bibfnamefont {S.}~\bibnamefont
				{Murakami}}, \bibinfo {author} {\bibfnamefont {N.}~\bibnamefont {Nagaosa}},\
			and\ \bibinfo {author} {\bibfnamefont {S.-C.}\ \bibnamefont {Zhang}},\
		}\bibfield  {title} {\bibinfo {title} {Spin-hall insulator},\ }\href
		{https://doi.org/10.1103/PhysRevLett.93.156804} {\bibfield  {journal}
			{\bibinfo  {journal} {Phys. Rev. Lett.}\ }\textbf {\bibinfo {volume} {93}},\
			\bibinfo {pages} {156804} (\bibinfo {year} {2004})}\BibitemShut {NoStop}%
		\bibitem [{\citenamefont {K{\"o}nig}\ \emph {et~al.}(2007)\citenamefont
			{K{\"o}nig}, \citenamefont {Wiedmann}, \citenamefont {Br{\"u}ne},
			\citenamefont {Roth}, \citenamefont {Buhmann}, \citenamefont {Molenkamp},
			\citenamefont {Qi},\ and\ \citenamefont {Zhang}}]{konig2007Science.318.766}%
		\BibitemOpen
		\bibfield  {author} {\bibinfo {author} {\bibfnamefont {M.}~\bibnamefont
				{K{\"o}nig}}, \bibinfo {author} {\bibfnamefont {S.}~\bibnamefont {Wiedmann}},
			\bibinfo {author} {\bibfnamefont {C.}~\bibnamefont {Br{\"u}ne}}, \bibinfo
			{author} {\bibfnamefont {A.}~\bibnamefont {Roth}}, \bibinfo {author}
			{\bibfnamefont {H.}~\bibnamefont {Buhmann}}, \bibinfo {author} {\bibfnamefont
				{L.~W.}\ \bibnamefont {Molenkamp}}, \bibinfo {author} {\bibfnamefont {X.-L.}\
				\bibnamefont {Qi}},\ and\ \bibinfo {author} {\bibfnamefont {S.-C.}\
				\bibnamefont {Zhang}},\ }\bibfield  {title} {\bibinfo {title} {Quantum spin
				hall insulator state in hgte quantum wells},\ }\href
		{https://doi.org/10.1126/science.1148047} {\bibfield  {journal} {\bibinfo
				{journal} {Science}\ }\textbf {\bibinfo {volume} {318}},\ \bibinfo {pages}
			{766} (\bibinfo {year} {2007})}\BibitemShut {NoStop}%
		\bibitem [{\citenamefont {Bernevig}\ \emph {et~al.}(2006)\citenamefont
			{Bernevig}, \citenamefont {Hughes},\ and\ \citenamefont
			{Zhang}}]{bernevig2006Science.314.1757}%
		\BibitemOpen
		\bibfield  {author} {\bibinfo {author} {\bibfnamefont {B.~A.}\ \bibnamefont
				{Bernevig}}, \bibinfo {author} {\bibfnamefont {T.~L.}\ \bibnamefont
				{Hughes}},\ and\ \bibinfo {author} {\bibfnamefont {S.-C.}\ \bibnamefont
				{Zhang}},\ }\bibfield  {title} {\bibinfo {title} {Quantum spin hall effect
				and topological phase transition in hgte quantum wells},\ }\href
		{https://doi.org/10.1126/science.1133734} {\bibfield  {journal} {\bibinfo
				{journal} {Science}\ }\textbf {\bibinfo {volume} {314}},\ \bibinfo {pages}
			{1757} (\bibinfo {year} {2006})}\BibitemShut {NoStop}%
		\bibitem [{\citenamefont {Schnyder}\ \emph {et~al.}(2008)\citenamefont
			{Schnyder}, \citenamefont {Ryu}, \citenamefont {Furusaki},\ and\
			\citenamefont {Ludwig}}]{schnyder2008PhysRevB.78.195125}%
		\BibitemOpen
		\bibfield  {author} {\bibinfo {author} {\bibfnamefont {A.~P.}\ \bibnamefont
				{Schnyder}}, \bibinfo {author} {\bibfnamefont {S.}~\bibnamefont {Ryu}},
			\bibinfo {author} {\bibfnamefont {A.}~\bibnamefont {Furusaki}},\ and\
			\bibinfo {author} {\bibfnamefont {A.~W.~W.}\ \bibnamefont {Ludwig}},\
		}\bibfield  {title} {\bibinfo {title} {Classification of topological
				insulators and superconductors in three spatial dimensions},\ }\href
		{https://doi.org/10.1103/PhysRevB.78.195125} {\bibfield  {journal} {\bibinfo
				{journal} {Phys. Rev. B}\ }\textbf {\bibinfo {volume} {78}},\ \bibinfo
			{pages} {195125} (\bibinfo {year} {2008})}\BibitemShut {NoStop}%
		\bibitem [{\citenamefont {Kitaev}(2009)}]{kitaev2009AIP.1134.22}%
		\BibitemOpen
		\bibfield  {author} {\bibinfo {author} {\bibfnamefont {A.}~\bibnamefont
				{Kitaev}},\ }\href@noop {} {\bibfield  {journal} {\bibinfo  {journal} {AIP
					Conference Proceedings}\ }\textbf {\bibinfo {volume} {1134}},\ \bibinfo
			{pages} {22} (\bibinfo {year} {2009})}\BibitemShut {NoStop}%
		\bibitem [{\citenamefont {Ryu}\ \emph {et~al.}(2010)\citenamefont {Ryu},
			\citenamefont {Schnyder}, \citenamefont {Furusaki},\ and\ \citenamefont
			{Ludwig}}]{Ryu2010NewJournalOfPhysics.12.065010}%
		\BibitemOpen
		\bibfield  {author} {\bibinfo {author} {\bibfnamefont {S.}~\bibnamefont
				{Ryu}}, \bibinfo {author} {\bibfnamefont {A.~P.}\ \bibnamefont {Schnyder}},
			\bibinfo {author} {\bibfnamefont {A.}~\bibnamefont {Furusaki}},\ and\
			\bibinfo {author} {\bibfnamefont {A.~W.~W.}\ \bibnamefont {Ludwig}},\
		}\bibfield  {title} {\bibinfo {title} {Topological insulators and
				superconductors: tenfold way and dimensional hierarchy},\ }\href
		{https://doi.org/10.1088/1367-2630/12/6/065010} {\bibfield  {journal}
			{\bibinfo  {journal} {New Journal of Physics}\ }\textbf {\bibinfo {volume}
				{12}},\ \bibinfo {pages} {065010} (\bibinfo {year} {2010})}\BibitemShut
		{NoStop}%
		\bibitem [{\citenamefont {Shen}()}]{shen2012}%
		\BibitemOpen
		\bibfield  {author} {\bibinfo {author} {\bibfnamefont {S.-Q.}\ \bibnamefont
				{Shen}},\ }\href@noop {} {\bibinfo  {journal} {\textit{Topological
					Insulators} (Springer Berlin, Germany, 2012)}\ }\BibitemShut {NoStop}%
		\bibitem [{\citenamefont {Bernevig}\ and\ \citenamefont
			{Hughes}()}]{bernevig2013}%
		\BibitemOpen
		\bibfield  {journal} {  }\bibfield  {author} {\bibinfo {author} {\bibfnamefont
				{B.~A.}\ \bibnamefont {Bernevig}}\ and\ \bibinfo {author} {\bibfnamefont
				{T.~L.}\ \bibnamefont {Hughes}},\ }\href@noop {} {\bibinfo  {journal}
			{\textit{Topological Insulators and Topological Superconductors} (Princeton
				University Press, Princeton, 2013)}\ }\BibitemShut {NoStop}%
		\bibitem [{\citenamefont {Haldane}\ and\ \citenamefont
			{Raghu}(2008)}]{haldane08PhysRevLett.100.013904}%
		\BibitemOpen
		\bibfield  {journal} {  }\bibfield  {author} {\bibinfo {author} {\bibfnamefont
				{F.~D.~M.}\ \bibnamefont {Haldane}}\ and\ \bibinfo {author} {\bibfnamefont
				{S.}~\bibnamefont {Raghu}},\ }\bibfield  {title} {\bibinfo {title} {Possible
				realization of directional optical waveguides in photonic crystals with
				broken time-reversal symmetry},\ }\href
		{https://doi.org/10.1103/PhysRevLett.100.013904} {\bibfield  {journal}
			{\bibinfo  {journal} {Phys. Rev. Lett.}\ }\textbf {\bibinfo {volume} {100}},\
			\bibinfo {pages} {013904} (\bibinfo {year} {2008})}\BibitemShut {NoStop}%
		\bibitem [{\citenamefont {Wang}\ \emph {et~al.}(2008)\citenamefont {Wang},
			\citenamefont {Chong}, \citenamefont {Joannopoulos},\ and\ \citenamefont
			{Solja\ifmmode \check{c}\else \v{c}\fi{}i\ifmmode~\acute{c}\else
				\'{c}\fi{}}}]{wang2008PhysRevLett.100.013905}%
		\BibitemOpen
		\bibfield  {author} {\bibinfo {author} {\bibfnamefont {Z.}~\bibnamefont
				{Wang}}, \bibinfo {author} {\bibfnamefont {Y.~D.}\ \bibnamefont {Chong}},
			\bibinfo {author} {\bibfnamefont {J.~D.}\ \bibnamefont {Joannopoulos}},\ and\
			\bibinfo {author} {\bibfnamefont {M.}~\bibnamefont {Solja\ifmmode
					\check{c}\else \v{c}\fi{}i\ifmmode~\acute{c}\else \'{c}\fi{}}},\ }\bibfield
		{title} {\bibinfo {title} {Reflection-free one-way edge modes in a
				gyromagnetic photonic crystal},\ }\href
		{https://doi.org/10.1103/PhysRevLett.100.013905} {\bibfield  {journal}
			{\bibinfo  {journal} {Phys. Rev. Lett.}\ }\textbf {\bibinfo {volume} {100}},\
			\bibinfo {pages} {013905} (\bibinfo {year} {2008})}\BibitemShut {NoStop}%
		\bibitem [{\citenamefont {Wang}\ \emph
			{et~al.}(2009{\natexlab{a}})\citenamefont {Wang}, \citenamefont {Chong},
			\citenamefont {Joannopoulos},\ and\ \citenamefont
			{Solja\v{c}i\'{c}}}]{wang2009Nature.461.772}%
		\BibitemOpen
		\bibfield  {author} {\bibinfo {author} {\bibfnamefont {Z.}~\bibnamefont
				{Wang}}, \bibinfo {author} {\bibfnamefont {Y.}~\bibnamefont {Chong}},
			\bibinfo {author} {\bibfnamefont {J.~D.}\ \bibnamefont {Joannopoulos}},\ and\
			\bibinfo {author} {\bibfnamefont {M.}~\bibnamefont {Solja\v{c}i\'{c}}},\
		}\bibfield  {title} {\bibinfo {title} {Observation of unidirectional
				backscattering-immune topological electromagnetic states},\ }\href
		{https://doi.org/10.1038/nature08293} {\bibfield  {journal} {\bibinfo
				{journal} {Nature}\ }\textbf {\bibinfo {volume} {461}},\ \bibinfo {pages}
			{772} (\bibinfo {year} {2009}{\natexlab{a}})}\BibitemShut {NoStop}%
		\bibitem [{\citenamefont {Hafezi}\ \emph {et~al.}(2011)\citenamefont {Hafezi},
			\citenamefont {Demler}, \citenamefont {Lukin},\ and\ \citenamefont
			{Taylor}}]{hafezi2011NaurePhysics.7.907}%
		\BibitemOpen
		\bibfield  {author} {\bibinfo {author} {\bibfnamefont {M.}~\bibnamefont
				{Hafezi}}, \bibinfo {author} {\bibfnamefont {E.~A.}\ \bibnamefont {Demler}},
			\bibinfo {author} {\bibfnamefont {M.~D.}\ \bibnamefont {Lukin}},\ and\
			\bibinfo {author} {\bibfnamefont {J.~M.}\ \bibnamefont {Taylor}},\ }\bibfield
		{title} {\bibinfo {title} {Robust optical delay lines with topological
				protection},\ }\href {https://doi.org/10.1038/nphys2063} {\bibfield
			{journal} {\bibinfo  {journal} {Nat. Phys.}\ }\textbf {\bibinfo {volume}
				{7}},\ \bibinfo {pages} {907} (\bibinfo {year} {2011})}\BibitemShut {NoStop}%
		\bibitem [{\citenamefont {Fang}\ \emph {et~al.}(2012)\citenamefont {Fang},
			\citenamefont {Yu},\ and\ \citenamefont {Fan}}]{kejie2012NatPhoton2012}%
		\BibitemOpen
		\bibfield  {author} {\bibinfo {author} {\bibfnamefont {K.}~\bibnamefont
				{Fang}}, \bibinfo {author} {\bibfnamefont {Z.}~\bibnamefont {Yu}},\ and\
			\bibinfo {author} {\bibfnamefont {S.}~\bibnamefont {Fan}},\ }\bibfield
		{title} {\bibinfo {title} {Realizing effective magnetic field for photons by
				controlling the phase of dynamic modulation},\ }\href
		{https://doi.org/10.1038/nphoton.2012.236} {\bibfield  {journal} {\bibinfo
				{journal} {Nat. Photon}\ }\textbf {\bibinfo {volume} {6}},\ \bibinfo {pages}
			{792} (\bibinfo {year} {2012})}\BibitemShut {NoStop}%
		\bibitem [{\citenamefont {Kane}\ and\ \citenamefont
			{Lubensky}(2013)}]{kane2013NaturePhysics.10.39}%
		\BibitemOpen
		\bibfield  {author} {\bibinfo {author} {\bibfnamefont {C.~L.}\ \bibnamefont
				{Kane}}\ and\ \bibinfo {author} {\bibfnamefont {T.~C.}\ \bibnamefont
				{Lubensky}},\ }\bibfield  {title} {\bibinfo {title} {Topological boundary
				modes in isostatic lattices},\ }\href {http://dx.doi.org/10.1038/nphys2835}
		{\bibfield  {journal} {\bibinfo  {journal} {Nat. Phys.}\ }\textbf {\bibinfo
				{volume} {10}},\ \bibinfo {pages} {39} (\bibinfo {year} {2013})}\BibitemShut
		{NoStop}%
		\bibitem [{\citenamefont {Prodan}\ and\ \citenamefont
			{Prodan}(2009)}]{prodan2009PhysRevLett.103.248101}%
		\BibitemOpen
		\bibfield  {author} {\bibinfo {author} {\bibfnamefont {E.}~\bibnamefont
				{Prodan}}\ and\ \bibinfo {author} {\bibfnamefont {C.}~\bibnamefont
				{Prodan}},\ }\bibfield  {title} {\bibinfo {title} {Topological phonon modes
				and their role in dynamic instability of microtubules},\ }\href
		{https://doi.org/10.1103/PhysRevLett.103.248101} {\bibfield  {journal}
			{\bibinfo  {journal} {Phys. Rev. Lett.}\ }\textbf {\bibinfo {volume} {103}},\
			\bibinfo {pages} {248101} (\bibinfo {year} {2009})}\BibitemShut {NoStop}%
		\bibitem [{\citenamefont {Fleury}\ \emph {et~al.}(2014)\citenamefont {Fleury},
			\citenamefont {Sounas}, \citenamefont {Sieck}, \citenamefont {Haberman},\
			and\ \citenamefont {Al{\`u}}}]{fleury2014Science.343.516}%
		\BibitemOpen
		\bibfield  {author} {\bibinfo {author} {\bibfnamefont {R.}~\bibnamefont
				{Fleury}}, \bibinfo {author} {\bibfnamefont {D.~L.}\ \bibnamefont {Sounas}},
			\bibinfo {author} {\bibfnamefont {C.~F.}\ \bibnamefont {Sieck}}, \bibinfo
			{author} {\bibfnamefont {M.~R.}\ \bibnamefont {Haberman}},\ and\ \bibinfo
			{author} {\bibfnamefont {A.}~\bibnamefont {Al{\`u}}},\ }\bibfield  {title}
		{\bibinfo {title} {Sound isolation and giant linear nonreciprocity in a
				compact acoustic circulator},\ }\href
		{https://doi.org/10.1126/science.1246957} {\bibfield  {journal} {\bibinfo
				{journal} {Science}\ }\textbf {\bibinfo {volume} {343}},\ \bibinfo {pages}
			{516} (\bibinfo {year} {2014})}\BibitemShut {NoStop}%
		\bibitem [{\citenamefont {Yang}\ \emph {et~al.}(2015)\citenamefont {Yang},
			\citenamefont {Gao}, \citenamefont {Shi}, \citenamefont {Lin}, \citenamefont
			{Gao}, \citenamefont {Chong},\ and\ \citenamefont
			{Zhang}}]{yang2015PhysRevLett.114.114301}%
		\BibitemOpen
		\bibfield  {author} {\bibinfo {author} {\bibfnamefont {Z.}~\bibnamefont
				{Yang}}, \bibinfo {author} {\bibfnamefont {F.}~\bibnamefont {Gao}}, \bibinfo
			{author} {\bibfnamefont {X.}~\bibnamefont {Shi}}, \bibinfo {author}
			{\bibfnamefont {X.}~\bibnamefont {Lin}}, \bibinfo {author} {\bibfnamefont
				{Z.}~\bibnamefont {Gao}}, \bibinfo {author} {\bibfnamefont {Y.}~\bibnamefont
				{Chong}},\ and\ \bibinfo {author} {\bibfnamefont {B.}~\bibnamefont {Zhang}},\
		}\bibfield  {title} {\bibinfo {title} {Topological acoustics},\ }\href
		{https://doi.org/10.1103/PhysRevLett.114.114301} {\bibfield  {journal}
			{\bibinfo  {journal} {Phys. Rev. Lett.}\ }\textbf {\bibinfo {volume} {114}},\
			\bibinfo {pages} {114301} (\bibinfo {year} {2015})}\BibitemShut {NoStop}%
		\bibitem [{\citenamefont {Peano}\ \emph {et~al.}(2015)\citenamefont {Peano},
			\citenamefont {Brendel}, \citenamefont {Schmidt},\ and\ \citenamefont
			{Marquardt}}]{peano2015PhysRevX.5.031011}%
		\BibitemOpen
		\bibfield  {author} {\bibinfo {author} {\bibfnamefont {V.}~\bibnamefont
				{Peano}}, \bibinfo {author} {\bibfnamefont {C.}~\bibnamefont {Brendel}},
			\bibinfo {author} {\bibfnamefont {M.}~\bibnamefont {Schmidt}},\ and\ \bibinfo
			{author} {\bibfnamefont {F.}~\bibnamefont {Marquardt}},\ }\bibfield  {title}
		{\bibinfo {title} {Topological phases of sound and light},\ }\href
		{https://doi.org/10.1103/PhysRevX.5.031011} {\bibfield  {journal} {\bibinfo
				{journal} {Phys. Rev. X}\ }\textbf {\bibinfo {volume} {5}},\ \bibinfo {pages}
			{031011} (\bibinfo {year} {2015})}\BibitemShut {NoStop}%
		\bibitem [{\citenamefont {He}\ \emph {et~al.}(2016)\citenamefont {He},
			\citenamefont {Ni}, \citenamefont {Ge}, \citenamefont {Sun}, \citenamefont
			{Chen}, \citenamefont {Lu}, \citenamefont {Liu},\ and\ \citenamefont
			{Chen}}]{he2016NatPhys.12.1124}%
		\BibitemOpen
		\bibfield  {author} {\bibinfo {author} {\bibfnamefont {C.}~\bibnamefont
				{He}}, \bibinfo {author} {\bibfnamefont {X.}~\bibnamefont {Ni}}, \bibinfo
			{author} {\bibfnamefont {H.}~\bibnamefont {Ge}}, \bibinfo {author}
			{\bibfnamefont {X.-C.}\ \bibnamefont {Sun}}, \bibinfo {author} {\bibfnamefont
				{Y.-B.}\ \bibnamefont {Chen}}, \bibinfo {author} {\bibfnamefont {M.-H.}\
				\bibnamefont {Lu}}, \bibinfo {author} {\bibfnamefont {X.-P.}\ \bibnamefont
				{Liu}},\ and\ \bibinfo {author} {\bibfnamefont {Y.-F.}\ \bibnamefont
				{Chen}},\ }\bibfield  {title} {\bibinfo {title} {Acoustic topological
				insulator and robust one-way sound transport},\ }\href
		{https://doi.org/10.1038/nphys3867} {\bibfield  {journal} {\bibinfo
				{journal} {Nat. Phys.}\ }\textbf {\bibinfo {volume} {12}},\ \bibinfo {pages}
			{1124} (\bibinfo {year} {2016})}\BibitemShut {NoStop}%
		\bibitem [{\citenamefont {Shindou}\ \emph
			{et~al.}(2013{\natexlab{a}})\citenamefont {Shindou}, \citenamefont {Ohe},
			\citenamefont {Matsumoto}, \citenamefont {Murakami},\ and\ \citenamefont
			{Saitoh}}]{shindou2013PhysRevB.87.174402}%
		\BibitemOpen
		\bibfield  {author} {\bibinfo {author} {\bibfnamefont {R.}~\bibnamefont
				{Shindou}}, \bibinfo {author} {\bibfnamefont {J.-i.}\ \bibnamefont {Ohe}},
			\bibinfo {author} {\bibfnamefont {R.}~\bibnamefont {Matsumoto}}, \bibinfo
			{author} {\bibfnamefont {S.}~\bibnamefont {Murakami}},\ and\ \bibinfo
			{author} {\bibfnamefont {E.}~\bibnamefont {Saitoh}},\ }\bibfield  {title}
		{\bibinfo {title} {Chiral spin-wave edge modes in dipolar magnetic thin
				films},\ }\href {https://doi.org/10.1103/PhysRevB.87.174402} {\bibfield
			{journal} {\bibinfo  {journal} {Phys. Rev. B}\ }\textbf {\bibinfo {volume}
				{87}},\ \bibinfo {pages} {174402} (\bibinfo {year}
			{2013}{\natexlab{a}})}\BibitemShut {NoStop}%
		\bibitem [{\citenamefont {Shindou}\ \emph
			{et~al.}(2013{\natexlab{b}})\citenamefont {Shindou}, \citenamefont
			{Matsumoto}, \citenamefont {Murakami},\ and\ \citenamefont
			{Ohe}}]{shindou2013PhysRevB.87.174427}%
		\BibitemOpen
		\bibfield  {author} {\bibinfo {author} {\bibfnamefont {R.}~\bibnamefont
				{Shindou}}, \bibinfo {author} {\bibfnamefont {R.}~\bibnamefont {Matsumoto}},
			\bibinfo {author} {\bibfnamefont {S.}~\bibnamefont {Murakami}},\ and\
			\bibinfo {author} {\bibfnamefont {J.-i.}\ \bibnamefont {Ohe}},\ }\bibfield
		{title} {\bibinfo {title} {Topological chiral magnonic edge mode in a
				magnonic crystal},\ }\href {https://doi.org/10.1103/PhysRevB.87.174427}
		{\bibfield  {journal} {\bibinfo  {journal} {Phys. Rev. B}\ }\textbf {\bibinfo
				{volume} {87}},\ \bibinfo {pages} {174427} (\bibinfo {year}
			{2013}{\natexlab{b}})}\BibitemShut {NoStop}%
		\bibitem [{\citenamefont {Atala}\ \emph {et~al.}(2013)\citenamefont {Atala},
			\citenamefont {Aidelsburger}, \citenamefont {Barreiro}, \citenamefont
			{Abanin}, \citenamefont {Kitagawa}, \citenamefont {Demler},\ and\
			\citenamefont {Bloch}}]{aidelsburger2013NatPhys.9.795}%
		\BibitemOpen
		\bibfield  {author} {\bibinfo {author} {\bibfnamefont {M.}~\bibnamefont
				{Atala}}, \bibinfo {author} {\bibfnamefont {M.}~\bibnamefont {Aidelsburger}},
			\bibinfo {author} {\bibfnamefont {J.~T.}\ \bibnamefont {Barreiro}}, \bibinfo
			{author} {\bibfnamefont {D.}~\bibnamefont {Abanin}}, \bibinfo {author}
			{\bibfnamefont {T.}~\bibnamefont {Kitagawa}}, \bibinfo {author}
			{\bibfnamefont {E.}~\bibnamefont {Demler}},\ and\ \bibinfo {author}
			{\bibfnamefont {I.}~\bibnamefont {Bloch}},\ }\bibfield  {title} {\bibinfo
			{title} {Direct measurement of the zak phase in topological bloch bands},\
		}\href {https://doi.org/10.1038/nphys2790} {\bibfield  {journal} {\bibinfo
				{journal} {Nat. Phys.}\ }\textbf {\bibinfo {volume} {9}},\ \bibinfo {pages}
			{795} (\bibinfo {year} {2013})}\BibitemShut {NoStop}%
		\bibitem [{\citenamefont {Aidelsburger}\ \emph {et~al.}(2013)\citenamefont
			{Aidelsburger}, \citenamefont {Atala}, \citenamefont {Lohse}, \citenamefont
			{Barreiro}, \citenamefont {Paredes},\ and\ \citenamefont
			{Bloch}}]{aidelsburger2013PhysRevLett.111.185301}%
		\BibitemOpen
		\bibfield  {author} {\bibinfo {author} {\bibfnamefont {M.}~\bibnamefont
				{Aidelsburger}}, \bibinfo {author} {\bibfnamefont {M.}~\bibnamefont {Atala}},
			\bibinfo {author} {\bibfnamefont {M.}~\bibnamefont {Lohse}}, \bibinfo
			{author} {\bibfnamefont {J.~T.}\ \bibnamefont {Barreiro}}, \bibinfo {author}
			{\bibfnamefont {B.}~\bibnamefont {Paredes}},\ and\ \bibinfo {author}
			{\bibfnamefont {I.}~\bibnamefont {Bloch}},\ }\bibfield  {title} {\bibinfo
			{title} {Realization of the hofstadter hamiltonian with ultracold atoms in
				optical lattices},\ }\href {https://doi.org/10.1103/PhysRevLett.111.185301}
		{\bibfield  {journal} {\bibinfo  {journal} {Phys. Rev. Lett.}\ }\textbf
			{\bibinfo {volume} {111}},\ \bibinfo {pages} {185301} (\bibinfo {year}
			{2013})}\BibitemShut {NoStop}%
		\bibitem [{\citenamefont {Miyake}\ \emph {et~al.}(2013)\citenamefont {Miyake},
			\citenamefont {Siviloglou}, \citenamefont {Kennedy}, \citenamefont {Burton},\
			and\ \citenamefont {Ketterle}}]{hirokazu2013PhysRevLett.111.185302}%
		\BibitemOpen
		\bibfield  {author} {\bibinfo {author} {\bibfnamefont {H.}~\bibnamefont
				{Miyake}}, \bibinfo {author} {\bibfnamefont {G.~A.}\ \bibnamefont
				{Siviloglou}}, \bibinfo {author} {\bibfnamefont {C.~J.}\ \bibnamefont
				{Kennedy}}, \bibinfo {author} {\bibfnamefont {W.~C.}\ \bibnamefont
				{Burton}},\ and\ \bibinfo {author} {\bibfnamefont {W.}~\bibnamefont
				{Ketterle}},\ }\bibfield  {title} {\bibinfo {title} {Realizing the harper
				hamiltonian with laser-assisted tunneling in optical lattices},\ }\href
		{https://doi.org/10.1103/PhysRevLett.111.185302} {\bibfield  {journal}
			{\bibinfo  {journal} {Phys. Rev. Lett.}\ }\textbf {\bibinfo {volume} {111}},\
			\bibinfo {pages} {185302} (\bibinfo {year} {2013})}\BibitemShut {NoStop}%
		\bibitem [{\citenamefont {Jotzu}\ \emph {et~al.}(2014)\citenamefont {Jotzu},
			\citenamefont {Messer}, \citenamefont {Desbuquois}, \citenamefont {Lebrat},
			\citenamefont {Uehlinger}, \citenamefont {Greif},\ and\ \citenamefont
			{Esslinger}}]{jotzu2014Nature.515.237}%
		\BibitemOpen
		\bibfield  {author} {\bibinfo {author} {\bibfnamefont {G.}~\bibnamefont
				{Jotzu}}, \bibinfo {author} {\bibfnamefont {M.}~\bibnamefont {Messer}},
			\bibinfo {author} {\bibfnamefont {R.}~\bibnamefont {Desbuquois}}, \bibinfo
			{author} {\bibfnamefont {M.}~\bibnamefont {Lebrat}}, \bibinfo {author}
			{\bibfnamefont {T.}~\bibnamefont {Uehlinger}}, \bibinfo {author}
			{\bibfnamefont {D.}~\bibnamefont {Greif}},\ and\ \bibinfo {author}
			{\bibfnamefont {T.}~\bibnamefont {Esslinger}},\ }\bibfield  {title} {\bibinfo
			{title} {Experimental realization of the topological haldane model with
				ultracold fermions},\ }\href {https://doi.org/10.1038/nature13915} {\bibfield
			{journal} {\bibinfo  {journal} {Nature}\ }\textbf {\bibinfo {volume}
				{515}},\ \bibinfo {pages} {237} (\bibinfo {year} {2014})}\BibitemShut
		{NoStop}%
		\bibitem [{\citenamefont {Stuhl}\ \emph {et~al.}(2015)\citenamefont {Stuhl},
			\citenamefont {Lu}, \citenamefont {Aycock}, \citenamefont {Genkina},\ and\
			\citenamefont {Spielman}}]{Stuhl2015Science.349.1514}%
		\BibitemOpen
		\bibfield  {author} {\bibinfo {author} {\bibfnamefont {B.~K.}\ \bibnamefont
				{Stuhl}}, \bibinfo {author} {\bibfnamefont {H.-I.}\ \bibnamefont {Lu}},
			\bibinfo {author} {\bibfnamefont {L.~M.}\ \bibnamefont {Aycock}}, \bibinfo
			{author} {\bibfnamefont {D.}~\bibnamefont {Genkina}},\ and\ \bibinfo {author}
			{\bibfnamefont {I.~B.}\ \bibnamefont {Spielman}},\ }\bibfield  {title}
		{\bibinfo {title} {Visualizing edge states with an atomic bose gas in the
				quantum hall regime},\ }\href {https://doi.org/10.1126/science.aaa8515}
		{\bibfield  {journal} {\bibinfo  {journal} {Science}\ }\textbf {\bibinfo
				{volume} {349}},\ \bibinfo {pages} {1514} (\bibinfo {year}
			{2015})}\BibitemShut {NoStop}%
		\bibitem [{\citenamefont {Cooper}\ \emph {et~al.}(2019)\citenamefont {Cooper},
			\citenamefont {Dalibard},\ and\ \citenamefont
			{Spielman}}]{cooper2019RevModPhys.91.015005}%
		\BibitemOpen
		\bibfield  {author} {\bibinfo {author} {\bibfnamefont {N.~R.}\ \bibnamefont
				{Cooper}}, \bibinfo {author} {\bibfnamefont {J.}~\bibnamefont {Dalibard}},\
			and\ \bibinfo {author} {\bibfnamefont {I.~B.}\ \bibnamefont {Spielman}},\
		}\bibfield  {title} {\bibinfo {title} {Topological bands for ultracold
				atoms},\ }\href {https://doi.org/10.1103/RevModPhys.91.015005} {\bibfield
			{journal} {\bibinfo  {journal} {Rev. Mod. Phys.}\ }\textbf {\bibinfo {volume}
				{91}},\ \bibinfo {pages} {015005} (\bibinfo {year} {2019})}\BibitemShut
		{NoStop}%
		\bibitem [{\citenamefont {Ozawa}\ \emph {et~al.}(2019)\citenamefont {Ozawa},
			\citenamefont {Price}, \citenamefont {Amo}, \citenamefont {Goldman},
			\citenamefont {Hafezi}, \citenamefont {Lu}, \citenamefont {Rechtsman},
			\citenamefont {Schuster}, \citenamefont {Simon}, \citenamefont {Zilberberg},\
			and\ \citenamefont {Carusotto}}]{ozawa2019RevModPhys.91.015006}%
		\BibitemOpen
		\bibfield  {author} {\bibinfo {author} {\bibfnamefont {T.}~\bibnamefont
				{Ozawa}}, \bibinfo {author} {\bibfnamefont {H.~M.}\ \bibnamefont {Price}},
			\bibinfo {author} {\bibfnamefont {A.}~\bibnamefont {Amo}}, \bibinfo {author}
			{\bibfnamefont {N.}~\bibnamefont {Goldman}}, \bibinfo {author} {\bibfnamefont
				{M.}~\bibnamefont {Hafezi}}, \bibinfo {author} {\bibfnamefont
				{L.}~\bibnamefont {Lu}}, \bibinfo {author} {\bibfnamefont {M.~C.}\
				\bibnamefont {Rechtsman}}, \bibinfo {author} {\bibfnamefont {D.}~\bibnamefont
				{Schuster}}, \bibinfo {author} {\bibfnamefont {J.}~\bibnamefont {Simon}},
			\bibinfo {author} {\bibfnamefont {O.}~\bibnamefont {Zilberberg}},\ and\
			\bibinfo {author} {\bibfnamefont {I.}~\bibnamefont {Carusotto}},\ }\bibfield
		{title} {\bibinfo {title} {Topological photonics},\ }\href
		{https://doi.org/10.1103/RevModPhys.91.015006} {\bibfield  {journal}
			{\bibinfo  {journal} {Rev. Mod. Phys.}\ }\textbf {\bibinfo {volume} {91}},\
			\bibinfo {pages} {015006} (\bibinfo {year} {2019})}\BibitemShut {NoStop}%
		\bibitem [{\citenamefont {Zhang}\ \emph {et~al.}(2018)\citenamefont {Zhang},
			\citenamefont {Xiao}, \citenamefont {Cheng}, \citenamefont {Lu},\ and\
			\citenamefont {Christensen}}]{zhang2018CommunPhys.1.97}%
		\BibitemOpen
		\bibfield  {author} {\bibinfo {author} {\bibfnamefont {X.}~\bibnamefont
				{Zhang}}, \bibinfo {author} {\bibfnamefont {M.}~\bibnamefont {Xiao}},
			\bibinfo {author} {\bibfnamefont {Y.}~\bibnamefont {Cheng}}, \bibinfo
			{author} {\bibfnamefont {M.-H.}\ \bibnamefont {Lu}},\ and\ \bibinfo {author}
			{\bibfnamefont {J.}~\bibnamefont {Christensen}},\ }\bibfield  {title}
		{\bibinfo {title} {Topological sound},\ }\href
		{https://doi.org/10.1038/s42005-018-0094-4} {\bibfield  {journal} {\bibinfo
				{journal} {Commun. Phys}\ }\textbf {\bibinfo {volume} {1}},\ \bibinfo {pages}
			{97} (\bibinfo {year} {2018})}\BibitemShut {NoStop}%
		\bibitem [{\citenamefont {Kondo}\ \emph {et~al.}(2020)\citenamefont {Kondo},
			\citenamefont {Akagi},\ and\ \citenamefont {Katsura}}]{kondo2020}%
		\BibitemOpen
		\bibfield  {author} {\bibinfo {author} {\bibfnamefont {H.}~\bibnamefont
				{Kondo}}, \bibinfo {author} {\bibfnamefont {Y.}~\bibnamefont {Akagi}},\ and\
			\bibinfo {author} {\bibfnamefont {H.}~\bibnamefont {Katsura}},\ }\bibfield
		{title} {\bibinfo {title} {Non-hermiticity and topological invariants of
				magnon bogoliubov–de gennes systems},\ }\href
		{https://doi.org/10.1093/ptep/ptaa151} {\bibfield  {journal} {\bibinfo
				{journal} {Prog. Theor. Exp. Phys.}\ }\textbf {\bibinfo {volume} {2020}},\
			\bibinfo {pages} {12A104} (\bibinfo {year} {2020})}\BibitemShut {NoStop}%
		\bibitem [{\citenamefont {Rudner}\ and\ \citenamefont
			{Levitov}(2009)}]{rudner2009PhysRevLett.102.065703}%
		\BibitemOpen
		\bibfield  {author} {\bibinfo {author} {\bibfnamefont {M.~S.}\ \bibnamefont
				{Rudner}}\ and\ \bibinfo {author} {\bibfnamefont {L.~S.}\ \bibnamefont
				{Levitov}},\ }\bibfield  {title} {\bibinfo {title} {Topological transition in
				a non-hermitian quantum walk},\ }\href
		{https://doi.org/10.1103/PhysRevLett.102.065703} {\bibfield  {journal}
			{\bibinfo  {journal} {Phys. Rev. Lett.}\ }\textbf {\bibinfo {volume} {102}},\
			\bibinfo {pages} {065703} (\bibinfo {year} {2009})}\BibitemShut {NoStop}%
		\bibitem [{\citenamefont {Esaki}\ \emph {et~al.}(2011)\citenamefont {Esaki},
			\citenamefont {Sato}, \citenamefont {Hasebe},\ and\ \citenamefont
			{Kohmoto}}]{esaki2011PhysRevB.84.205128}%
		\BibitemOpen
		\bibfield  {author} {\bibinfo {author} {\bibfnamefont {K.}~\bibnamefont
				{Esaki}}, \bibinfo {author} {\bibfnamefont {M.}~\bibnamefont {Sato}},
			\bibinfo {author} {\bibfnamefont {K.}~\bibnamefont {Hasebe}},\ and\ \bibinfo
			{author} {\bibfnamefont {M.}~\bibnamefont {Kohmoto}},\ }\bibfield  {title}
		{\bibinfo {title} {Edge states and topological phases in non-hermitian
				systems},\ }\href {https://doi.org/10.1103/PhysRevB.84.205128} {\bibfield
			{journal} {\bibinfo  {journal} {Phys. Rev. B}\ }\textbf {\bibinfo {volume}
				{84}},\ \bibinfo {pages} {205128} (\bibinfo {year} {2011})}\BibitemShut
		{NoStop}%
		\bibitem [{\citenamefont {Liang}\ and\ \citenamefont
			{Huang}(2013)}]{liang2013PhysRevA.87.012118}%
		\BibitemOpen
		\bibfield  {author} {\bibinfo {author} {\bibfnamefont {S.-D.}\ \bibnamefont
				{Liang}}\ and\ \bibinfo {author} {\bibfnamefont {G.-Y.}\ \bibnamefont
				{Huang}},\ }\bibfield  {title} {\bibinfo {title} {Topological invariance and
				global berry phase in non-hermitian systems},\ }\href
		{https://doi.org/10.1103/PhysRevA.87.012118} {\bibfield  {journal} {\bibinfo
				{journal} {Phys. Rev. A}\ }\textbf {\bibinfo {volume} {87}},\ \bibinfo
			{pages} {012118} (\bibinfo {year} {2013})}\BibitemShut {NoStop}%
		\bibitem [{\citenamefont {Lee}(2016)}]{lee2016PhysRevLett.116.133903}%
		\BibitemOpen
		\bibfield  {author} {\bibinfo {author} {\bibfnamefont {T.~E.}\ \bibnamefont
				{Lee}},\ }\bibfield  {title} {\bibinfo {title} {Anomalous edge state in a
				non-hermitian lattice},\ }\href
		{https://doi.org/10.1103/PhysRevLett.116.133903} {\bibfield  {journal}
			{\bibinfo  {journal} {Phys. Rev. Lett.}\ }\textbf {\bibinfo {volume} {116}},\
			\bibinfo {pages} {133903} (\bibinfo {year} {2016})}\BibitemShut {NoStop}%
		\bibitem [{\citenamefont {Leykam}\ and\ \citenamefont
			{Chong}(2016)}]{leykam2016PhysRevLett.117.143901}%
		\BibitemOpen
		\bibfield  {author} {\bibinfo {author} {\bibfnamefont {D.}~\bibnamefont
				{Leykam}}\ and\ \bibinfo {author} {\bibfnamefont {Y.~D.}\ \bibnamefont
				{Chong}},\ }\bibfield  {title} {\bibinfo {title} {Edge solitons in
				nonlinear-photonic topological insulators},\ }\href
		{https://doi.org/10.1103/PhysRevLett.117.143901} {\bibfield  {journal}
			{\bibinfo  {journal} {Phys. Rev. Lett.}\ }\textbf {\bibinfo {volume} {117}},\
			\bibinfo {pages} {143901} (\bibinfo {year} {2016})}\BibitemShut {NoStop}%
		\bibitem [{\citenamefont {Xiong}(2018)}]{xiong2018JPhysCommun.2.035043}%
		\BibitemOpen
		\bibfield  {author} {\bibinfo {author} {\bibfnamefont {Y.}~\bibnamefont
				{Xiong}},\ }\bibfield  {title} {\bibinfo {title} {Why does bulk boundary
				correspondence fail in some non-hermitian topological models},\ }\href
		{https://doi.org/10.1088/2399-6528/aab64a} {\bibfield  {journal} {\bibinfo
				{journal} {J. Phys. Commun}\ }\textbf {\bibinfo {volume} {2}},\ \bibinfo
			{pages} {035043} (\bibinfo {year} {2018})}\BibitemShut {NoStop}%
		\bibitem [{\citenamefont {Shen}\ \emph {et~al.}(2018)\citenamefont {Shen},
			\citenamefont {Zhen},\ and\ \citenamefont
			{Fu}}]{shen2018PhysRevLett.120.146402}%
		\BibitemOpen
		\bibfield  {author} {\bibinfo {author} {\bibfnamefont {H.}~\bibnamefont
				{Shen}}, \bibinfo {author} {\bibfnamefont {B.}~\bibnamefont {Zhen}},\ and\
			\bibinfo {author} {\bibfnamefont {L.}~\bibnamefont {Fu}},\ }\bibfield
		{title} {\bibinfo {title} {Topological band theory for non-hermitian
				hamiltonians},\ }\href {https://doi.org/10.1103/PhysRevLett.120.146402}
		{\bibfield  {journal} {\bibinfo  {journal} {Phys. Rev. Lett.}\ }\textbf
			{\bibinfo {volume} {120}},\ \bibinfo {pages} {146402} (\bibinfo {year}
			{2018})}\BibitemShut {NoStop}%
		\bibitem [{\citenamefont {Yao}\ and\ \citenamefont
			{Wang}(2018)}]{yao2018PhysRevLett.121.086803}%
		\BibitemOpen
		\bibfield  {author} {\bibinfo {author} {\bibfnamefont {S.}~\bibnamefont
				{Yao}}\ and\ \bibinfo {author} {\bibfnamefont {Z.}~\bibnamefont {Wang}},\
		}\bibfield  {title} {\bibinfo {title} {Edge states and topological invariants
				of non-hermitian systems},\ }\href
		{https://doi.org/10.1103/PhysRevLett.121.086803} {\bibfield  {journal}
			{\bibinfo  {journal} {Phys. Rev. Lett.}\ }\textbf {\bibinfo {volume} {121}},\
			\bibinfo {pages} {086803} (\bibinfo {year} {2018})}\BibitemShut {NoStop}%
		\bibitem [{\citenamefont {Gong}\ \emph {et~al.}(2018)\citenamefont {Gong},
			\citenamefont {Ashida}, \citenamefont {Kawabata}, \citenamefont {Takasan},
			\citenamefont {Higashikawa},\ and\ \citenamefont
			{Ueda}}]{gong2018PhysRevX.8.031079}%
		\BibitemOpen
		\bibfield  {author} {\bibinfo {author} {\bibfnamefont {Z.}~\bibnamefont
				{Gong}}, \bibinfo {author} {\bibfnamefont {Y.}~\bibnamefont {Ashida}},
			\bibinfo {author} {\bibfnamefont {K.}~\bibnamefont {Kawabata}}, \bibinfo
			{author} {\bibfnamefont {K.}~\bibnamefont {Takasan}}, \bibinfo {author}
			{\bibfnamefont {S.}~\bibnamefont {Higashikawa}},\ and\ \bibinfo {author}
			{\bibfnamefont {M.}~\bibnamefont {Ueda}},\ }\bibfield  {title} {\bibinfo
			{title} {Topological phases of non-hermitian systems},\ }\href
		{https://doi.org/10.1103/PhysRevX.8.031079} {\bibfield  {journal} {\bibinfo
				{journal} {Phys. Rev. X}\ }\textbf {\bibinfo {volume} {8}},\ \bibinfo {pages}
			{031079} (\bibinfo {year} {2018})}\BibitemShut {NoStop}%
		\bibitem [{\citenamefont {Kawabata}\ \emph {et~al.}(2019)\citenamefont
			{Kawabata}, \citenamefont {Shiozaki}, \citenamefont {Ueda},\ and\
			\citenamefont {Sato}}]{kawabata2019PhysRevX.9.041015}%
		\BibitemOpen
		\bibfield  {author} {\bibinfo {author} {\bibfnamefont {K.}~\bibnamefont
				{Kawabata}}, \bibinfo {author} {\bibfnamefont {K.}~\bibnamefont {Shiozaki}},
			\bibinfo {author} {\bibfnamefont {M.}~\bibnamefont {Ueda}},\ and\ \bibinfo
			{author} {\bibfnamefont {M.}~\bibnamefont {Sato}},\ }\bibfield  {title}
		{\bibinfo {title} {Symmetry and topology in non-hermitian physics},\ }\href
		{https://doi.org/10.1103/PhysRevX.9.041015} {\bibfield  {journal} {\bibinfo
				{journal} {Phys. Rev. X}\ }\textbf {\bibinfo {volume} {9}},\ \bibinfo {pages}
			{041015} (\bibinfo {year} {2019})}\BibitemShut {NoStop}%
		\bibitem [{\citenamefont {Zhou}\ and\ \citenamefont
			{Lee}(2019)}]{zhou2019PhysRevB.99.235112}%
		\BibitemOpen
		\bibfield  {author} {\bibinfo {author} {\bibfnamefont {H.}~\bibnamefont
				{Zhou}}\ and\ \bibinfo {author} {\bibfnamefont {J.~Y.}\ \bibnamefont {Lee}},\
		}\bibfield  {title} {\bibinfo {title} {Periodic table for topological bands
				with non-hermitian symmetries},\ }\href
		{https://doi.org/10.1103/PhysRevB.99.235112} {\bibfield  {journal} {\bibinfo
				{journal} {Phys. Rev. B}\ }\textbf {\bibinfo {volume} {99}},\ \bibinfo
			{pages} {235112} (\bibinfo {year} {2019})}\BibitemShut {NoStop}%
		\bibitem [{\citenamefont {Ghatak}\ and\ \citenamefont
			{Das}(2019)}]{ghatak2019JOfPhysCondMatt.31.263001}%
		\BibitemOpen
		\bibfield  {author} {\bibinfo {author} {\bibfnamefont {A.}~\bibnamefont
				{Ghatak}}\ and\ \bibinfo {author} {\bibfnamefont {T.}~\bibnamefont {Das}},\
		}\bibfield  {title} {\bibinfo {title} {New topological invariants in
				non-hermitian systems},\ }\href {https://doi.org/10.1088/1361-648x/ab11b3}
		{\bibfield  {journal} {\bibinfo  {journal} {Journal of Physics: Condensed
					Matter}\ }\textbf {\bibinfo {volume} {31}},\ \bibinfo {pages} {263001}
			(\bibinfo {year} {2019})}\BibitemShut {NoStop}%
		\bibitem [{\citenamefont {Ashida}\ \emph {et~al.}(2020)\citenamefont {Ashida},
			\citenamefont {Gong},\ and\ \citenamefont {Ueda}}]{ashida2020}%
		\BibitemOpen
		\bibfield  {author} {\bibinfo {author} {\bibfnamefont {Y.}~\bibnamefont
				{Ashida}}, \bibinfo {author} {\bibfnamefont {Z.}~\bibnamefont {Gong}},\ and\
			\bibinfo {author} {\bibfnamefont {M.}~\bibnamefont {Ueda}},\ }\bibfield
		{title} {\bibinfo {title} {Non-hermitian physics},\ }\href
		{https://doi.org/10.1080/00018732.2021.1876991} {\bibfield  {journal}
			{\bibinfo  {journal} {Advances in Physics}\ }\textbf {\bibinfo {volume}
				{69}},\ \bibinfo {pages} {249} (\bibinfo {year} {2020})}\BibitemShut
		{NoStop}%
		\bibitem [{\citenamefont {Rudner}\ and\ \citenamefont
			{Levitov}(2019)}]{kawabata2019NatureCommunications.10.297}%
		\BibitemOpen
		\bibfield  {author} {\bibinfo {author} {\bibfnamefont {M.~S.}\ \bibnamefont
				{Rudner}}\ and\ \bibinfo {author} {\bibfnamefont {L.~S.}\ \bibnamefont
				{Levitov}},\ }\bibfield  {title} {\bibinfo {title} {Topological unification
				of time-reversal and particle-hole symmetries in non-hermitian physics},\
		}\href {https://doi.org/10.1038/s41467-018-08254-y} {\bibfield  {journal}
			{\bibinfo  {journal} {Nature Communications}\ }\textbf {\bibinfo {volume}
				{10}},\ \bibinfo {pages} {297} (\bibinfo {year} {2019})}\BibitemShut
		{NoStop}%
		\bibitem [{\citenamefont {Altland}\ and\ \citenamefont
			{Zirnbauer}(1997)}]{altland1997PhysRevB.55.1142}%
		\BibitemOpen
		\bibfield  {author} {\bibinfo {author} {\bibfnamefont {A.}~\bibnamefont
				{Altland}}\ and\ \bibinfo {author} {\bibfnamefont {M.~R.}\ \bibnamefont
				{Zirnbauer}},\ }\bibfield  {title} {\bibinfo {title} {Nonstandard symmetry
				classes in mesoscopic normal-superconducting hybrid structures},\ }\href
		{https://doi.org/10.1103/PhysRevB.55.1142} {\bibfield  {journal} {\bibinfo
				{journal} {Phys. Rev. B}\ }\textbf {\bibinfo {volume} {55}},\ \bibinfo
			{pages} {1142} (\bibinfo {year} {1997})}\BibitemShut {NoStop}%
		\bibitem [{\citenamefont {Dyson}(1962)}]{dyson1962JMathPhys.3.1199}%
		\BibitemOpen
		\bibfield  {author} {\bibinfo {author} {\bibfnamefont {F.~J.}\ \bibnamefont
				{Dyson}},\ }\href@noop {} {\bibfield  {journal} {\bibinfo  {journal} {J.
					Math. Phys.}\ }\textbf {\bibinfo {volume} {3}},\ \bibinfo {pages} {1199}
			(\bibinfo {year} {1962})}\BibitemShut {NoStop}%
		\bibitem [{\citenamefont {Mehta}()}]{mehta1991}%
		\BibitemOpen
		\bibfield  {author} {\bibinfo {author} {\bibfnamefont {M.~L.}\ \bibnamefont
				{Mehta}},\ }\href@noop {} {\bibinfo  {journal} {\textit{ Random Matrices}.
				(Academic Press, San Diego, 1991)}\ }\BibitemShut {NoStop}%
		\bibitem [{\citenamefont {Bernard}\ and\ \citenamefont
			{LeClair}()}]{bernardAmdLeClair2002.207}%
		\BibitemOpen
		\bibfield  {journal} {  }\bibfield  {author} {\bibinfo {author} {\bibfnamefont
				{D.}~\bibnamefont {Bernard}}\ and\ \bibinfo {author} {\bibfnamefont
				{A.}~\bibnamefont {LeClair}},\ }\bibfield  {title} {\bibinfo {title} {A
				classification of non-hermitian random matrices},\ }\href
		{https://arxiv.org/abs/cond-mat/0110649} {\bibinfo  {journal} {in
				\textit{Statistical Field Theories}, edited by A. Cappelli and G. Mussardo
				(Springer Netherland, Dordrecht, 2002)}\ ,\ \bibinfo {pages} {p.
				207}}\BibitemShut {NoStop}%
		\bibitem [{\citenamefont {Sato}\ \emph {et~al.}(2012)\citenamefont {Sato},
			\citenamefont {Hasebe}, \citenamefont {Esaki},\ and\ \citenamefont
			{Kohmoto}}]{sato2012ProgressOfTheoreticalPhysics.127.937}%
		\BibitemOpen
		\bibfield  {journal} {  }\bibfield  {author} {\bibinfo {author} {\bibfnamefont
				{M.}~\bibnamefont {Sato}}, \bibinfo {author} {\bibfnamefont {K.}~\bibnamefont
				{Hasebe}}, \bibinfo {author} {\bibfnamefont {K.}~\bibnamefont {Esaki}},\ and\
			\bibinfo {author} {\bibfnamefont {M.}~\bibnamefont {Kohmoto}},\ }\bibfield
		{title} {\bibinfo {title} {{Time-Reversal Symmetry in Non-Hermitian
					Systems}},\ }\href {https://doi.org/10.1143/PTP.127.937} {\bibfield
			{journal} {\bibinfo  {journal} {Progress of Theoretical Physics}\ }\textbf
			{\bibinfo {volume} {127}},\ \bibinfo {pages} {937} (\bibinfo {year}
			{2012})}\BibitemShut {NoStop}%
		\bibitem [{\citenamefont {Lieu}(2018)}]{lieu2018PhysRevB.98.115135}%
		\BibitemOpen
		\bibfield  {author} {\bibinfo {author} {\bibfnamefont {S.}~\bibnamefont
				{Lieu}},\ }\bibfield  {title} {\bibinfo {title} {Topological symmetry classes
				for non-hermitian models and connections to the bosonic bogoliubov--de gennes
				equation},\ }\href {https://doi.org/10.1103/PhysRevB.98.115135} {\bibfield
			{journal} {\bibinfo  {journal} {Phys. Rev. B}\ }\textbf {\bibinfo {volume}
				{98}},\ \bibinfo {pages} {115135} (\bibinfo {year} {2018})}\BibitemShut
		{NoStop}%
		\bibitem [{\citenamefont {Barnett}(2013)}]{barnett13PhysRevA.88.063631}%
		\BibitemOpen
		\bibfield  {author} {\bibinfo {author} {\bibfnamefont {R.}~\bibnamefont
				{Barnett}},\ }\bibfield  {title} {\bibinfo {title} {Edge-state instabilities
				of bosons in a topological band},\ }\href
		{https://doi.org/10.1103/PhysRevA.88.063631} {\bibfield  {journal} {\bibinfo
				{journal} {Phys. Rev. A}\ }\textbf {\bibinfo {volume} {88}},\ \bibinfo
			{pages} {063631} (\bibinfo {year} {2013})}\BibitemShut {NoStop}%
		\bibitem [{\citenamefont {Galilo}\ \emph {et~al.}(2015)\citenamefont {Galilo},
			\citenamefont {Lee},\ and\ \citenamefont
			{Barnett}}]{galilo15PhysRevLett.115.245302}%
		\BibitemOpen
		\bibfield  {author} {\bibinfo {author} {\bibfnamefont {B.}~\bibnamefont
				{Galilo}}, \bibinfo {author} {\bibfnamefont {D.~K.~K.}\ \bibnamefont {Lee}},\
			and\ \bibinfo {author} {\bibfnamefont {R.}~\bibnamefont {Barnett}},\
		}\bibfield  {title} {\bibinfo {title} {Selective population of edge states in
				a 2d topological band system},\ }\href
		{https://doi.org/10.1103/PhysRevLett.115.245302} {\bibfield  {journal}
			{\bibinfo  {journal} {Phys. Rev. Lett.}\ }\textbf {\bibinfo {volume} {115}},\
			\bibinfo {pages} {245302} (\bibinfo {year} {2015})}\BibitemShut {NoStop}%
		\bibitem [{\citenamefont {Engelhardt}\ \emph {et~al.}(2016)\citenamefont
			{Engelhardt}, \citenamefont {Benito}, \citenamefont {Platero},\ and\
			\citenamefont {Brandes}}]{engelhardt2016PhysRevLett.117.045302}%
		\BibitemOpen
		\bibfield  {author} {\bibinfo {author} {\bibfnamefont {G.}~\bibnamefont
				{Engelhardt}}, \bibinfo {author} {\bibfnamefont {M.}~\bibnamefont {Benito}},
			\bibinfo {author} {\bibfnamefont {G.}~\bibnamefont {Platero}},\ and\ \bibinfo
			{author} {\bibfnamefont {T.}~\bibnamefont {Brandes}},\ }\bibfield  {title}
		{\bibinfo {title} {Topological instabilities in ac-driven bosonic systems},\
		}\href {https://doi.org/10.1103/PhysRevLett.117.045302} {\bibfield  {journal}
			{\bibinfo  {journal} {Phys. Rev. Lett.}\ }\textbf {\bibinfo {volume} {117}},\
			\bibinfo {pages} {045302} (\bibinfo {year} {2016})}\BibitemShut {NoStop}%
		\bibitem [{\citenamefont {Peano}\ \emph {et~al.}(2016)\citenamefont {Peano},
			\citenamefont {Houde}, \citenamefont {Marquardt},\ and\ \citenamefont
			{Clerk}}]{peano2016PhysRevX.6.041026}%
		\BibitemOpen
		\bibfield  {author} {\bibinfo {author} {\bibfnamefont {V.}~\bibnamefont
				{Peano}}, \bibinfo {author} {\bibfnamefont {M.}~\bibnamefont {Houde}},
			\bibinfo {author} {\bibfnamefont {F.}~\bibnamefont {Marquardt}},\ and\
			\bibinfo {author} {\bibfnamefont {A.~A.}\ \bibnamefont {Clerk}},\ }\bibfield
		{title} {\bibinfo {title} {Topological quantum fluctuations and traveling
				wave amplifiers},\ }\href {https://doi.org/10.1103/PhysRevX.6.041026}
		{\bibfield  {journal} {\bibinfo  {journal} {Phys. Rev. X}\ }\textbf {\bibinfo
				{volume} {6}},\ \bibinfo {pages} {041026} (\bibinfo {year}
			{2016})}\BibitemShut {NoStop}%
		\bibitem [{\citenamefont {Ling}\ and\ \citenamefont
			{Kain}(2021)}]{ling2021PhysRevA.104.013305}%
		\BibitemOpen
		\bibfield  {author} {\bibinfo {author} {\bibfnamefont {H.~Y.}\ \bibnamefont
				{Ling}}\ and\ \bibinfo {author} {\bibfnamefont {B.}~\bibnamefont {Kain}},\
		}\bibfield  {title} {\bibinfo {title} {Selection rule for topological
				amplifiers in bogoliubov--de gennes systems},\ }\href
		{https://doi.org/10.1103/PhysRevA.104.013305} {\bibfield  {journal} {\bibinfo
				{journal} {Phys. Rev. A}\ }\textbf {\bibinfo {volume} {104}},\ \bibinfo
			{pages} {013305} (\bibinfo {year} {2021})}\BibitemShut {NoStop}%
		\bibitem [{\citenamefont {Schrieffer}(1964)}]{schrieffer64Book}%
		\BibitemOpen
		\bibfield  {author} {\bibinfo {author} {\bibfnamefont {J.~R.}\ \bibnamefont
				{Schrieffer}},\ }\bibfield  {title} {\bibinfo {title} {Theory of
				superconductivity},\ }\href@noop {} {\bibfield  {journal} {\bibinfo
				{journal} {Benjamin, Reading, Massachusetts}\ } (\bibinfo {year}
			{1964})}\BibitemShut {NoStop}%
		\bibitem [{\citenamefont {Streda}(1982)}]{streda1982JPhysC.15.L717}%
		\BibitemOpen
		\bibfield  {author} {\bibinfo {author} {\bibfnamefont {P.}~\bibnamefont
				{Streda}},\ }\bibfield  {title} {\bibinfo {title} {Theory of quantised hall
				conductivity in two dimensions},\ }\href
		{https://doi.org/10.1088/0022-3719/15/22/005} {\bibfield  {journal} {\bibinfo
				{journal} {Journal of Physics C: Solid State Physics}\ }\textbf {\bibinfo
				{volume} {15}},\ \bibinfo {pages} {L717} (\bibinfo {year}
			{1982})}\BibitemShut {NoStop}%
		\bibitem [{\citenamefont {Hatsugai}(1993)}]{hatsugai1993PhysRevLett.71.3697}%
		\BibitemOpen
		\bibfield  {author} {\bibinfo {author} {\bibfnamefont {Y.}~\bibnamefont
				{Hatsugai}},\ }\bibfield  {title} {\bibinfo {title} {Chern number and edge
				states in the integer quantum hall effect},\ }\href
		{https://doi.org/10.1103/PhysRevLett.71.3697} {\bibfield  {journal} {\bibinfo
				{journal} {Phys. Rev. Lett.}\ }\textbf {\bibinfo {volume} {71}},\ \bibinfo
			{pages} {3697} (\bibinfo {year} {1993})}\BibitemShut {NoStop}%
		\bibitem [{\citenamefont {Asb\'{o}th}\ \emph {et~al.}()\citenamefont
			{Asb\'{o}th}, \citenamefont {Oroszl\'{a}ny},\ and\ \citenamefont
			{P\'{a}lyi}}]{asboth2016}%
		\BibitemOpen
		\bibfield  {author} {\bibinfo {author} {\bibfnamefont {J.~K.}\ \bibnamefont
				{Asb\'{o}th}}, \bibinfo {author} {\bibfnamefont {L.}~\bibnamefont
				{Oroszl\'{a}ny}},\ and\ \bibinfo {author} {\bibfnamefont {A.}~\bibnamefont
				{P\'{a}lyi}},\ }\href@noop {} {\bibinfo  {journal} {\textit{ A Short Course
					on Topological insulators} (Springer Heidelberg, Germany, 2016)}\
		}\BibitemShut {NoStop}%
		\bibitem [{\citenamefont {Creutz}\ and\ \citenamefont
			{Horv\'ath}(1994)}]{creutz1994PhysRevD.50.2297}%
		\BibitemOpen
		\bibfield  {journal} {  }\bibfield  {author} {\bibinfo {author} {\bibfnamefont
				{M.}~\bibnamefont {Creutz}}\ and\ \bibinfo {author} {\bibfnamefont
				{I.}~\bibnamefont {Horv\'ath}},\ }\bibfield  {title} {\bibinfo {title}
			{Surface states and chiral symmetry on the lattice},\ }\href
		{https://doi.org/10.1103/PhysRevD.50.2297} {\bibfield  {journal} {\bibinfo
				{journal} {Phys. Rev. D}\ }\textbf {\bibinfo {volume} {50}},\ \bibinfo
			{pages} {2297} (\bibinfo {year} {1994})}\BibitemShut {NoStop}%
		\bibitem [{\citenamefont {K\"{o}nig}\ \emph {et~al.}(2008)\citenamefont
			{K\"{o}nig}, \citenamefont {Buhmann}, \citenamefont {Molenkamp},
			\citenamefont {Hughes}, \citenamefont {C.-X.~Liu},\ and\ \citenamefont
			{Zhang}}]{konig2008JPhysSocJapan.77.031007}%
		\BibitemOpen
		\bibfield  {author} {\bibinfo {author} {\bibfnamefont {M.}~\bibnamefont
				{K\"{o}nig}}, \bibinfo {author} {\bibfnamefont {H.}~\bibnamefont {Buhmann}},
			\bibinfo {author} {\bibfnamefont {L.~W.}\ \bibnamefont {Molenkamp}}, \bibinfo
			{author} {\bibfnamefont {T.}~\bibnamefont {Hughes}}, \bibinfo {author}
			{\bibfnamefont {X.-L.~Q.}\ \bibnamefont {C.-X.~Liu}},\ and\ \bibinfo {author}
			{\bibfnamefont {S.-C.}\ \bibnamefont {Zhang}},\ }\href@noop {} {\bibfield
			{journal} {\bibinfo  {journal} {J. Phys. Soc. Japan}\ }\textbf {\bibinfo
				{volume} {77}},\ \bibinfo {pages} {031007} (\bibinfo {year}
			{2008})}\BibitemShut {NoStop}%
		\bibitem [{\citenamefont {Wang}\ \emph
			{et~al.}(2009{\natexlab{b}})\citenamefont {Wang}, \citenamefont {Hao},\ and\
			\citenamefont {Zhang}}]{wang2009PhysRevB.80.115420}%
		\BibitemOpen
		\bibfield  {author} {\bibinfo {author} {\bibfnamefont {Z.}~\bibnamefont
				{Wang}}, \bibinfo {author} {\bibfnamefont {N.}~\bibnamefont {Hao}},\ and\
			\bibinfo {author} {\bibfnamefont {P.}~\bibnamefont {Zhang}},\ }\bibfield
		{title} {\bibinfo {title} {Topological winding properties of spin edge states
				in the kane-mele graphene model},\ }\href
		{https://doi.org/10.1103/PhysRevB.80.115420} {\bibfield  {journal} {\bibinfo
				{journal} {Phys. Rev. B}\ }\textbf {\bibinfo {volume} {80}},\ \bibinfo
			{pages} {115420} (\bibinfo {year} {2009}{\natexlab{b}})}\BibitemShut
		{NoStop}%
		\bibitem [{\citenamefont {Huang}\ and\ \citenamefont
			{Arovas}(2012)}]{huang2012arXiv:1205.6266}%
		\BibitemOpen
		\bibfield  {author} {\bibinfo {author} {\bibfnamefont {Z.}~\bibnamefont
				{Huang}}\ and\ \bibinfo {author} {\bibfnamefont {D.~P.}\ \bibnamefont
				{Arovas}},\ }\bibfield  {title} {\bibinfo {title} {Edge states, entanglement
				spectra, and wannier functions in haldane's honeycomb lattice model and its
				bilayer generalization},\ }\href@noop {} {\bibfield  {journal} {\bibinfo
				{journal} {arXiv:1205.6266}\ } (\bibinfo {year} {2012})}\BibitemShut
		{NoStop}%
		\bibitem [{\citenamefont {Doh}\ and\ \citenamefont
			{Jeon}(2013)}]{doh2013PhysRevB.88.245115}%
		\BibitemOpen
		\bibfield  {author} {\bibinfo {author} {\bibfnamefont {H.}~\bibnamefont
				{Doh}}\ and\ \bibinfo {author} {\bibfnamefont {G.~S.}\ \bibnamefont {Jeon}},\
		}\bibfield  {title} {\bibinfo {title} {Bifurcation of the edge-state width in
				a two-dimensional topological insulator},\ }\href
		{https://doi.org/10.1103/PhysRevB.88.245115} {\bibfield  {journal} {\bibinfo
				{journal} {Phys. Rev. B}\ }\textbf {\bibinfo {volume} {88}},\ \bibinfo
			{pages} {245115} (\bibinfo {year} {2013})}\BibitemShut {NoStop}%
		\bibitem [{\citenamefont {Pantale{\'{o}}n}\ and\ \citenamefont
			{Xian}(2017)}]{pantaleon2017JPhysCondensMatter.29.295701}%
		\BibitemOpen
		\bibfield  {author} {\bibinfo {author} {\bibfnamefont {P.~A.}\ \bibnamefont
				{Pantale{\'{o}}n}}\ and\ \bibinfo {author} {\bibfnamefont {Y.}~\bibnamefont
				{Xian}},\ }\bibfield  {title} {\bibinfo {title} {Analytical study of the edge
				states in the bosonic haldane model},\ }\href
		{https://doi.org/10.1088/1361-648x/aa75bf} {\bibfield  {journal} {\bibinfo
				{journal} {J. Phys.: Condens. Matter}\ }\textbf {\bibinfo {volume} {29}},\
			\bibinfo {pages} {295701} (\bibinfo {year} {2017})}\BibitemShut {NoStop}%
		\bibitem [{\citenamefont {Mizoguchi}\ and\ \citenamefont
			{Koma}(2021)}]{tomonari2021PhysRevB.103.195310}%
		\BibitemOpen
		\bibfield  {author} {\bibinfo {author} {\bibfnamefont {T.}~\bibnamefont
				{Mizoguchi}}\ and\ \bibinfo {author} {\bibfnamefont {T.}~\bibnamefont
				{Koma}},\ }\bibfield  {title} {\bibinfo {title} {Bulk-edge correspondence in
				two-dimensional topological semimetals: A transfer matrix study of antichiral
				edge modes},\ }\href {https://doi.org/10.1103/PhysRevB.103.195310} {\bibfield
			{journal} {\bibinfo  {journal} {Phys. Rev. B}\ }\textbf {\bibinfo {volume}
				{103}},\ \bibinfo {pages} {195310} (\bibinfo {year} {2021})}\BibitemShut
		{NoStop}%
		\bibitem [{\citenamefont {Blaizot}\ and\ \citenamefont
			{Ripka}()}]{blaizot96QuantumTheoryBook}%
		\BibitemOpen
		\bibfield  {author} {\bibinfo {author} {\bibfnamefont {J.-P.}\ \bibnamefont
				{Blaizot}}\ and\ \bibinfo {author} {\bibfnamefont {G.}~\bibnamefont
				{Ripka}},\ }\href@noop {} {\bibinfo  {journal} {\textit{Quanum theory of
					finite systems} (The MIT Press, Cambridge, MA, 1986)}\ }\BibitemShut
		{NoStop}%
		\bibitem [{\citenamefont {Garrison}\ and\ \citenamefont
			{Wright}(1988)}]{garrison1988PhysLettA.128.177}%
		\BibitemOpen
		\bibfield  {journal} {  }\bibfield  {author} {\bibinfo {author} {\bibfnamefont
				{J.~C.}\ \bibnamefont {Garrison}}\ and\ \bibinfo {author} {\bibfnamefont
				{E.~M.}\ \bibnamefont {Wright}},\ }\href
		{https://www.sciencedirect.com/science/article/abs/pii/037596018890905X}
		{\bibfield  {journal} {\bibinfo  {journal} {Phys. Lett. A}\ }\textbf
			{\bibinfo {volume} {128}},\ \bibinfo {pages} {177} (\bibinfo {year}
			{1988})}\BibitemShut {NoStop}%
		\bibitem [{\citenamefont {Fukui}\ \emph {et~al.}(2005)\citenamefont {Fukui},
			\citenamefont {Hatsugai},\ and\ \citenamefont
			{Suzuki}}]{fukui2005JPSJ.74.1674}%
		\BibitemOpen
		\bibfield  {author} {\bibinfo {author} {\bibfnamefont {T.}~\bibnamefont
				{Fukui}}, \bibinfo {author} {\bibfnamefont {Y.}~\bibnamefont {Hatsugai}},\
			and\ \bibinfo {author} {\bibfnamefont {H.}~\bibnamefont {Suzuki}},\
		}\bibfield  {title} {\bibinfo {title} {Chern numbers in discretized brillouin
				zone: Efficient method of computing (spin) hall conductances},\ }\href
		{https://doi.org/10.1143/JPSJ.74.1674} {\bibfield  {journal} {\bibinfo
				{journal} {Journal of the Physical Society of Japan}\ }\textbf {\bibinfo
				{volume} {74}},\ \bibinfo {pages} {1674} (\bibinfo {year}
			{2005})}\BibitemShut {NoStop}%
		\bibitem [{\citenamefont {Berry}(1984)}]{berry1984BerryPhase}%
		\BibitemOpen
		\bibfield  {author} {\bibinfo {author} {\bibfnamefont {M.~V.}\ \bibnamefont
				{Berry}},\ }\href {http://www.physics.mcgill.ca/~keshav/555/berryphase2.pdf}
		{\bibfield  {journal} {\bibinfo  {journal} {Proc. R. Soc. A}\ }\textbf
			{\bibinfo {volume} {392}},\ \bibinfo {pages} {45} (\bibinfo {year}
			{1984})}\BibitemShut {NoStop}%
		\bibitem [{\citenamefont {Fruchart}\ and\ \citenamefont
			{Carpentier}(2013)}]{fruchart2013ComptesRendusPhysique.14.779}%
		\BibitemOpen
		\bibfield  {author} {\bibinfo {author} {\bibfnamefont {M.}~\bibnamefont
				{Fruchart}}\ and\ \bibinfo {author} {\bibfnamefont {D.}~\bibnamefont
				{Carpentier}},\ }\bibfield  {title} {\bibinfo {title} {An introduction to
				topological insulators},\ }\href
		{https://doi.org/https://doi.org/10.1016/j.crhy.2013.09.013} {\bibfield
			{journal} {\bibinfo  {journal} {Comptes Rendus Physique}\ }\textbf {\bibinfo
				{volume} {14}},\ \bibinfo {pages} {779 } (\bibinfo {year}
			{2013})}\BibitemShut {NoStop}%
		\bibitem [{\citenamefont {Furukawa}\ and\ \citenamefont
			{Ueda}(2015)}]{furukawa2017NewJournalOfPhysics.17.115014}%
		\BibitemOpen
		\bibfield  {author} {\bibinfo {author} {\bibfnamefont {S.}~\bibnamefont
				{Furukawa}}\ and\ \bibinfo {author} {\bibfnamefont {M.}~\bibnamefont
				{Ueda}},\ }\bibfield  {title} {\bibinfo {title} {Excitation band topology and
				edge matter waves in bose-einstein condensates in optical lattices},\ }\href
		{http://stacks.iop.org/1367-2630/17/i=11/a=115014} {\bibfield  {journal}
			{\bibinfo  {journal} {New Journal of Physics}\ }\textbf {\bibinfo {volume}
				{17}},\ \bibinfo {pages} {115014} (\bibinfo {year} {2015})}\BibitemShut
		{NoStop}%
		\bibitem [{\citenamefont {Schomerus}(2013)}]{schomerus2013OptLett.38.1912}%
		\BibitemOpen
		\bibfield  {author} {\bibinfo {author} {\bibfnamefont {H.}~\bibnamefont
				{Schomerus}},\ }\bibfield  {title} {\bibinfo {title} {Topologically protected
				midgap states in complex photonic lattices},\ }\href
		{https://doi.org/10.1364/OL.38.001912} {\bibfield  {journal} {\bibinfo
				{journal} {Opt. Lett.}\ }\textbf {\bibinfo {volume} {38}},\ \bibinfo {pages}
			{1912} (\bibinfo {year} {2013})}\BibitemShut {NoStop}%
		\bibitem [{\citenamefont {Poli}\ \emph {et~al.}(2015)\citenamefont {Poli},
			\citenamefont {Bellec}, \citenamefont {Kuhl}, \citenamefont {Mortessagne},\
			and\ \citenamefont {Schomerus}}]{poli2015NatCommun.6.6710}%
		\BibitemOpen
		\bibfield  {author} {\bibinfo {author} {\bibfnamefont {C.}~\bibnamefont
				{Poli}}, \bibinfo {author} {\bibfnamefont {M.}~\bibnamefont {Bellec}},
			\bibinfo {author} {\bibfnamefont {U.}~\bibnamefont {Kuhl}}, \bibinfo {author}
			{\bibfnamefont {F.}~\bibnamefont {Mortessagne}},\ and\ \bibinfo {author}
			{\bibfnamefont {H.}~\bibnamefont {Schomerus}},\ }\bibfield  {title} {\bibinfo
			{title} {Selective enhancement of topologically induced interface states in a
				dielectric resonator chain},\ }\href {https://doi.org/10.1038/ncomms7710}
		{\bibfield  {journal} {\bibinfo  {journal} {Nat. Commun.}\ }\textbf {\bibinfo
				{volume} {6}},\ \bibinfo {pages} {6710} (\bibinfo {year} {2015})}\BibitemShut
		{NoStop}%
		\bibitem [{\citenamefont {St-Jean}\ \emph {et~al.}(2017)\citenamefont
			{St-Jean}, \citenamefont {Goblot}, \citenamefont {Galopin}, \citenamefont
			{Lemaître}, \citenamefont {Ozawa}, \citenamefont {Le~Gratiet}, \citenamefont
			{Sagnes}, \citenamefont {Bloch},\ and\ \citenamefont
			{Amo}}]{st-jean2017NatPhoton.11.651}%
		\BibitemOpen
		\bibfield  {author} {\bibinfo {author} {\bibfnamefont {P.}~\bibnamefont
				{St-Jean}}, \bibinfo {author} {\bibfnamefont {V.}~\bibnamefont {Goblot}},
			\bibinfo {author} {\bibfnamefont {E.}~\bibnamefont {Galopin}}, \bibinfo
			{author} {\bibfnamefont {A.}~\bibnamefont {Lemaître}}, \bibinfo {author}
			{\bibfnamefont {T.}~\bibnamefont {Ozawa}}, \bibinfo {author} {\bibfnamefont
				{L.}~\bibnamefont {Le~Gratiet}}, \bibinfo {author} {\bibfnamefont
				{I.}~\bibnamefont {Sagnes}}, \bibinfo {author} {\bibfnamefont
				{J.}~\bibnamefont {Bloch}},\ and\ \bibinfo {author} {\bibfnamefont
				{A.}~\bibnamefont {Amo}},\ }\bibfield  {title} {\bibinfo {title} {Lasing in
				topological edge states of a one-dimensional lattice},\ }\href
		{https://doi.org/10.1038/s41566-017-0006-2} {\bibfield  {journal} {\bibinfo
				{journal} {Nat. Photon}\ }\textbf {\bibinfo {volume} {11}},\ \bibinfo {pages}
			{651} (\bibinfo {year} {2017})}\BibitemShut {NoStop}%
		\bibitem [{\citenamefont {Zhao}\ \emph {et~al.}(2018)\citenamefont {Zhao},
			\citenamefont {Miao}, \citenamefont {Teimourpour}, \citenamefont {Malzard},
			\citenamefont {El-Ganainy}, \citenamefont {Schomerus},\ and\ \citenamefont
			{Feng}}]{zhao2018NatCommun.9.981}%
		\BibitemOpen
		\bibfield  {author} {\bibinfo {author} {\bibfnamefont {H.}~\bibnamefont
				{Zhao}}, \bibinfo {author} {\bibfnamefont {P.}~\bibnamefont {Miao}}, \bibinfo
			{author} {\bibfnamefont {M.~H.}\ \bibnamefont {Teimourpour}}, \bibinfo
			{author} {\bibfnamefont {S.}~\bibnamefont {Malzard}}, \bibinfo {author}
			{\bibfnamefont {R.}~\bibnamefont {El-Ganainy}}, \bibinfo {author}
			{\bibfnamefont {H.}~\bibnamefont {Schomerus}},\ and\ \bibinfo {author}
			{\bibfnamefont {L.}~\bibnamefont {Feng}},\ }\bibfield  {title} {\bibinfo
			{title} {Topological hybrid silicon microlasers},\ }\href
		{https://doi.org/10.1038/s41467-018-03434-2} {\bibfield  {journal} {\bibinfo
				{journal} {Nat. Commun.}\ }\textbf {\bibinfo {volume} {9}},\ \bibinfo {pages}
			{981} (\bibinfo {year} {2018})}\BibitemShut {NoStop}%
		\bibitem [{\citenamefont {Harari}\ \emph {et~al.}(2018)\citenamefont {Harari},
			\citenamefont {Bandres}, \citenamefont {Lumer}, \citenamefont {Rechtsman},
			\citenamefont {Chong}, \citenamefont {Khajavikhan}, \citenamefont
			{Christodoulides},\ and\ \citenamefont
			{Segev}}]{harari2018Science.359.eaar4003}%
		\BibitemOpen
		\bibfield  {author} {\bibinfo {author} {\bibfnamefont {G.}~\bibnamefont
				{Harari}}, \bibinfo {author} {\bibfnamefont {M.~A.}\ \bibnamefont {Bandres}},
			\bibinfo {author} {\bibfnamefont {Y.}~\bibnamefont {Lumer}}, \bibinfo
			{author} {\bibfnamefont {M.~C.}\ \bibnamefont {Rechtsman}}, \bibinfo {author}
			{\bibfnamefont {Y.~D.}\ \bibnamefont {Chong}}, \bibinfo {author}
			{\bibfnamefont {M.}~\bibnamefont {Khajavikhan}}, \bibinfo {author}
			{\bibfnamefont {D.~N.}\ \bibnamefont {Christodoulides}},\ and\ \bibinfo
			{author} {\bibfnamefont {M.}~\bibnamefont {Segev}},\ }\bibfield  {title}
		{\bibinfo {title} {Topological insulator laser: Theory},\ }\href
		{https://doi.org/10.1126/science.aar4003} {\bibfield  {journal} {\bibinfo
				{journal} {Science}\ }\textbf {\bibinfo {volume} {359}},\ \bibinfo {pages}
			{eaar4003} (\bibinfo {year} {2018})}\BibitemShut {NoStop}%
		\bibitem [{\citenamefont {Bandres}\ \emph {et~al.}(2018)\citenamefont
			{Bandres}, \citenamefont {Wittek}, \citenamefont {Harari}, \citenamefont
			{Parto}, \citenamefont {Ren}, \citenamefont {Segev}, \citenamefont
			{Christodoulides},\ and\ \citenamefont
			{Khajavikhan}}]{bandres2018Science.359.eaar4005}%
		\BibitemOpen
		\bibfield  {author} {\bibinfo {author} {\bibfnamefont {M.~A.}\ \bibnamefont
				{Bandres}}, \bibinfo {author} {\bibfnamefont {S.}~\bibnamefont {Wittek}},
			\bibinfo {author} {\bibfnamefont {G.}~\bibnamefont {Harari}}, \bibinfo
			{author} {\bibfnamefont {M.}~\bibnamefont {Parto}}, \bibinfo {author}
			{\bibfnamefont {J.}~\bibnamefont {Ren}}, \bibinfo {author} {\bibfnamefont
				{M.}~\bibnamefont {Segev}}, \bibinfo {author} {\bibfnamefont {D.~N.}\
				\bibnamefont {Christodoulides}},\ and\ \bibinfo {author} {\bibfnamefont
				{M.}~\bibnamefont {Khajavikhan}},\ }\bibfield  {title} {\bibinfo {title}
			{Topological insulator laser: Experiments},\ }\href
		{https://doi.org/10.1126/science.aar4005} {\bibfield  {journal} {\bibinfo
				{journal} {Science}\ }\textbf {\bibinfo {volume} {359}},\ \bibinfo {pages}
			{eaar4005} (\bibinfo {year} {2018})}\BibitemShut {NoStop}%
		\bibitem [{\citenamefont {Bahari}\ \emph {et~al.}(2017)\citenamefont {Bahari},
			\citenamefont {Ndao}, \citenamefont {Vallini}, \citenamefont {El~Amili},
			\citenamefont {Fainman},\ and\ \citenamefont
			{Kant{\'e}}}]{bahari2017Science.358.636}%
		\BibitemOpen
		\bibfield  {author} {\bibinfo {author} {\bibfnamefont {B.}~\bibnamefont
				{Bahari}}, \bibinfo {author} {\bibfnamefont {A.}~\bibnamefont {Ndao}},
			\bibinfo {author} {\bibfnamefont {F.}~\bibnamefont {Vallini}}, \bibinfo
			{author} {\bibfnamefont {A.}~\bibnamefont {El~Amili}}, \bibinfo {author}
			{\bibfnamefont {Y.}~\bibnamefont {Fainman}},\ and\ \bibinfo {author}
			{\bibfnamefont {B.}~\bibnamefont {Kant{\'e}}},\ }\bibfield  {title} {\bibinfo
			{title} {Nonreciprocal lasing in topological cavities of arbitrary
				geometries},\ }\href {https://doi.org/10.1126/science.aao4551} {\bibfield
			{journal} {\bibinfo  {journal} {Science}\ }\textbf {\bibinfo {volume}
				{358}},\ \bibinfo {pages} {636} (\bibinfo {year} {2017})}\BibitemShut
		{NoStop}%
		\bibitem [{\citenamefont {Wu}\ and\ \citenamefont
			{Niu}(2001)}]{wu2001PhysRevA.64.061603}%
		\BibitemOpen
		\bibfield  {author} {\bibinfo {author} {\bibfnamefont {B.}~\bibnamefont
				{Wu}}\ and\ \bibinfo {author} {\bibfnamefont {Q.}~\bibnamefont {Niu}},\
		}\bibfield  {title} {\bibinfo {title} {Landau and dynamical instabilities of
				the superflow of bose-einstein condensates in optical lattices},\ }\href
		{https://doi.org/10.1103/PhysRevA.64.061603} {\bibfield  {journal} {\bibinfo
				{journal} {Phys. Rev. A}\ }\textbf {\bibinfo {volume} {64}},\ \bibinfo
			{pages} {061603} (\bibinfo {year} {2001})}\BibitemShut {NoStop}%
		\bibitem [{\citenamefont {Kawaguchi}\ and\ \citenamefont
			{Ohmi}(2004)}]{kawaguchi2004PhysRevA.70.043610}%
		\BibitemOpen
		\bibfield  {author} {\bibinfo {author} {\bibfnamefont {Y.}~\bibnamefont
				{Kawaguchi}}\ and\ \bibinfo {author} {\bibfnamefont {T.}~\bibnamefont
				{Ohmi}},\ }\bibfield  {title} {\bibinfo {title} {Splitting instability of a
				multiply charged vortex in a bose-einstein condensate},\ }\href
		{https://doi.org/10.1103/PhysRevA.70.043610} {\bibfield  {journal} {\bibinfo
				{journal} {Phys. Rev. A}\ }\textbf {\bibinfo {volume} {70}},\ \bibinfo
			{pages} {043610} (\bibinfo {year} {2004})}\BibitemShut {NoStop}%
		\bibitem [{\citenamefont {Nakamura}\ \emph {et~al.}(2008)\citenamefont
			{Nakamura}, \citenamefont {Mine}, \citenamefont {Okumura},\ and\
			\citenamefont {Yamanaka}}]{nakamura2008PhysRevA.77.043601}%
		\BibitemOpen
		\bibfield  {author} {\bibinfo {author} {\bibfnamefont {Y.}~\bibnamefont
				{Nakamura}}, \bibinfo {author} {\bibfnamefont {M.}~\bibnamefont {Mine}},
			\bibinfo {author} {\bibfnamefont {M.}~\bibnamefont {Okumura}},\ and\ \bibinfo
			{author} {\bibfnamefont {Y.}~\bibnamefont {Yamanaka}},\ }\bibfield  {title}
		{\bibinfo {title} {Condition for emergence of complex eigenvalues in the
				bogoliubov--de gennes equations},\ }\href
		{https://doi.org/10.1103/PhysRevA.77.043601} {\bibfield  {journal} {\bibinfo
				{journal} {Phys. Rev. A}\ }\textbf {\bibinfo {volume} {77}},\ \bibinfo
			{pages} {043601} (\bibinfo {year} {2008})}\BibitemShut {NoStop}%
		\bibitem [{\citenamefont {Weimann}\ \emph {et~al.}(2017)\citenamefont
			{Weimann}, \citenamefont {Kremer}, \citenamefont {Plotnik}, \citenamefont
			{Lumer}, \citenamefont {Nolte}, \citenamefont {Makris}, \citenamefont
			{Segev}, \citenamefont {Rechtsman},\ and\ \citenamefont
			{Szameit}}]{weimann2017NatureMaterials.16.433}%
		\BibitemOpen
		\bibfield  {author} {\bibinfo {author} {\bibfnamefont {S.}~\bibnamefont
				{Weimann}}, \bibinfo {author} {\bibfnamefont {M.}~\bibnamefont {Kremer}},
			\bibinfo {author} {\bibfnamefont {Y.}~\bibnamefont {Plotnik}}, \bibinfo
			{author} {\bibfnamefont {Y.}~\bibnamefont {Lumer}}, \bibinfo {author}
			{\bibfnamefont {S.}~\bibnamefont {Nolte}}, \bibinfo {author} {\bibfnamefont
				{K.~G.}\ \bibnamefont {Makris}}, \bibinfo {author} {\bibfnamefont
				{M.}~\bibnamefont {Segev}}, \bibinfo {author} {\bibfnamefont {M.~C.}\
				\bibnamefont {Rechtsman}},\ and\ \bibinfo {author} {\bibfnamefont
				{A.}~\bibnamefont {Szameit}},\ }\bibfield  {title} {\bibinfo {title}
			{Topologically protected bound states in photonic parity–time-symmetric
				crystals},\ }\href {https://doi.org/10.1038/nmat4811} {\bibfield  {journal}
			{\bibinfo  {journal} {Nature Materials}\ }\textbf {\bibinfo {volume} {16}},\
			\bibinfo {pages} {433} (\bibinfo {year} {2017})}\BibitemShut {NoStop}%
		\bibitem [{\citenamefont {Parto}\ \emph {et~al.}(2018)\citenamefont {Parto},
			\citenamefont {Wittek}, \citenamefont {Hodaei}, \citenamefont {Harari},
			\citenamefont {Bandres}, \citenamefont {Ren}, \citenamefont {Rechtsman},
			\citenamefont {Segev}, \citenamefont {Christodoulides},\ and\ \citenamefont
			{Khajavikhan}}]{parto2018PhysRevLett.120.113901}%
		\BibitemOpen
		\bibfield  {author} {\bibinfo {author} {\bibfnamefont {M.}~\bibnamefont
				{Parto}}, \bibinfo {author} {\bibfnamefont {S.}~\bibnamefont {Wittek}},
			\bibinfo {author} {\bibfnamefont {H.}~\bibnamefont {Hodaei}}, \bibinfo
			{author} {\bibfnamefont {G.}~\bibnamefont {Harari}}, \bibinfo {author}
			{\bibfnamefont {M.~A.}\ \bibnamefont {Bandres}}, \bibinfo {author}
			{\bibfnamefont {J.}~\bibnamefont {Ren}}, \bibinfo {author} {\bibfnamefont
				{M.~C.}\ \bibnamefont {Rechtsman}}, \bibinfo {author} {\bibfnamefont
				{M.}~\bibnamefont {Segev}}, \bibinfo {author} {\bibfnamefont {D.~N.}\
				\bibnamefont {Christodoulides}},\ and\ \bibinfo {author} {\bibfnamefont
				{M.}~\bibnamefont {Khajavikhan}},\ }\bibfield  {title} {\bibinfo {title}
			{Edge-mode lasing in 1d topological active arrays},\ }\href
		{https://doi.org/10.1103/PhysRevLett.120.113901} {\bibfield  {journal}
			{\bibinfo  {journal} {Phys. Rev. Lett.}\ }\textbf {\bibinfo {volume} {120}},\
			\bibinfo {pages} {113901} (\bibinfo {year} {2018})}\BibitemShut {NoStop}%
		\bibitem [{\citenamefont {Su}\ \emph {et~al.}(1979)\citenamefont {Su},
			\citenamefont {Schrieffer},\ and\ \citenamefont
			{Heeger}}]{su1979PhysRevLett.42.1698}%
		\BibitemOpen
		\bibfield  {author} {\bibinfo {author} {\bibfnamefont {W.~P.}\ \bibnamefont
				{Su}}, \bibinfo {author} {\bibfnamefont {J.~R.}\ \bibnamefont {Schrieffer}},\
			and\ \bibinfo {author} {\bibfnamefont {A.~J.}\ \bibnamefont {Heeger}},\
		}\bibfield  {title} {\bibinfo {title} {Solitons in polyacetylene},\ }\href
		{https://doi.org/10.1103/PhysRevLett.42.1698} {\bibfield  {journal} {\bibinfo
				{journal} {Phys. Rev. Lett.}\ }\textbf {\bibinfo {volume} {42}},\ \bibinfo
			{pages} {1698} (\bibinfo {year} {1979})}\BibitemShut {NoStop}%
		\bibitem [{\citenamefont {Sargent~III}\ \emph {et~al.}()\citenamefont
			{Sargent~III}, \citenamefont {Scully},\ and\ \citenamefont
			{Lamb~Jr.}}]{lamb1974laserPhysics}%
		\BibitemOpen
		\bibfield  {author} {\bibinfo {author} {\bibfnamefont {M.}~\bibnamefont
				{Sargent~III}}, \bibinfo {author} {\bibfnamefont {M.~O.}\ \bibnamefont
				{Scully}},\ and\ \bibinfo {author} {\bibfnamefont {W.~E.}\ \bibnamefont
				{Lamb~Jr.}},\ }\href@noop {} {\bibinfo  {journal} {\textit{Laser Physics}
				(CRC Press, New York, 1974)}\ }\BibitemShut {NoStop}%
		\bibitem [{\citenamefont {Ho}(1998)}]{ho1998PhysRevLett.81.742}%
		\BibitemOpen
		\bibfield  {journal} {  }\bibfield  {author} {\bibinfo {author} {\bibfnamefont
				{T.-L.}\ \bibnamefont {Ho}},\ }\bibfield  {title} {\bibinfo {title} {Spinor
				bose condensates in optical traps},\ }\href
		{https://doi.org/10.1103/PhysRevLett.81.742} {\bibfield  {journal} {\bibinfo
				{journal} {Phys. Rev. Lett.}\ }\textbf {\bibinfo {volume} {81}},\ \bibinfo
			{pages} {742} (\bibinfo {year} {1998})}\BibitemShut {NoStop}%
		\bibitem [{\citenamefont {Ohmi}\ and\ \citenamefont
			{Machida}(1998)}]{ohmi1998doi:10.1143/JPSJ.67.1822}%
		\BibitemOpen
		\bibfield  {author} {\bibinfo {author} {\bibfnamefont {T.}~\bibnamefont
				{Ohmi}}\ and\ \bibinfo {author} {\bibfnamefont {K.}~\bibnamefont {Machida}},\
		}\bibfield  {title} {\bibinfo {title} {Bose-einstein condensation with
				internal degrees of freedom in alkali atom gases},\ }\href
		{https://doi.org/10.1143/JPSJ.67.1822} {\bibfield  {journal} {\bibinfo
				{journal} {Journal of the Physical Society of Japan}\ }\textbf {\bibinfo
				{volume} {67}},\ \bibinfo {pages} {1822} (\bibinfo {year}
			{1998})}\BibitemShut {NoStop}%
		\bibitem [{\citenamefont {Law}\ \emph {et~al.}(1998)\citenamefont {Law},
			\citenamefont {Pu},\ and\ \citenamefont
			{Bigelow}}]{law1998PhysRevLett.81.5257}%
		\BibitemOpen
		\bibfield  {author} {\bibinfo {author} {\bibfnamefont {C.~K.}\ \bibnamefont
				{Law}}, \bibinfo {author} {\bibfnamefont {H.}~\bibnamefont {Pu}},\ and\
			\bibinfo {author} {\bibfnamefont {N.~P.}\ \bibnamefont {Bigelow}},\
		}\bibfield  {title} {\bibinfo {title} {Quantum spins mixing in spinor
				bose-einstein condensates},\ }\href
		{https://doi.org/10.1103/PhysRevLett.81.5257} {\bibfield  {journal} {\bibinfo
				{journal} {Phys. Rev. Lett.}\ }\textbf {\bibinfo {volume} {81}},\ \bibinfo
			{pages} {5257} (\bibinfo {year} {1998})}\BibitemShut {NoStop}%
		\bibitem [{\citenamefont {Stamper-Kurn}\ and\ \citenamefont
			{Ueda}(2013)}]{stamperKurn2013RevModPhys.85.1191}%
		\BibitemOpen
		\bibfield  {author} {\bibinfo {author} {\bibfnamefont {D.~M.}\ \bibnamefont
				{Stamper-Kurn}}\ and\ \bibinfo {author} {\bibfnamefont {M.}~\bibnamefont
				{Ueda}},\ }\bibfield  {title} {\bibinfo {title} {Spinor bose gases:
				Symmetries, magnetism, and quantum dynamics},\ }\href
		{https://doi.org/10.1103/RevModPhys.85.1191} {\bibfield  {journal} {\bibinfo
				{journal} {Rev. Mod. Phys.}\ }\textbf {\bibinfo {volume} {85}},\ \bibinfo
			{pages} {1191} (\bibinfo {year} {2013})}\BibitemShut {NoStop}%
		\bibitem [{\citenamefont {Gerbier}\ \emph {et~al.}(2006)\citenamefont
			{Gerbier}, \citenamefont {Widera}, \citenamefont {F\"olling}, \citenamefont
			{Mandel},\ and\ \citenamefont {Bloch}}]{gerbier2006PhysRevA.73.041602}%
		\BibitemOpen
		\bibfield  {author} {\bibinfo {author} {\bibfnamefont {F.}~\bibnamefont
				{Gerbier}}, \bibinfo {author} {\bibfnamefont {A.}~\bibnamefont {Widera}},
			\bibinfo {author} {\bibfnamefont {S.}~\bibnamefont {F\"olling}}, \bibinfo
			{author} {\bibfnamefont {O.}~\bibnamefont {Mandel}},\ and\ \bibinfo {author}
			{\bibfnamefont {I.}~\bibnamefont {Bloch}},\ }\bibfield  {title} {\bibinfo
			{title} {Resonant control of spin dynamics in ultracold quantum gases by
				microwave dressing},\ }\href {https://doi.org/10.1103/PhysRevA.73.041602}
		{\bibfield  {journal} {\bibinfo  {journal} {Phys. Rev. A}\ }\textbf {\bibinfo
				{volume} {73}},\ \bibinfo {pages} {041602} (\bibinfo {year}
			{2006})}\BibitemShut {NoStop}%
		\bibitem [{\citenamefont {Kain}\ and\ \citenamefont
			{Ling}(2014)}]{kain2014PhysRevA.90.063626}%
		\BibitemOpen
		\bibfield  {author} {\bibinfo {author} {\bibfnamefont {B.}~\bibnamefont
				{Kain}}\ and\ \bibinfo {author} {\bibfnamefont {H.~Y.}\ \bibnamefont
				{Ling}},\ }\bibfield  {title} {\bibinfo {title} {Nonequilibrium states of a
				quenched bose gas},\ }\href {https://doi.org/10.1103/PhysRevA.90.063626}
		{\bibfield  {journal} {\bibinfo  {journal} {Phys. Rev. A}\ }\textbf {\bibinfo
				{volume} {90}},\ \bibinfo {pages} {063626} (\bibinfo {year}
			{2014})}\BibitemShut {NoStop}%
	\end{thebibliography}
	
\end{document}